\documentclass{article}
\pdfoutput=1

\usepackage{natbib}
\usepackage{epubtk}
\usepackage{graphicx}
\usepackage{rotate}
\usepackage{amsmath}
\usepackage{wasysym}
\usepackage{amssymb}
\usepackage{yhmath} 
\usepackage{booktabs}
\usepackage{textcomp}
\usepackage{xspace}

\newcommand{\ve}[1]{\mathbf{#1}}

\newcommand{\hmc}[1]{\hat{\mathcal{#1}}}

\def\depp{(\ve{X}[\tau_c])}

\def\dtau{(\tau_c)}

\def\ex{\mathbf{e_x}}
\def\ez{\mathbf{e_z}}
\def\ey{\mathbf{e_y}}
\def\ep{\mathbf{e_{\rho}}}
\def\ea{\mathbf{e_{\alpha}}}
\def\eb{\mathbf{e_{\beta}}}
\def\exs{\mathbf{e_{x}^{*}}}

\def\eys{\mathbf{e_{y}^{*}}}
\def\els{\mathbf{e_{l}^{*}}}
\def\deg{^{\circ}}

\newcommand{\kms}{km~s\super{-1}\xspace}

\newcommand{\I}{\textit{I}\xspace}
\newcommand{\Q}{\textit{Q}\xspace}
\newcommand{\U}{\textit{U}\xspace}
\newcommand{\V}{\textit{V}\xspace}

\begin{document}

\title{Magnetic Structure of Sunspots}

\author{\epubtkAuthorData{Juan M.\ Borrero}{%
Kiepenheuer-Institut f\"ur Sonnenphysik \\
Sch\"oneckstr.\ 6, D-79104 Freiburg, Germany}{%
borrero@kis.uni-freiburg.de}{}
\and
\epubtkAuthorData{Kiyoshi Ichimoto}{%
Kwasan and Hida Observatories \\
Kyoto University, Yamashina, Kyoto 607-8471, Japan}{%
ichimoto@kwasan.kyoto-u.ac.jp}{}
}

\date{}
\maketitle

\begin{abstract}
In this review we give an overview about the current
state-of-knowledge of the magnetic field in sunspots from an
observational point of view. We start by offering a brief description
of tools that are most commonly employed to infer the magnetic field
in the solar atmosphere with emphasis in the photosphere of
sunspots. We then address separately the global and local magnetic
structure of sunspots, focusing on the implications of the current
observations for the different sunspots models, energy transport
mechanisms, extrapolations of the magnetic field towards the corona,
and other issues.
\end{abstract}

\epubtkKeywords{Solar physics, Magnetic fields, Sunspots}

\newpage
\tableofcontents

\newpage
\section{Introduction}
\label{section:introduction}

\subsection{Role of magnetic field in cosmic bodies}
\label{subsection:rolefield}

The role of the magnetic field has become firmly recognized in astrophysics
as humans discover the rich variety of phenomena present in the universe.
The largest constituent of cosmic bodies is in the state of plasma,
i.e., ionized gas, which interacts with magnetic fields. Through this
interaction the magnetic field is responsible for many of the structures
and dynamics that we observe with modern instrumentation. The fundamental
role of magnetic fields can be summarized as follows:

\begin{itemize}

\item By trapping charged particles, the magnetic field guides the
  plasma motions (for example, it confines plasma or suppresses the
  convective fluid motions) and generates a variety of density
  structures in universe.
\item The magnetic pressure causes the plasma to expand and makes it
  buoyant, thereby driving the emergence of magnetic loops into the
  tenuous `outer' atmosphere against the action of gravity.
\item The magnetic field guides (MHD) waves and transports energy and
  disturbances from a site of energy injection to other locations.
\item Magnetic fields inhibit the thermal conduction across them, and
  make the presence of multi-temperature structures possible in
  tenuous, high conductivity gas like stellar corona.
\item The magnetic field stores and releases energy that produces
  transient dynamic phenomena like flare explosions and plasma
  ejections.
\item The magnetic field also plays a crucial role in accelerating
  non-thermal particles to the relativistic regime in tenuous plasmas.

The magnetic field is, therefore, one of the fundamental ingredients of
the universe. In the case of the Sun, our nearest star, we observe
spectacular active and dynamic phenomena driven by magnetic fields
with their spatial, temporal, and spectral structures in
detail. Because of this, the Sun serves as an excellent plasma
laboratory and provides us a unique opportunity to study the
fundamental processes of the cosmical magnetohydrodynamics. Let us
make an addition on the list of the role of the magnetic field, i.e.:

\item The magnetic field produces or modifies the polarization
  property of the light emitted from, or absorbed by, cosmical bodies,
  making themselves measurable.

\end{itemize}

\subsection{Discovery of sunspot's magnetic field}
\label{subsection:discovery}

The key physical process that makes human be aware of the existence of magnetic fields in 
sunspots is the interaction between atoms and the magnetic field, i.e., the so called Zeeman effect,
discovered in laboratory by the young physicist Pieter Zeeman in The Netherlands by the end of the
19th century \citep{Zeeman1897a}. This effect describes how the electronic energy levels of 
an atom split in the presence of a magnetic field, giving raise to several absorption/emission
spectral lines where there was only one spectral line in the absence of the magnetic field.
In addition, the magnetic field modifies the polarization properties of the emitted/absorbed photons
in a manner that depends on the viewing angle between the observer and the magnetic field vector.

An initial hint of the presence of the Zeeman effect in the spectra of sunspots 
is actually found in a historic record prior to the discovery of the Zeeman effect 
by Lockyer in 1866, describing ``thick spectral lines in sunspots''. Cortie in 1896 mentioned 
a reversal (bright core) of an absorption line in sunspot spectra, 
which would obviously be a manifestation of the Zeeman effect under a strong magnetic field.

A more concrete evidence of the presence of magnetic field in sunspots was established
by George Hale in a paper entitled ``On the Probable Existence of a Magnetic Field
in Sun-Spots'' \citep{Hale1908}. He observed line splitting and polarization in sunspot spectra observed by
the newly constructed 35-feet solar tower at the Mount Wilson Observatory. By comparing the separation 
between the spectral components in the observed lines in sunspots and in sparks in laboratory experiments, 
he deduced that the magnetic field strength in sunspots was about 2600\,--\,2900 Gauss.
This was the first detection of the extraterrestrial magnetic field, which opened the way for 
measuring the magnetic field on the Sun and on other astronomical objects. A more detailed description
of the discovery of magnetic fields in sunspots can also be found in \cite{Iniesta1996}.


\subsection{Current tools to infer sunspot's magnetic field}
\label{subsection:tools}

Not much has changed since Hale's discovery of magnetic fields in
sunspots \citep{Hale1908}. The broadening of the intensity profiles of
spectral lines he saw on his photographic plates was produced by the
Zeeman splitting of the atomic energy levels in the presence of the
sunspot's magnetic field. Hale estimated a magnetic field strength of
about 2600\,--\,2900~Gauss. This basic technique is still widely used
nowadays. The addition of the polarization profiles: Stokes \Q, \U,
and \V, besides the intensity or Stokes~\I, allows us to determine
not only the strength of the magnetic field but the full magnetic
field vector $\ve{B}$. This is done thanks to the radiative transfer
equation (RTE):
\begin{eqnarray}
\label{equation:rte}
\frac{d\ve{I}_{\lambda}\depp}{d\tau_c} = \hmc{K}_{\lambda}\depp [\ve{I}_{\lambda}\depp-\ve{S}_{\lambda}\depp],
\end{eqnarray}
where $\ve{I}_{\lambda}\depp=(I,Q,U,V)^{\dag}$\epubtkFootnote{The symbol
  $^{\dag}$ indicates the transpose.} is the Stokes vector at a given
wavelength $\lambda$. The variation of the Stokes vector with optical
depth $\tau_c$ appears on the right-hand side of
Equation~(\ref{equation:rte}). The dependence of $\ve{I}_{\lambda}$ with
$\tau_c$ arises from the fact that $\ve{X}$ is a function of the
optical depth itself: $\ve{X}=\ve{X}[\tau_c]$. Here $\ve{X}$
represents the physical parameters that describe the solar atmosphere:
\begin{eqnarray}
\label{equation:x}
\ve{X}\dtau = [\ve{B}\dtau,T\dtau,P_g\dtau,P_e\dtau,\rho\dtau,V_{\mathrm{los}}\dtau,V_{\mathrm{mic}}\dtau,V_{\mathrm{mac}}\dtau],
\end{eqnarray}
where $\ve{B}\dtau$ is the magnetic field vector, $T\dtau$ is the
temperature stratification, $P_g\dtau$ and $P_e\dtau$ are the gas and
electron pressure stratification, $\rho\dtau$ is the density
stratification, and $V_{\mathrm{los}}\dtau$ is the stratification with
optical depth of the line-of-sight velocity. In addition,
macro-turbulent $V_{\mathrm{mac}}\dtau$ and micro-turbulent
$V_{\mathrm{mic}}\dtau$ velocities are often employed to model
velocity fields occurring at spatial scales much smaller than the
resolution element. Finally, on the right-hand side of
Equation~(\ref{equation:rte}) we have the propagation  matrix
$\hmc{K}_{\lambda}\depp$ and the source function $\ve{S}_{\lambda}\depp$
at a wavelength $\lambda$.  The latter is always non-polarized and,
therefore, only contributes to Stokes~\I:
\begin{eqnarray}
\label{equation:source_nonpolarized}
\ve{S}_{\lambda}\depp = (S_{\lambda}\depp,0,0,0)\dag.
\end{eqnarray}

The radiative transfer equation has a formal solution in the form:
\begin{eqnarray}
\label{equation:rtesol}
\ve{I}_{\lambda}\depp = \int_0^{\infty} \hmc{O}_{\lambda}\depp \hmc{K}_{c\lambda}\depp \ve{S}_{\lambda}\depp d\tau_c ,
\end{eqnarray}
where $\hmc{O}_{\lambda}(0,\tau_c)$ is the evolution operator, which
needs to be evaluated at every layer in order to perform the
integration. During the 1960s and early 1970s, the first  numerical
solutions to the radiative transfer equation for polarized light
became available \citep{Beckers1969a,Beckers1969b, Stenflo1971,
  Egidio1972,Wittmann1974,Auer1977}. Techniques to solve
Equation~(\ref{equation:rte}) have continued to be developed even during
the past two decades
\citep{Rees1989,Luis1998,Arturo1999a,Arturo1999b}.

Figure~\ref{figure:StokesMovie} shows an example of how the Stokes
vector varies when the magnetic field vector changes. In that movie we
use spherical coordinates to represent the three components of the
magnetic field vector: $\ve{B}=(B,\gamma,\varphi)$, where $B$ is the
strength of the magnetic field, $\gamma$ is the inclination of the
magnetic field with respect to the observer, and $\varphi$ is the
azimuth of the magnetic field in the plane perpendicular to the
observer's line-of-sight. In Figure~\ref{figure:StokesMovie} we assume
that the observer looks down along the \textit{z}-axis, but this does not
need to be always the case.

A major milestone was reached when these methods to solve the
RTEs~(\ref{equation:rte}) and (\ref{equation:rtesol}) 
were implemented into efficient minimization
algorithms that allow for the retrieval  of magnetic field vector in
an automatic way \citep{basilio1992, basilio2007,
  Hector_Inversion2002, JC2003, Bellot_Inversion2006}. This retrieval
is usually done by means of non-linear minimization algorithms that
iterate the free parameters of the model $\ve{X}\dtau$
(Equation~(\ref{equation:x})) while minimizing the difference between
the observed and theoretical Stokes profiles (measured by the merit
function $\chi^2$). The $\ve{X}\dtau$ that minimizes this difference
is assumed to correspond to the physical parameters present in the
solar atmosphere:
%
\begin{equation}
\label{equation:chi2}
\chi^2 = \frac{1}{4M-L} \sum_{i=1}^{4} \sum_{k=1}^{M}
\Bigg[
\frac{I_i^{\mathrm{obs}}(\lambda_k)-I_i^{\mathrm{syn}}(\lambda_k,\ve{X}[\tau_c])}
{\sigma_{ik}}
\Bigg]^2.
\end{equation}

Here $I_i^{\mathrm{obs}}(\lambda_k)$ and
$I_i^{\mathrm{syn}}(\lambda_k,\ve{X}[\tau_c])$ represent the observed
and theoretical (i.e., synthetic) Stokes vector, respectively. The latter
is obtained from the solution of the RTE~(\ref{equation:rtesol}) 
given a particular set of free parameters $\ve{X}$
(Equation~(\ref{equation:x})). The letter $L$ represents the total
number of free parameters in $\ve{X}$ and, thus, the term $4M-L$
represents the total number of degrees of freedom of the problem
(number of data points minus the number of free parameters). In
Equation~(\ref{equation:chi2}), indexes $i$ and $k$ run for the four
components of the Stokes vector (\I,\Q,\U,\V) and for all wavelengths,
respectively. 
Finally, $\sigma_{ik}$ represents the error (e.g., noise) in the
observations $I_i^{\mathrm{obs}}(\lambda_k)$.

Traditionally, the $\chi^2$-minimization has been carried out by
minimization algorithms such as the Levenberg--Marquardt method
\citep{Press1986}. However, more elaborated methods have also been
employed in recent years: genetic algorithms
\citep{Charbonneau1995,Andreas2003}, Principal-Component Analysis
\citep{Rees2000,HectorPCA}, and Artifical Neural Networks
\citep{Carroll2001,HectorANN}.

\epubtkMovie{stokes.mpg}{stokes.png}{%
  \begin{figure}[htbp]
    \centerline{\includegraphics[width=10cm]{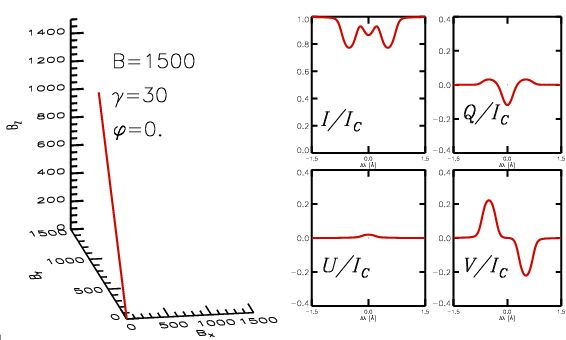}}
    \caption{The change of the synthetic emergent Stokes profiles
      (\I,\Q,\U,\V) when the magnetic field present in the solar plasma
      varies. The magnetic field vector is expressed in spherical
      coordinates: $B$ moduli of the magnetic field vector, $\gamma$
      inclination of the magnetic field vector with respect to the
      observer's line-of-sight (\textit{z}-axis in this case), and
      $\varphi$ azimuth of the magnetic field vector in the plane
      perpendicular to the observer's line-of-sight. Results have been
      obtained under the Milne--Eddington approximation.}
    \label{figure:StokesMovie}
\end{figure}}

Due to our limited knowledge of the line-formation theory, these
investigations have usually been limited to the study of the
photospheric magnetic field, where Local Thermodynamic Equilibrium and
Zeeman effect apply for the most part. However, recent advancements
in the line-formation-theory under NLTE conditions
\citep{Mihalas_book}, scattering polarization
\citep{Javier2002,Rafa2003,Egidio_book}, etc., allow us to
extend these techniques to the study of the chromosphere
\citep{Hector2000a,Andres2008,Trujillo2009,Casini2009}. Indeed, some
recent works have appeared where the magnetic structure of sunspots in
the chromosphere is being investigated
\citep{Hector2000b,Hector_Ca2005a,Hector_Ca2005b,Orozco2005}.

Techniques to study the coronal magnetic field from polarimetric
measurements of spectral lines are also becoming available nowadays
\citep{Steve2007,Steve2008}. These observations, carried out mostly
with near-infrared spectral lines, are recorded using coronographs (to
block the large photospheric contribution coming from the solar disk)
and, therefore, limited to the solar limb. Other possibilities to
observe polarization on the solar disk involve EUV (Extreme Ultra
Violet) lines, which are only accessible from space, and radio
observations of Gyroresonance and Gyrosynchrotron emissions, which can
show large polarization signals:
\cite{White2001,White2005}, \cite{Brosius2002}, and \cite{Brosius2006}. 
Unfortunately, so
far radio measurements have allowed only to infer the magnetic field
strength in the solar corona. Interestingly, opacity effects in the
gyroresonance emission (see Equations~(1) and (2) in \citealp{White2001})
might also permit to infer the inclination of the magnetic field
vector with respect to the observer's line-of-sight, i.e.,
$\gamma$. However, this possibility has not been yet successfully
exploited.

\subsubsection{Formation heights}
\label{subsubsection:formationheights}

According to Equations~(\ref{equation:rte}) and (\ref{equation:x}) the
solution to the radiative transfer equation depends on the
stratification with optical depth $\tau_c$ of the physical
parameters. The range of optical depths in which the solution
$\ve{X}(\tau_c)$ will be valid depends on the region of the photosphere in
which the analyzed spectral lines are formed. In the future we will
refer to this range as $\bar{\tau} = [\tau_{c,\mathrm{min}} ,
  \tau_{c,\mathrm{max}}]$. $\bar{\tau}$ can be determined by means of the
so-called \textit{contribution functions}
\citep{grossmann1988,solanki1994} and the \textit{response functions}
\citep{egidio1977,basilio1994}. In the literature, it is usually
considered that the range of optical depths, that a given spectral
lines is sensitive to, is so narrow that the physical parameters do
not change significantly over
$[\tau_{c,\mathrm{min}},\tau_{c,\mathrm{max}}]$. This can be
mathematically expressed as:
\begin{equation}
\label{equation:mecondition}
\frac{X_f(\tau_{c,\mathrm{max}})-X_f(\tau_{c,\mathrm{min}})} {X_f(\tau_{c,\mathrm{max}})+X_f(\tau_{c,\mathrm{min}})} << 1 ,
\end{equation}
where $X_f$ refers to the $f$-component of $\ve{X}$
(Equation~(\ref{equation:x})). When the conditions in
Equation~(\ref{equation:mecondition}) are met for all $f$'s, a
Milne--Eddington-like (ME) inversion can be applied. The advantage of
ME-codes is that an analytical solution for the
RTE~(\ref{equation:rte}) exists in this case. ME-codes assume that the
physical parameters are constant in the range $\bar{\tau}$. One way to
determine the magnetic field at different heights in the solar
atmosphere is to perform ME-inversions  of spectropolarimetric data in
several spectral lines that are formed at different average optical
depths $\bar{\tau}$'s, with each spectral line yielding information in
a plane at a different height above the solar surface.

As an example of the results retrieved by a Milne--Eddington-like
inversion code we show, in Figures~\ref{figure:sunspotmap1} and
\ref{figure:sunspotmap2}, the three components of the magnetic field
vector, for two different sunspots, in the observer's reference
frame. $B$ or magnetic field strength is shown in the upper-right
panels, $\gamma$ or the inclination of the magnetic field vector with
respect to the observer's line-of-sight in the lower-left panels, and
finally, $\varphi$ or the azimuthal angle of the magnetic field vector
in the plane perpendicular to the observer's light-of-sight in the
lower-right panels. The first sunspot, AR~10923
(Figure~\ref{figure:sunspotmap1}), was observed very close to disk
center ($\Theta \simeq 9\deg$) on November 14, 2006. The second
sunspot, AR~10933 (Figure~\ref{figure:sunspotmap2}), was observed on
January 9, 2007 very close to the solar limb ($\Theta \simeq
50\deg$). In both cases, the magnetic field vector was obtained from
the VFISV Milne--Eddington-type inversion \citep{vfisv2010} of the
Stokes vector recorded with the spectropolarimeter on-board the
Japanese spacecraft Hinode \citep{hinode1, hinode2, hinode3}. The
observed Stokes vector corresponds to the Fe\,{\sc i} line pair at
630~nm, which are formed in the photosphere. As explained above,
Milne--Eddington inversion codes assume that, among others, the
magnetic field vector does not change with optical depth: $\ve{B} \ne
f(\tau_c)$ (see Equation~(\ref{equation:x})). Therefore,
Figures~\ref{figure:sunspotmap1} and \ref{figure:sunspotmap2} should
be interpreted as the averaged magnetic field vector over the region
in which the employed spectral lines are formed: $\bar{\tau} \simeq
[1,10^{-3}]$.

When the conditions in Equation~(\ref{equation:mecondition}) are not
met, it is not possible to perform a ME-line inversion. If we do, the
results should be interpreted accordingly, that is, the inferred
values for $\ve{X}$ correspond to an average over the region
$\bar{\tau} \in [\tau_{c,\mathrm{min}}, \tau_{c,\mathrm{max}}]$ where the spectral line
is formed. A different approach consists in the application of
inversion codes for the radiative transfer equation that consider the
full $\tau_c$ dependence of the physical parameters $\ve{X}$. In this
case, the solution of the radiative transfer equation can only be
found numerically \citep[cf.][]{Arturo1999b}. Examples of these codes
are: SIR \citep{basilio1992}, SPINOR \citep{frutiger1999}, and LILIA
\citep{Hector_Inversion2002}. This allows to obtain the optical depth
dependence ($\tau_c$-dependence) of the physical parameters with one
single spectral line.  Ideally, in order to increase the range of
validity of the inferred models, one still wants to employ different
spectral lines.

\epubtkImage{ic-bfield-gamma-phi_14nov06.jpg}{%
  \begin{figure}[htbp]
    \centerline{\includegraphics[width=7cm]{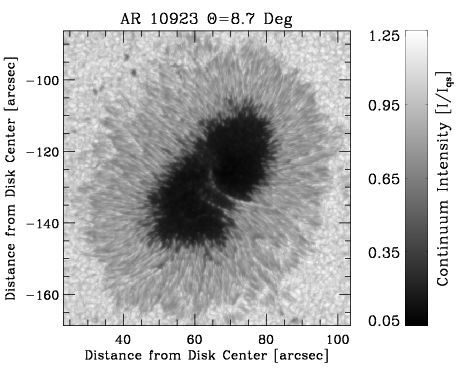}\qquad
      \includegraphics[width=7cm]{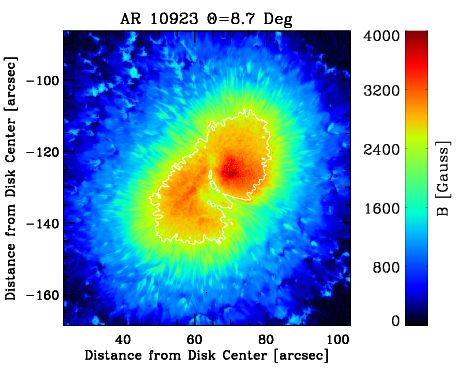}}
     \centerline{\includegraphics[width=7cm]{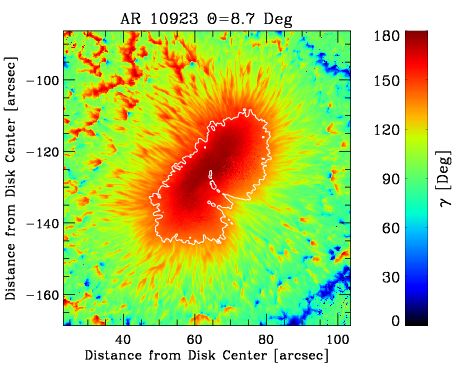}\qquad
      \includegraphics[width=7cm]{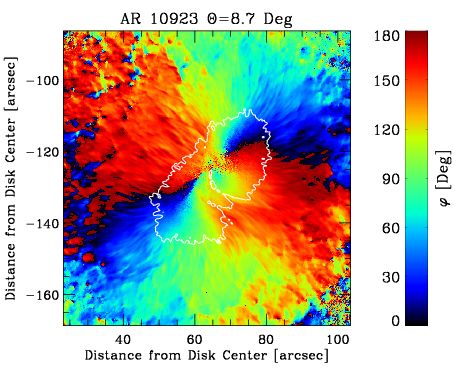}}
    \caption{These plots show the magnetic field vector in the sunspot
      AR~10923, observed on November 14, 2006 close to disk center
      ($\Theta$~=~8.7\textdegree\ at the umbral center). The upper-left panel
      displays the normalized (to the quiet Sun value) continuum
      intensity at 630~nm. The upper-right panel displays the total
      magnetic field strength, whereas the lower-left and lower-right
      panels show the inclination of the magnetic field vector
      $\gamma$ with respect to the observer's line-of-sight, and the
      azimuth of the magnetic field vector in the plane perpendicular
      to the line-of-sight $\varphi$, respectively. The white contours
      on the colored panels indicate the umbral boundary, defined as
      the region in the top-left panel where $I/I_{\mathrm{qs}} <
      0.3$. These maps should be interpreted as the average over the
      optical depth range in which the employed spectral lines are
      formed: $\bar{\tau} \simeq [1,10^{-3}]$.}
    \label{figure:sunspotmap1}
\end{figure}}

\epubtkImage{ic-bfield-gamma-phi_09jan07.jpg}{%
  \begin{figure}[htbp]
    \centerline{\includegraphics[width=7cm]{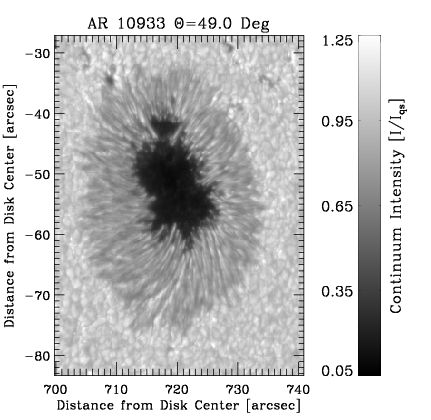}\qquad
      \includegraphics[width=7cm]{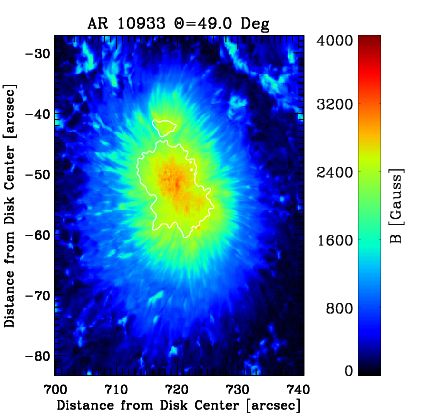}}
     \centerline{\includegraphics[width=7cm]{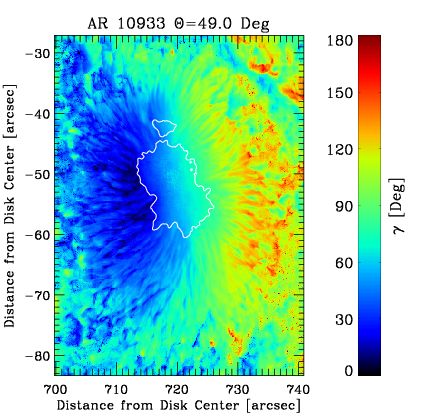}\qquad
      \includegraphics[width=7cm]{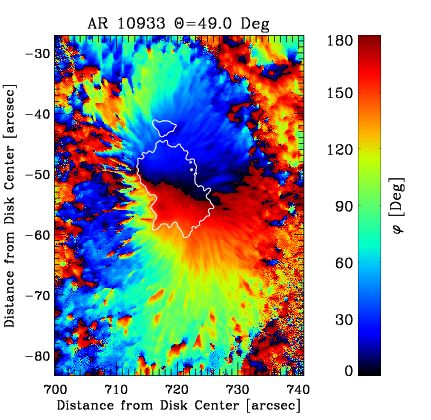}}
    \caption{Same as Figure~\ref{figure:sunspotmap1} but for the
      sunspot AR~10933, observed on January 9, 2007 close to the limb
      ($\Theta$~=~49.0\textdegree\ at the umbral center).}
    \label{figure:sunspotmap2}
\end{figure}}

\subsubsection{Azimuth ambiguity}
\label{subsubsection:180deg}

The elements of the propagation matrix $\hmc{K}_{\lambda}\depp$
(Equation~(\ref{equation:rte})) for the linear polarization 
\cite[see, e.g.,][Chapter~7.5]{JCbook} can be written as:
\begin{eqnarray}
\label{equation:etaq}
\eta_Q \propto \cos2\varphi \\
\label{equation:etau}
\eta_U \propto \sin2\varphi ,
\end{eqnarray}
where $\varphi$ corresponds to the azimuthal angle of the magnetic
field vector in the plane perpendicular to the observer's
line-of-sight. Equations~(\ref{equation:etaq}) and (\ref{equation:etau}) also
hold for the dispersion profiles (magneto-optical effects) $\rho_Q$
and $\rho_U$ present in the propagation matrix $\hmc{K}_{\lambda}$. Note
that these matrix elements remain unchanged if we take $\varphi+\pi$
instead of $\varphi$. Because of this the radiative transfer equation
cannot distinguish between these two possible solutions for the
azimuth: [$\varphi,\varphi+\pi$]. This is the so-called
180\textdegree-ambiguity problem in the azimuth of the magnetic
field. Because of this ambiguity, the azimuthal angle of the magnetic
field $\varphi$ (as retrieved from the inversion of spectrolarmetric
data) in Figures~\ref{figure:sunspotmap1} and \ref{figure:sunspotmap2}
(lower-right panels) is displayed only between 0\textdegree and
180\textdegree. A number of techniques have been developed to solve
this problem. These techniques can be classified in terms of the
auxiliary physical quantity that is employed:

\begin{itemize}

\item \textit{Acute-angle} methods: these techniques minimize the
  angle between the magnetic field vector inferred from the
  observations (see Section~\ref{subsection:tools}) and the magnetic
  field vector obtained from a given model. The question is, therefore,
  how is the model magnetic field obtained. Traditionally, it is
  obtained from potential or force-free extrapolations of the observed
  longitudinal component of the magnetic field: 
  $B_{\mathrm{los}}=B\cos\gamma$, which is $\varphi$ independent. The
  extrapolation yields the horizontal component of the magnetic field,
  which is then compared with the two possible ambiguous
  solutions: $\varphi$ and $\varphi+\pi$. Whichever is closer to the
  extrapolated horizontal component is then considered to be the
  correct, ambiguity-free, solution. Potential field and force-free
  extrapolations can be obtained employing Fourier transforms
  \citep{alissandrakis1981, gary1989}. Some methods that solve the
  180\textdegree-ambiguity employing this technique have been presented by
  \cite{wang1997} and \cite{wang2001}. In addition, Green's function
  can also be used for the extrapolations and to solve the ambiguity
  \citep{sakurai1982,abramenko1986, cuperman1990, cuperman1992}.

\item \textit{Current free and null divergence} methods: these methods
  select the solution, $\varphi$ or $\varphi+\pi$, that minimizes the
  current vector $\ve{j}$ and/or the divergence of the magnetic field:
  $\nabla \cdot \ve{B}$. The calculation of these quantities makes
  use of the derivatives of the three components of the magnetic field
  vector. Because the vertical (\textit{z}-axis) derivatives are usually not
  available through a Milne--Eddington inversion (see
  Sections~\ref{subsubsection:formationheights},
  \ref{subsection:constanttau}, and \ref{subsection:potential}) only the
  vertical component of the current $j_z$ is employed. In addition,
  the term $\partial B_z/\partial z$ is neglected in the calculation
  of the divergence of the magnetic field. The minimization of the
  aforementioned quantities can be done locally or globally. Finally,
  note that current free and null divergence methods usually rely on
  initial solutions given by acute-angle methods and potential  field
  extrapolations.

\end{itemize}

In recent reviews by \cite{Metcalf2006} and \cite{Leka2009} several of
these techniques are compared against each other, employing previously
known magnetic field configurations and measuring their degree of
success employing different metrics when recovering the original
one. It is important to mention that in these reviews, some other very
successful methods (which do not necessarily fall into the
aforementioned categories) are also employed\epubtkFootnote{Many of
  the codes that have been compared in these papers are publicly
  available to the community. Minimum energy method:
  \url{http://www.cora.nwra.com/AMBIG}; Non-Potential field calculation:
  \url{http://sd-www.jhuapl.edu/FlareGenesis/Team/Manolis/codes/ambiguity_resolution/}; and
  AZAM: \url{http://www.csac.hao.ucar.edu/csac/visualize.jsp}}: the
non-potential magnetic field calculation method by \cite{Manolis2005}
and the manual utility AZAM by Lites \textit{et al.} (\textit{private
  communication}), which is part of the ASP routines
\citep{elmore1992}. In those reviews it is found that acute-angle
methods perform well only if the configuration of the magnetic field
is simple, whereas interactive methods (AZAM) tend to fail in the
presence of unresolved structures below the resolution element of the
observations. Current free and null divergence methods tend to work
better when both conditions \citep{canfield1993,Metcalf1994} are
applied instead of only one \citep{gary1995,Crouch2008}, with local
minimization being more prone to propagate errors than global
minimization techniques.

Several of these techniques are very suitable to study complex
regions, in particular outside sunspots. However, in regular sunspots
(excluding those with prominent light bridges or
$\delta$-sunspots\epubtkFootnote{$\delta$-sunspots are commonly
  defined as those where the umbra possesses two difference
  polarities.}) the magnetic field is highly organized, with filaments
that are radially aligned in the penumbra. We can use this fact to
resolve the 180\textdegree-ambiguity in the determination of the
azimuthal angle $\varphi$. This is done by finding the coordinates of
the magnetic field vector $\ve{B}$ in the \textit{local reference
  frame}: $\{\ea,\eb,\ep\}$\epubtkFootnote{The unit vectors of the
  \textit{local reference frame} are defined as follows: $\ep$ is the
  unit vector that is perpendicular to the tangential plane on the
  solar surface at the point of observation, while $\ea$ and $\eb$ are
  inside this plane (see Figure~\ref{figure:earthsun}).} and taking
whichever solution, $\ve{B}(\varphi)$ or $\ve{B}(\varphi+\pi)$,
minimizes the following quantity:
\begin{equation}
\mathrm{min} \left(\frac{\ve{B} \cdot \ve{r}}{|\ve{B} \cdot \ve{r}|}\pm 1\right) \;,
\label{equation:parallel}
\end{equation}
where the vector $\ve{r}$ corresponds to the radial direction in the
sunspot or, in other words, $\ve{r}$ is the vector that connects the
center of the umbra with the point of observation. Because the
condition of radial magnetic fields (Equation~(\ref{equation:parallel}))
can only be safely applied  in the \textit{local reference frame}, it
is important to describe how $\ve{B}$ and $\ve{r}$ are obtained. A
detailed account is provided in Appendix~\ref{section:appendix} of
this paper.

Because we aim at minimizing the above value
(Equation~(\ref{equation:parallel})) this method can be considered as an
\textit{acute-angle} method where the reference magnetic field is not
obtained from a potential extrapolation but rather assumed to be
radial. Note that if the sunspot has positive polarity, the magnetic
field vector and the radial vector tend to be parallel:
$\ve{B}\ve{r}>0$ and, therefore, the $-$ (minus) sign should be used in
Equation~(\ref{equation:parallel}). If the sunspot has negative
polarity, then the magnetic field vector and the radial vector are
anti-parallel and, therefore, the sign $+$ (plus) should be
employed. However, this is only a convention: we can choose to
represent the magnetic field vector as if a sunspot had a different
polarity as the one indicated by Stokes~\V.

As an example of the method depicted here we show, in
Figures~\ref{figure:fig_amb1}, \ref{figure:fig_amb2}, and
\ref{figure:fig_amb3}, the vertical $B_{\rho}$ and horizontal
$B_{\beta}$ and $B_{\alpha}$ components of the magnetic field vector
(Equation~(\ref{equation:bfieldinloc})), once the
180\textdegree-ambiguity has been resolved for two sunspots: AR~10923
and AR~10933 (same as in Figures~\ref{figure:sunspotmap1} and
\ref{figure:sunspotmap2}). $B_{\rho}$, $B_{\beta}$, and $B_{\alpha}$ are
the components of the magnetic field vector in the \textit{local
  reference frame}. Note that strictly speaking, the unit vectors
$\eb$ and $\ea$ shown in these figures correspond to the unit vectors
at the umbral center. Although differences are small, at other points
in the image the unit vectors have different directions since those
points have different ($X_c$,$Y_c$) and ($\alpha,\beta$) coordinates
(Equations~(\ref{equation:sinalfa})\,--\,(\ref{equation:cosbeta4})). Once
the 180\textdegree-ambiguity has been solved we can obtain, in the
\textit{local reference frame}, the inclination and the azimuth of the
magnetic field, $\zeta$ (Figure~\ref{figure:fig_amb4}) and $\Psi$
(Figure~\ref{figure:fig_amb5}) as:
\begin{eqnarray}
\label{equation:zeta}
\zeta &=& \cos^{-1}\left[\frac{B_{\rho}}{\sqrt{B_{\alpha}^2+B_{\beta}^2}}\right] \;,\\
\label{equation:psi}
\Psi  &=& \tan^{-1}\left[\frac{B_{\beta}}{B_{\alpha}}\right] \;.
\end{eqnarray}

It is important to notice that because the ambiguity has now been
solved, the angle $\Psi$ varies between 0\textdegree\ and
360\textdegree\ (see Figure~\ref{figure:fig_amb5}), whereas before,
lower-right panels in Figures~\ref{figure:sunspotmap1} and
\ref{figure:sunspotmap2}, $\varphi$ ranged only between
0\textdegree\ and 180\textdegree.

As already mentioned, the method we have described here works very
well for regular (e.g., round) sunspots. There is, however, one
important caveat: when the retrieved inclination $\gamma$ (in the
\textit{observer's reference frame}) is close to 0, the azimuth
$\varphi$ is not well defined. In this case, applying
Equation~(\ref{equation:parallel}) does not make much sense. Here we
must resort to other techniques \citep{Metcalf2006} to solve the
ambiguity. The region where $\gamma$~=~0\textdegree\ occurs usually at the
center of the umbra for sunspots close to disk center, and it shifts
towards the center-side penumbra as the sunspot is closer to the
limb. A similar coordinate transformation as the one depicted here
have been described in \cite{Hagyard1987} and \cite{Venka1988}, with the
difference that no attempt to solve the 180\textdegree-ambiguity was
made. \cite{BellotRubio2004} and \cite{Jorge2005} employ a smoothness
condition to solve the 180\textdegree-ambiguity, however their
coordinate transform is done in two dimensions, whereas here we
consider the Sun's spherical shape. In addition, only one heliocentric
angle $\Theta$ was considered in their transformation, whereas here
$\Theta$ changes for each point on the solar surface
(Equation~(\ref{equation:helioangle})). One might think that the
variation of the angle $\Theta$ across the field-of-view (FOV) are
negligible. However, for a FOV with 100~\texttimes~100~arcsec\super{2}
this variation can be as large as 4\,--\,5\textdegree. These differences
can be important, for instance, when searching for regions in the
sunspot penumbra where the magnetic field points down into the solar
surface: $B_{\rho}<0$.

\epubtkImage{br_14nov06-br_09jan07.png}{%
  \begin{figure}[htbp]
    \centerline{\includegraphics[width=9.6cm]{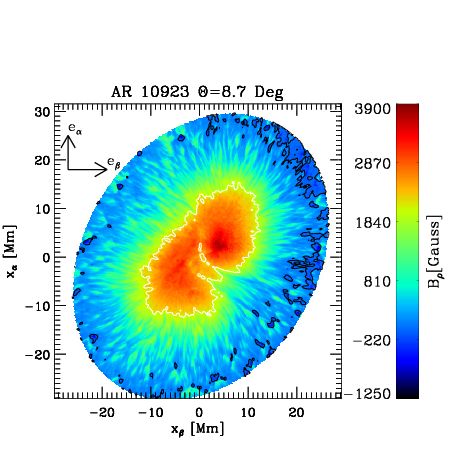}}
    \centerline{\includegraphics[width=9.6cm]{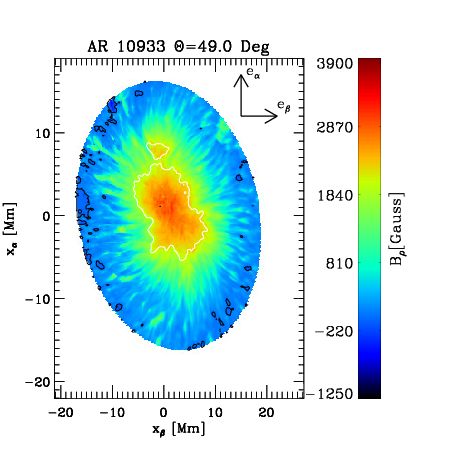}}
    \caption{Vertical component of the magnetic field $B_{\rho}$ in the
      \emph{local reference frame} in two different sunspots: AR~10923
      (top; $\Theta$~=~8.7\textdegree) and AR~10933 (bottom:
      $\Theta$~=~49.0\textdegree). The black contours highlight the regions
      where the magnetic field points downwards towards the solar
      center: $B_{\rho} < 0$. The white contours surround the umbral
      region, defined as the region where the continuum intensity
      (normalized to the quiet Sun intensity) $I/I_{\mathrm{qs}}<0.3$. 
      The horizontal and vertical
      directions in these plots correspond to the $\eb$ and $\ea$
      directions, respectively.}
    \label{figure:fig_amb1}
\end{figure}}

\epubtkImage{bb_14nov06-bb_09jan07.png}{%
  \begin{figure}[htbp]
    \centerline{\includegraphics[width=9.6cm]{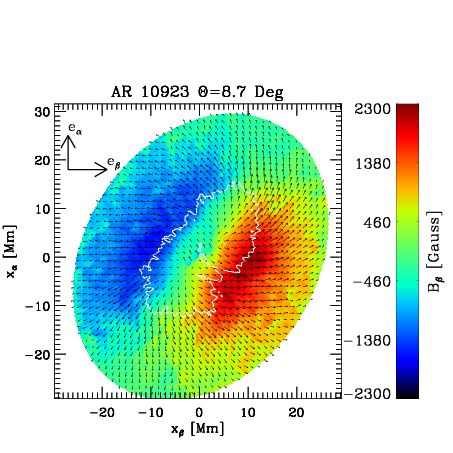}}
    \centerline{\includegraphics[width=9.6cm]{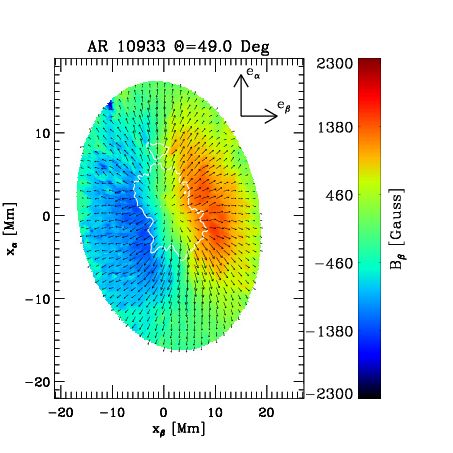}}
    \caption{Same as Figure~\ref{figure:fig_amb1} but for the
      $B_{\beta}$ component of the magnetic field vector in the
      \emph{local reference frame}. The
      arrow field indicates the direction of the magnetic field vector
      in the plane tangential to the solar surface.}
    \label{figure:fig_amb2}
\end{figure}}

\epubtkImage{ba_14nov06-ba_09jan07.png}{%
  \begin{figure}[htbp]
    \centerline{\includegraphics[width=9.6cm]{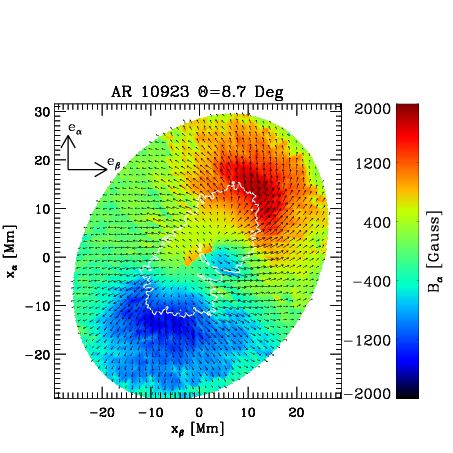}}
    \centerline{\includegraphics[width=9.6cm]{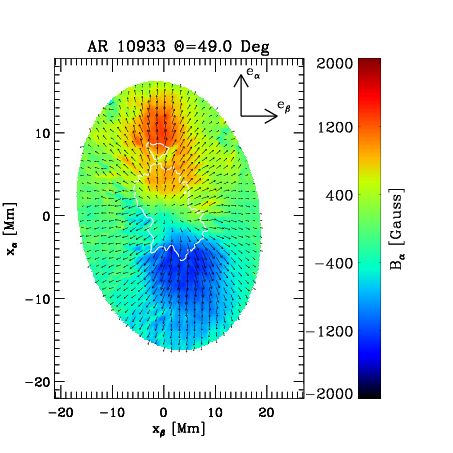}}
    \caption{Same as Figure~\ref{figure:fig_amb2} but for the
      $B_{\alpha}$ component of the magnetic field vector.}
    \label{figure:fig_amb3}
\end{figure}}

\epubtkImage{inc_14nov06-inc_09jan07.png}{%
  \begin{figure}[htbp]
    \centerline{\includegraphics[width=9.6cm]{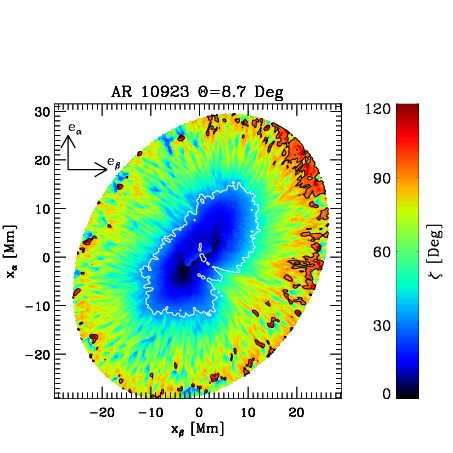}}
    \centerline{\includegraphics[width=9.6cm]{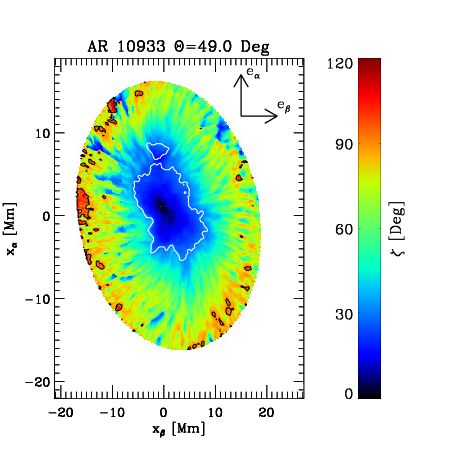}}
    \caption{Same as Figure~\ref{figure:fig_amb1} but for the
      inclination of the magnetic field with respect to the normal
      vector to the solar surface $\ep$: $\zeta$
      (see Equation~(\ref{equation:zeta})). The black contours indicate the
      regions where $\zeta > 90\deg$ and coincide with the regions, in
      Figure~\ref{figure:fig_amb1}, where $B_{\rho} < 0$.}
    \label{figure:fig_amb4}
\end{figure}}

\epubtkImage{psi_14nov06-psi_09jan07.png}{%
  \begin{figure}[htbp]
    \centerline{\includegraphics[width=9.6cm]{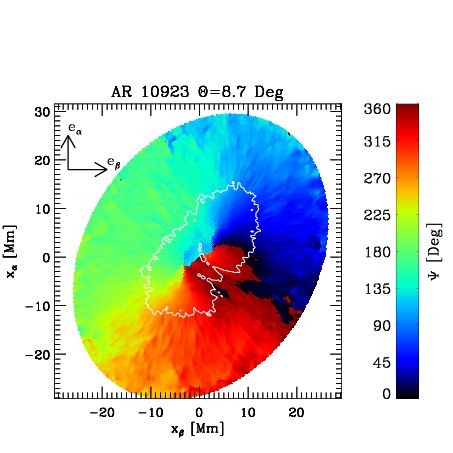}}
    \centerline{\includegraphics[width=9.6cm]{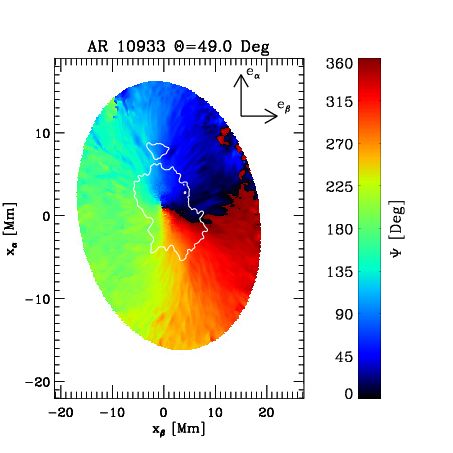}}
    \caption{Same as Figure~\ref{figure:fig_amb4} but for the
      azimuthal angle of the magnetic field in the plane of the solar
      surface: $\Psi$ (see Equation~(\ref{equation:psi})).}
    \label{figure:fig_amb5}
\end{figure}}

\clearpage
\subsubsection{Geometrical height and optical depth scales}
\label{subsubsection:tau2z}

Traditionally, inversion codes for the RTE~(\ref{equation:rte}) 
such as: SIR \citep{basilio1992} and
SPINOR \citep{frutiger1999}, provide the physical parameters as a
function of the optical depth, $\ve{X}(\tau_c)$
(Equation~(\ref{equation:x})). The optical depth is evaluated at some
wavelength where there are no spectral lines (continuum), hence the
sub-index \textit{c}. When this is done for each pixel in an observed
two-dimensional map, the inversion code yields
$\ve{X}(x_{\beta},x_{\alpha},\tau_c)$. However, it is oftentimes
convenient to express them as a function of the geometrical height
$x_{\rho}$. To that end, the following relationship is employed:
\begin{equation}
\label{equation:tauz}
d\tau_c = -\rho(x_{\rho}) \chi_c[T(x_{\rho}),P_g(x_{\rho}),P_e(x_{\rho})] dx_{\rho} ,
\end{equation}
where $\chi_c$ is the opacity evaluated at a continuum wavelength and
depends on the temperature, gas pressure, and electron pressure. Now,
these thermodynamic parameters barely affect the emergent Stokes
profiles $\ve{I}_{\lambda}$ and, therefore, are usually not obtained from
the inversion of the polarization profiles themselves. Instead, other
kind of constraints are usually employed to determine them, being the
most common one, the application of the vertical hydrostatic
equilibrium equation:
\begin{equation}
\label{equation:hydro}
\frac{d P_g(x_{\rho})}{dx_{\rho}} = -g \rho(x_{\rho}) ,
\end{equation}
which after applying Equation~(\ref{equation:tauz}) becomes:
\begin{equation}
\label{equation:hydrotau}
\frac{d P_g(\tau_c)}{d\tau_c} = \frac{g}{\chi_c(\tau_c)}.
\end{equation}

Note that, since Equations~(\ref{equation:hydro}) and
(\ref{equation:hydrotau}) do not depend on $(x_{\beta},x_{\alpha})$, they can
be applied independently for each pixel in the map. Hence, the
geometrical height scale (at each pixel) can be obtained by following
the next steps:

\begin{enumerate}
\item Given a boundary condition for the gas pressure in the uppermost
  layer of the atmosphere, $P_g(\tau_{\min})$, we can employ the
  fixed-point iteration described in \cite{Wittmann1974} and
    \cite{Mihalas_book} to obtain the electron pressure in this layer:
  $P_e(\tau_{\min})$.

\item From the inversion, the full temperature stratification
  $T(\tau_c$) and, thus, $T(\tau_{\min})$ are known. Since the
  continuum opacity $\chi_c$ depends on the electron pressure, gas
  pressure, and temperature, it is therefore possible to obtain
  $\chi_c(\tau_{\min})$.

\item A predictor-corrector method is employed to integrate downwards
  Equation~(\ref{equation:hydrotau}) and obtain
  $P_g(\tau_{\min-1})$. This is done by first assuming that
  $\chi_c$ is constant between $\tau_{\min}$ and
  $\tau_{\min-1}$:
\begin{equation}
P_{g,1}(\tau_{\min-1}) = P_g(\tau_{\min}) - \frac{g}{\chi_c(\tau_{\min})} [\tau_{\min}-\tau_{\min-1}]
\end{equation}
and with $P_{g,1}(\tau_{\min-1})$, we apply step~\#1 to calculate
$P_{e,1}(\tau_{\min-1})$.

\item Since we also know $T(\tau_{\min-1})$, we repeat step~\#2 to
  recalculate $\chi_c(\tau_{\min-1})$, which is then employed to
  re-integrate Equation~(\ref{equation:hydrotau}) as:
\begin{equation}
P_{g,2}(\tau_{\min-1}) = P_g(\tau_{\min}) - \frac{2g}{[\chi_c(\tau_{\min}+\chi_c(\tau_{\min-1})]} [\tau_{\min}-\tau_{\min-1}].
\end{equation}
Step~\#4 is repeated $k$-times until convergence:
$|P_{g,k}(\tau_{\min-1})-P_{g,k-1}(\tau_{\min-1})| <
\epsilon$.

\item We now have $P_{g}(\tau_{\min-1})$. In addition,
  $T(\tau_c$) and, thus, $T(\tau_{\min})$ are known. Consequently, we can
  repeat steps~\#1 to \#3 in order to infer $P_{g}(\tau_{\min-2})$.

\item Thus, repeating steps~\#1 through \#5 yields: $P_g(\tau_c)$,
  $P_e(\tau_c)$, and $\chi_c(\tau_c)$.

\item The equation of ideal gases can be now employed to determine
  $\rho(\tau_c)$. And, finally, the integration of
  Equation~(\ref{equation:tauz}) yields the geometrical depth scale as:
  $\tau_c(x_{\rho})$. To integrate this equation, a boundary condition
  is needed. This is usually taken as $x_{\rho}(\tau_c=1)=0$, which sets
  an offset to the geometrical height such that the continuum level
  $\tau_c=1$ coincides with $x_{\rho}=0$.

\end{enumerate}

Applying the condition of hydrostatic equilibrium to obtain the
density, gas pressure, and the geometrical height scale \textit{z} is strictly
valid only when the Lorentz force are small and the velocities are
much smaller than the speed of sound. In the chromosphere and corona
this is certainly not the case. In the solar photosphere the
assumption of hydrostatic equilibrium is, in general, well
justified. One exception are sunspots, where the large velocities and
magnetic fields might break down this assumption. In these case, a
more general momentum (force balance) equation must be
employed\epubtkFootnote{Note that the most general momentum equation
  would also include, in the right-hand term of
  Equation~(\ref{equation:momentum}), the terms corresponding to the
  viscous forces: $\mu_1 \nabla^2 \ve{v}$ and
  $[\mu_2+(1/3)\mu_1]\ve{\nabla}(\ve{\nabla} \ve{v})$, where the
  coefficients $\mu_1$ and $\mu_2$ are often referred to as shear
  viscosity and bulk viscosity, respectively.}:
\begin{equation}
\label{equation:momentum}
\rho (\ve{v} \ve{\nabla})\ve{v}= -\ve{\nabla} P_g +\frac{1}{c} \ve{j}\times \ve{B} + \rho \ve{g}.
\end{equation}

Trying to solve this equation to obtain the gas pressure, density, and
geometrical height scale is not an easy task. In the hydrostatic case,
the horizontal derivatives did not play any role, thus simplifying
Equation~(\ref{equation:momentum}) into:
\[
\mathrm{hydrostatic: } 
\left \{ \begin{array}{ll} 
  d P_g / dx_{\rho}   &= -\rho g \\ 
  d P_g / dx_{\beta}  &= 0 \\ 
  d P_g / dx_{\alpha} &= 0 .
\end{array}\right.
\]
However, if the Lorentz force $\ve{j}\times \ve{B}$ and the advection
term $(\ve{v}\ve{\nabla})\ve{v}$ cannot be neglected, the horizontal
components of the momentum equation must be considered. In addition,
the horizontal derivatives of the gas pressure mix the results of the
magnetic field and velocity from nearby pixels. Thus, the
determination of the gas pressure, density, and geometrical height
scale cannot be achieved individually for each pixel of the
map. Instead, a global technique must be employed. This can be done by
shifting the \textit{z}-scale at each pixel in the map (effectively changing
the boundary condition mentioned in step~\#7 above) in order to
globally minimize the imbalances in the three components of the
momentum equation and the term $\ve{\nabla} \cdot \ve{B}$. The shift
at each pixel, $\mathrm{Z}_{\mathrm{w}}(x_{\beta},x_{\alpha})$, represents
the Wilson depression. This kind of approach has been followed by
\cite{maltby1977}, \cite{solanki1993}, \cite{valentin1993}, and 
\cite{shibu2004}. However,
changing the boundary condition in step~\#7 does not change the fact
that the vertical stratification of the gas pressure still complies
with hydrostatic equilibrium (Equation~(\ref{equation:hydro})). A way
out of this problem has not been figured out until very recently with
the work of \cite{Puschmann2008, Puschmann2010a}, who have devised a
technique that takes into account the general momentum
equation~(\ref{equation:momentum}) when determining the gas pressure
and establishing a common
\textit{z}-scale. Figure~\ref{figure:zw_klaus} shows a map for the
Wilson depression in a small region of the inner penumbra of a sunspot
\cite[adapted from][]{Puschmann2010a}. Another interesting technique
has been proposed recently by \cite{thorsten2008}, where the vertical
height scale can be obtained, instead of a posteriori as in
\cite{Puschmann2010a}, directly during the inversion of the Stokes
profiles. This is achieved by performing the inversion employing
Artificial Neural Networks \citep[ANNs;][see
  Section~\ref{subsection:tools}]{Carroll2001} that have been
previously trained with snapshots of MHD simulations, which are given
in the \textit{z}-scale.

\epubtkImage{zw_klaus.png}{%
  \begin{figure}[htbp]
  \centerline{\includegraphics[width=9.6cm]{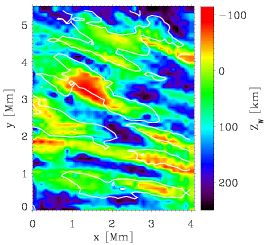}}
      \caption{Map of the Wilson depression
        $\mathrm{Z}_{\mathrm{w}}(x,y)$ in a small region of the inner
        penumbra in AR~10953 observed on May 1, 2007 with
        Hinode/SP. The white contours enclose regions where upflows
        are present: $V_{\mathrm{los}}$~\textgreater~0.3~\kms. Negative values of
        $\mathrm{Z}_{\mathrm{w}}$ correspond to elevated
        structures. In this figure $x$ and $y$ correspond to our
        coordinates $x_{\beta}$ and $x_{\alpha}$,
        respectively \citep[from][reproduced by permission of the
          AAS]{Puschmann2010a}.}
    \label{figure:zw_klaus}
\end{figure}}
\newpage
\section{Global Magnetic Structure}
\label{section:global}

In this section, we will discuss the global structure of the magnetic
field vector in sunspots. Even though sunspot's magnetic fields are
organized at very small scales (see, for example,
Figures~\ref{figure:fig_amb1}\,--\,\ref{figure:fig_amb5}), there are
many questions that can be addressed considering mainly its global
structure: wave propagation \citep{lena2008,moradi2008},
helioseismology \citep{moradi2010,Cameron2011}, extrapolations to
obtain the coronal magnetic field
\citep{schrijver2008,metcalf2008,derosa2009}. In the former cases,
small-scale magnetic structures do not interact with typical
helioseimology waves (p and f-modes) because their wavelengths are
much larger than the typical sizes of the magnetic structures. In the
latter case, small-scale horizontal magnetic structures do not affect
the coronal magnetic structure because they produce loops that close
at photospheric and chromospheric levels \citep{wiegelmann2010}.

Other branch where observational inferences of sunspot's global
magnetic structure are needed is in theoretical modeling of sunspots
\citep[i.e., magneto-hydrostatic;][]{low1975, low1980, osherovich1983,
  pizzo1986, pizzo1990, jahn1994}. These
models employ the magnetic field configuration inferred from
observations as boundary conditions in their equations, as well as
employing the observations to validate their final results.

The first section of this chapter will be devoted to study the
magnetic field configuration as seen at a constant optical depth or
$\tau$-level, whereas the second section will study the vertical
variations of the magnetic field. These two can be employed as
Dirichlet or Neumann boundary conditions, respectively, in theoretical
models and extrapolations. The rest of the sections in this chapter
will focus on other issues such as the plasma-$\beta$, potentiality of
the magnetic field, thermal-magnetic relation, and so forth.

\subsection[As seen at constant $\tau$-level]{As seen at constant
  {\boldmath$\tau$}-level}
\label{subsection:constanttau}

In Figures~\ref{figure:sunspotmap1} and \ref{figure:sunspotmap2} in
Section~\ref{subsection:tools}, and
Figures~\ref{figure:fig_amb1}\,--\,\ref{figure:fig_amb5} in
Section~\ref{subsubsection:180deg}, we have presented the 3 components
of the magnetic field both in the \textit{observer's reference frame}
and in the \textit{local reference frame}. Those maps were obtained
from the inversion of spectropolarimetric observations employing a
Milne--Eddington (ME) atmospheric model (see
Section~\ref{subsubsection:formationheights}). This means that the
results from a ME inversion should be interpreted as an average of the
magnetic field  vector over the region where the lines are formed
$\bar{\tau}$: $B_\beta(x_\beta,x_\alpha,\bar{\tau})$,
$B_\alpha(x_\beta,x_\alpha,\bar{\tau})$, and
$B_\rho(x_\beta,x_\alpha,\bar{\tau})$. This makes the results from the
ME inversion ideal to study the magnetic field at a constant
$\tau$-level. The coordinates $x_\alpha$ and $x_\beta$ refer to the
\textit{local reference frame}: $\{\eb,\ea,\ep\}$ as described in
Section~\ref{subsubsection:180deg}. Note that the optical depth $\tau$
is employed instead of $x_\rho$, which is the coordinate representing
the geometrical height. For convenience let us now consider polar
coordinates in the $\alpha\beta$-plane: $(r,\theta)$ with $r$ being
the radial distance between any point in the sunspot to the center of
the umbra. $\theta$ is defined as the angle between the radial vector
that connects this point with the umbra center and the $\eb$ axis (see,
for example, Figure~\ref{figure:fig_amb1}). With this transformation we
now have:
$B_{h}(r,\theta,\bar{\tau})=\sqrt{B_\beta^2(r,\theta,\bar{\tau})+
  B_\alpha^2(r,\theta,\bar{\tau})}$ (horizontal component of the
magnetic field) and $B_\rho(r,\theta,\bar{\tau})$ (vertical component
of the magnetic field).

We will now focus on the radial variations of the $\Psi$-azimuthally
averaged (see Equation~(\ref{equation:psi})) components of the magnetic field
vector. Since sunspots are not usually axisymmetric we will employ
ellipses, as illustrated in Figure~\ref{figure:fig7}, to calculate
those averages. The ellipses are determined by first obtaining the
coordinates of the center of the umbra:
$\{x_{\beta,u};x_{\alpha,u}\}$, and then fitting ellipses with
different major and minor semi-axes, such that the outermost blue
ellipses in Figure~\ref{figure:fig7} provides a good match to the
boundary between the penumbra and the quiet Sun. The upper panel in
Figure~\ref{figure:fig7} shows the ellipses for AR~10923 observed on
November 14, 2006 at $\Theta$~=~8.7\textdegree, whereas the lower panel shows
AR~10933 observed on January 9, 2007 at $\Theta$~=~49.0\textdegree.

The radial variation of the azimuthal averages is presented in
Figure~\ref{figure:fig8}. The vertical bars in this figure represent
the standard deviation for all considered points along each ellipse's
perimeter. Note that, although the scatter is significant, the radial
variation of the different components of the magnetic field vector are
very well defined. Furthermore, both sunspots (AR~10923 in the upper
panels; AR~10933 in the lower panels) show very similar behaviors of
the magnetic field vector with $r/R_s$ ($R_s$ refers to the total
sunspot radius). This happens for all relevant physical quantities:
the total magnetic field strength $B_{\mathrm{tot}}(r,\bar{\tau})$ (green
curve), the vertical component of the magnetic field vector
$B_\rho(r,\bar{\tau})$ (red curve), as well as for the horizontal
component of the magnetic field vector
$B_{h}(r)=\sqrt{B_\beta^2(r,\bar{\tau})+B_{\alpha}^2(r,\bar{\tau})}$
(blue curve). Consequently, the radial variation of the inclination of
the magnetic  field vector with respect to the vertical on the solar
surface $\zeta$, which can be obtained from $B_\rho$ and $B_h$
(Equation~(\ref{equation:zeta})) is also very similar for both sunspots
(right panels in Figure~\ref{figure:fig8}).

The vertical component of the magnetic field vector $B_\rho$
monotonously decreases with the radial distance from the center of the
umbra, while the transverse component of the magnetic field, $B_{h}$,
first increases until $r/R_s \sim 0.5$, and decreases afterward. In
both sunspots the vertical and the transverse component become equally
strong close to $r/R_s \sim 0.5$, which results in an inclination for
the magnetic field vector of $\zeta \simeq 45\deg$ exactly in the
middle of the sunspot radius (right panels in
Figure~\ref{figure:fig8}). This location is very close to the
umbra-penumbra boundary, which occurs at approximately $r/R_s \simeq
0.4$ (vertical dashed lines in Figure~\ref{figure:fig8}). The
inclination of the magnetic field $\zeta$ monotonously increases from
the center of the sunspot, where it is considerably vertical ($\zeta
\simeq 10\mbox{\,--\,}20\deg$), to the outer penumbra, where it becomes almost
horizontal ($\zeta \simeq 80\deg$). Furthermore, the inclination at
individual regions at large radial distances from the sunspot's center
can be truly horizontal ($\zeta = 90\deg$) or, as indicated by the
vertical bars in Figure~\ref{figure:fig8}, the magnetic field vector
can even point downwards in the solar surface, with $B_\rho<0$ at certain
locations. This is also clearly noticeable in the black contours in
Figures~\ref{figure:fig_amb1} and \ref{figure:fig_amb4}. Before
Hinode/SP data became available, detecting these patches where the
magnetic field returns into the solar surface \citep{luis2007} was not
possible unless more complex inversions (not ME-like) were carried out
(see Section~\ref{subsection:verticaltau}). Nowadays with Hinode's
0.32" resolution, these patches which sometimes can be as long as
3\,--\,4~Mm, are detected routinely \citep[see
  Figure~\ref{figure:fig_amb1}; also Figure~4 in][]{luis2007}. Note
that theoretical models for the sunspot magnetic field allow for the
possibility of returning-flux at the edge of the sunspot
\citep{osherovich1982,osherovich1983,osherovich1984}.

All these results are consistent with previous results obtained from
Milne--Eddington inversions such as:
\cite{lites1993,stanchfield1997,luis2002, luis2007}. Although most of
these inversions were also obtained from the analysis of
spectropolarimetric data in the Fe\,{\sc i} line pair at 630~nm, a few
of them also present maps of the magnetic field vector obtained from
other spectral lines such as C\,{\sc i} 538.0~nm and Fe\,{\sc i} 537.9~nm
\citep{stanchfield1997}, or Fe\,{\sc i} 1548~nm in
\cite{luis2002}. Analysis of spectropolarimetric data employing other
techniques such as the magnetogram equation, which yields the vertical
component of the magnetic field at a constant $\tau$-level, have also
been carried out by other authors \citep{nazaret2005}. The consistency
between all the aforementioned results is remarkable, specially if we
consider that each work studied different sunspots and employed
different spectral lines.

The picture of a sunspot that one draws from these radial variations
is that of a vertical flux tube, with a diameter of 30\,--\,40~Mm
(judging from Figure~\ref{figure:fig7}), where the magnetic field is
very strong and vertical at the flux tube's axis (umbral center),
while it becomes weaker and more horizontal as we move towards the
edges of the flux tube. Even though these results were obtained only
for a fixed $\tau$-level on the solar photosphere, they clearly
indicate that the flux tube is expanding with height as the magnetic
field encounters a lower density plasma. The overall radial variations
of the components of the magnetic field seem to be independent of the
sunspot size, although the maximum field strength (which occurs at the
sunspot's center) clearly does, as illustrated by
Figure~\ref{figure:fig8}, where the magnetic field strength for
AR~10923 peaks at about 3300~Gauss (large sunspot), whereas for
AR~10933 (small sunspot) it peaks at around 2900~Gauss. This has been
further demonstrated by several works that employed data from many
different sunspots \citep{ringnes1960, brants1982, kopp1992,
  collados1994, livingston2002, jin2006}.

\epubtkImage{ics_14nov06-ics_09jan07.png}{%
  \begin{figure}[htbp]
  \centerline{\includegraphics[width=10cm]{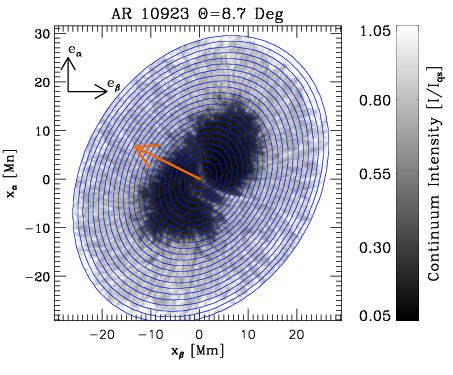}}
  \centerline{\includegraphics[width=10cm]{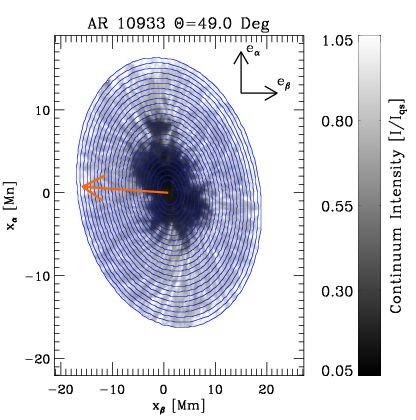}}
    \caption{Map of the continuum intensity for two sunspots. The top
      panel shows AR~10923 observed at $\Theta$~=~8.7\textdegree, whereas the
      bottom panel shows AR~10933, observed at
      $\Theta$~=~49.0\textdegree. These are the same sunspots as discussed in
      Sections~\ref{subsection:tools} and
      \ref{subsubsection:180deg}. The blue ellipses are employed to
      determine the azimuthal averages ($\Psi$-averages) of the
      magnetic field vector. Note that the outermost ellipse tries to
      match the boundary between the penumbra and the quiet Sun. The
      orange arrow points towards the center of the solar disk.}
    \label{figure:fig7}
\end{figure}}

\epubtkImage{brad-incrad_14nov06-brad-incrad_09jan07.png}{%
  \begin{figure}[htbp]
    \centerline{\includegraphics[width=6.6cm]{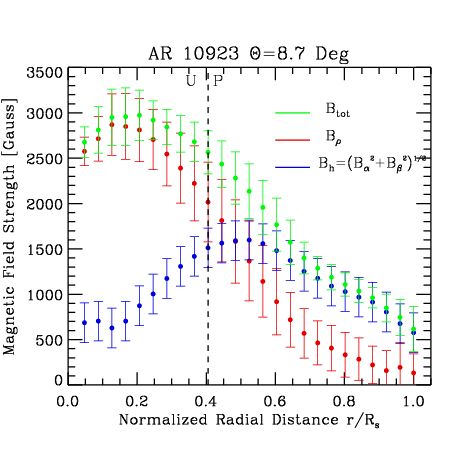}\qquad
      \includegraphics[width=6.6cm]{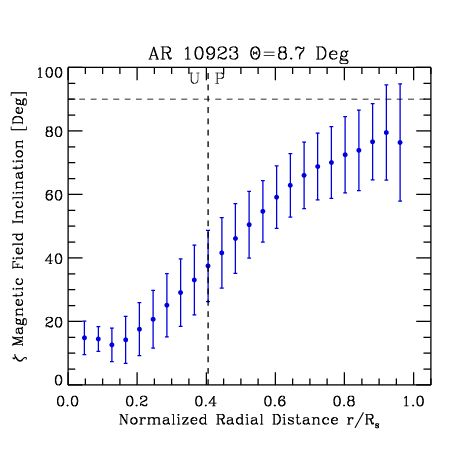}}
     \centerline{\includegraphics[width=6.6cm]{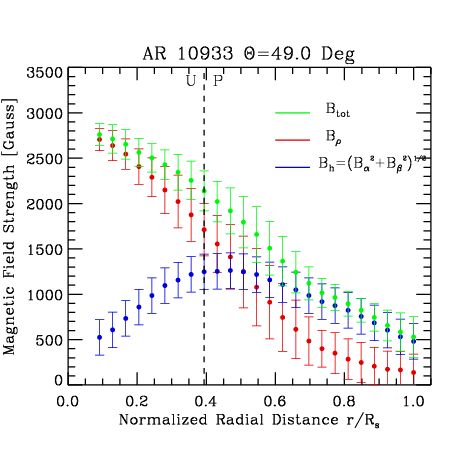}\qquad
      \includegraphics[width=6.6cm]{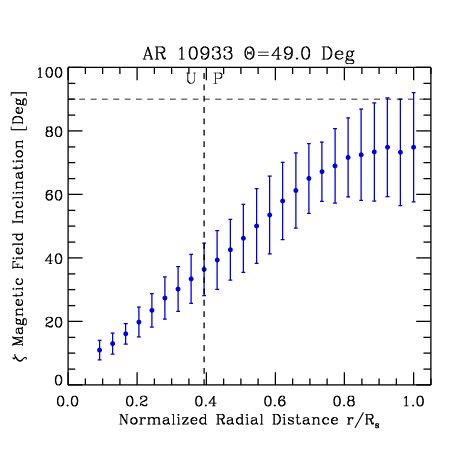}}
    \caption{\emph{Left panels}: Azimuthally averaged components of
      the magnetic field vector as a function of the normalized radial
      distance $r/R_s$ from the sunspot's center. The magnetic field
      corresponds to a constant $\tau$-level. In green the total
      magnetic field strength $B_{\mathrm{tot}}(r,\bar{\tau})$ is presented
      while red and blue refer to the vertical $B_\rho$ and horizontal
      $B_h$ components of the magnetic field. Top
      panel shows the radial variations for AR~10923 and the bottom
      panel refers to AR~10933 (see Figure~\ref{figure:fig7} for
      details). \emph{Right panels}: inclination at a constant
      $\tau$-level of the magnetic field vector with respect to the
      vertical direction on the solar surface, as a function of the
      normalized radial distance from the sunspot's center:
      $\zeta(r,\bar{\tau})$ (see Equation~(\ref{equation:zeta})). The
      horizontal dashed line is placed at $\zeta$~=~90\textdegree, indicating
      when the magnetic field points downwards on the solar
      surface. The vertical dashed line at $r/R_s \simeq 0.4$ is
      placed at the boundary between the umbra and the penumbra.}
    \label{figure:fig8}
\end{figure}}

As explained in Section~\ref{subsubsection:formationheights},
Figures~\ref{figure:fig_amb1}, \ref{figure:fig_amb2},
\ref{figure:fig_amb3}, and \ref{figure:fig8} refer to the average
magnetic field vector in the photosphere: $\bar{\tau} \in
[1,10^{-3}]$. This is because they were obtained from the
MilneEddington inversion of spectropolarimetric data for the line
Fe\,{\sc i} pair at 630~nm. The investigations of the magnetic field
vector in the chromosphere is far more complicated, since Non-Local
Thermodynamic Equilibrium (NLTE) conditions make the interpretation of
the Stokes parameters more difficult. However, in the last years a
number of works have addressed some of these issues. For example,
\cite{Orozco2005} analyzes data from the Si\,{\sc i} and He\,{\sc i}
spectral lines at 1083~nm, which are formed in the mid-photosphere and
upper-chromosphere, respectively. They find very similar radial
variations of the magnetic field vector in  the chromosphere and the
photosphere, with the main difference being a reduction in the total
magnetic field strength. Furthermore, \cite{Hector_Ca2005a} has
presented an actual NLTE inversion of the Ca\,{\sc ii} lines at 849.8
and 854.2~nm. These two spectral lines are formed in the photosphere
and chromosphere: $\bar{\tau} \in [1,10^{-6}]$.

\clearpage

\subsection[Vertical-$\tau$ variations]{Vertical-{\boldmath$\tau$}
  variations}
\label{subsection:verticaltau}

The determination of the vertical variations of the magnetic field in
sunspots has been a recurrent topic in Solar Physics for
decades. Traditionally, this determination had been done through a
combination of spectropolarimetric observations, where the magnetic
field is measured at different heights in the solar atmosphere
\citep{franz1972,wittmann1974b}, and theoretical considerations such
as employing a given sunspot model, applying the $\nabla \cdot
\ve{B}=0$ condition, etcetera \citep[][and references
  therein]{hagyard1983,osherovich1984}. Those first attempts were
usually limited to the vertical component of the magnetic field
$B_\rho$:
\begin{equation}
\frac{d B_\rho}{d x_\rho} \approx \frac{B_{\rho,2}-B_{\rho,1}}{x_{\rho,2}-x_{\rho,1}} \;\; \mathrm{[G\ km^{-1}]} ,
\end{equation}
where $x_\rho$ is the coordinate along the direction that is
perpendicular to the solar surface (see Figure~\ref{figure:earthsun})
and has been referred to as $z$ in
Section~\ref{subsubsection:tau2z}. In those early works, inferences of
the vertical gradient of the vertical component of the magnetic field
could differ by as much as an order of magnitude:
1\,--\,10~G~km\super{-1} \citep{kotov1970}, 0.5\,--\,2~G~km\super{-1}
\citep{makita1976}. Here we will refer, however, to the gradients of the
magnetic field in terms of the optical depth scale
(Equation~(\ref{equation:tauz})):
\begin{equation}
\frac{d B_\rho}{d \tau} \approx \frac{B_{\rho,2}-B_{\rho,1}}{\bar{\tau}_{2}-\bar{\tau}_{1}} \;\; \mathrm{[G]}.
\end{equation}

If $x_{\rho,2}$ and $x_{\rho,1}$ (or alternatively $\bar{\tau}_2$ and
$\bar{\tau}_1$) are sufficiently far apart ($>$~1000~km), the gradient
refers to the average gradient between the chromosphere and the
photosphere. This can be done, for example, employing pairs of lines
where one of them is photospheric and another one is chromospheric:
Fe\,{\sc i} 525.0~nm and C\,{\sc iv} 154.8~nm \citep{hagyard1983},
Fe\,{\sc i} 1082.8~nm and He\,{\sc i} 1083.0~nm \citep{kozlova2009},
Fe\,{\sc i} 630.2~nm and Na\,{\sc i} 589.6~nm
\citep{Leka2003}. Through a Milne--Eddington-like inversion (or
applying a magnetrogram calibration) the vertical component of the
magnetic field can be inferred separately for each line and, thus,
separately for $x_{\rho 1}$ and $x_{\rho,2}$. Another way is to employ
a single spectral line whose formation range is very wide. Examples of
such lines are: Ca\,{\sc ii} 393.3~nm or Ca\,{\sc ii} 854.2~nm. These
lines are sensitive to $\bar{\tau} \in [1,10^{-6}]$ , with $\tau_c=1$
being the photosphere and $\tau_c=10^{-6}$ the chromosphere
\citep{Hector_Ca2005a,Hector_Ca2005b}.

Since the theory of spectral line formation in the chromosphere is not
currently fully understood (see Section~\ref{subsection:tools}), in
this review we will focus mostly in the photospheric gradient of the
magnetic field. To that end, we will employ the Fe\,{\sc i} line pair
at 630~nm observed with Hinode/SP. These two spectral lines are both
formed within a range of optical depths of $\bar{\tau} \in
[1,10^{-3}]$. We perform an inversion of the Stokes vector in these
two spectral lines, assuming that each of the physical parameters in
$\ve{X}$ (Equation~(\ref{equation:x})) change linearly with the logarithm of
the optical depth:
\begin{equation}
\label{equation:eq2nodes}
X_k(\tau_c) = X_k(\log\tau_c=0) + \log\tau_c \left.\frac{dX_k}{d\tau_c}\right|_{\log\tau_c=0} ,
\end{equation}
where $X_k$ refers to the $k$-component of $\ve{X}$. Note that the
inversion cannot be carried out with a Milne--Eddington-like inversion
code, since those assume that the physical parameters do not change
with optical depth: $X_k \ne f(\tau_c)$ (see
Sections~\ref{subsubsection:formationheights} and
\ref{subsubsection:180deg}). Instead, we employ an inversion code that
allows for the inclusion of gradients in the physical parameters. In
this case we have used the SIR inversion code \citep{basilio1992}, but
we could have also employed SPINOR \citep{frutiger1999} or LILIA
\citep{Hector_Inversion2002}. Applying this inversion code allows us
to determine two-dimensional maps of the three components of the
magnetic field vector $B(x_\beta,x_\alpha)$,
$\gamma(x_\beta,x_\alpha)$, and $\varphi(x_\beta,x_\alpha)$ at
different optical depths $\tau_c$
(cf.\ Figures~\ref{figure:sunspotmap1} and \ref{figure:sunspotmap2}).

\epubtkImage{bfrad-bzrad-btrad-inrad_14nov06_2nodes.png}{%
  \begin{figure}[htbp]
    \centerline{\includegraphics[width=6.6cm]{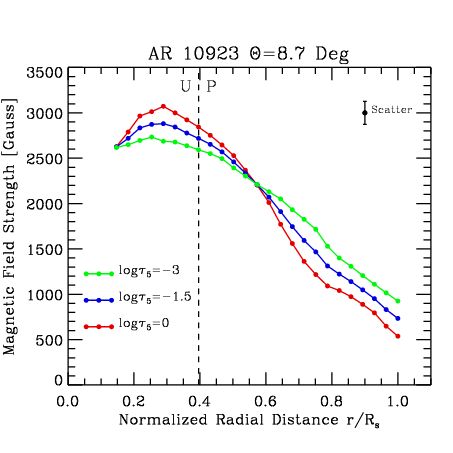}\qquad
      \includegraphics[width=6.6cm]{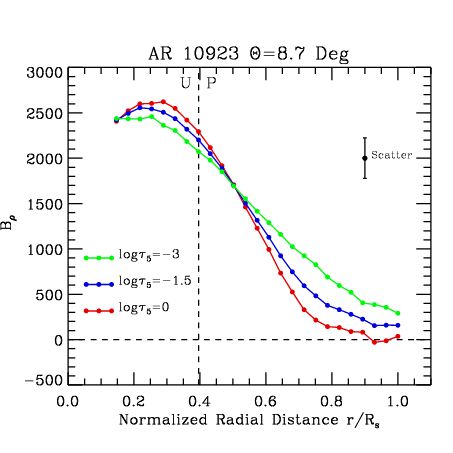}}
     \centerline{\includegraphics[width=6.6cm]{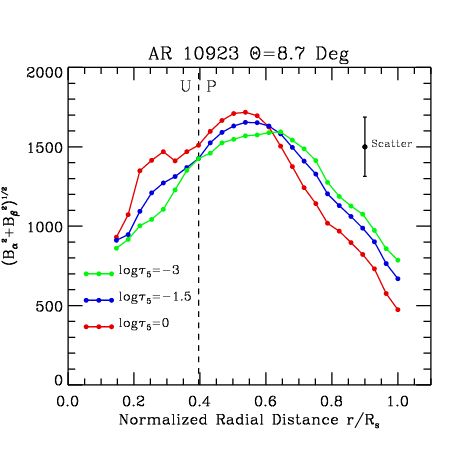}\qquad
      \includegraphics[width=6.6cm]{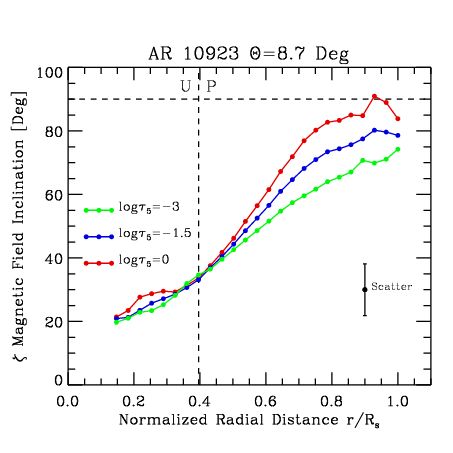}}
    \caption{Azimuthally averaged components of the magnetic field
      vector as a function of the normalized radial distance in the
      sunspot $r/R_s$: total magnetic field strength $B$ (upper-left),
      vertical component of the magnetic field $B_\rho$ (upper-right),
      horizontal component of the magnetic field $B_h$ (lower-left),
      inclination of the magnetic field vector with respect to the
      vertical direction on the solar surface $\zeta$
      (lower-right). Each panel contains three curves, representing
      different optical depths: red is for the deep photosphere or
      continuum level ($\log\tau_c=0$), blue is the mid-photosphere
      ($\log\tau_c=-1.5$), and green is the upper-photosphere
      ($\log\tau_c=-3$). The vertical dashed line at $r/R_s \approx
      0.4$ indicates the separation between the umbra and the
      penumbra. These results correspond to the sunspot AR~10923
      observed on November 14, 2006 at $\Theta$~=~8.7\textdegree\ (see also
      Figure~\ref{figure:sunspotmap1}; upper panels in
      Figures~\ref{figure:fig_amb1}, \ref{figure:fig_amb2},
      \ref{figure:fig_amb3}, and \ref{figure:fig8}).}
    \label{figure:fig9}
\end{figure}}

\clearpage

Once those maps are obtained, the 180\textdegree-ambiguity in the
azimuth of the magnetic field $\varphi$ can be resolved at each
optical depth following the prescriptions given in
Section~\ref{subsubsection:180deg}. This allows to obtain
$B_\rho(x_\beta,x_\alpha,\tau_c)$ (vertical component of the magnetic
field on the solar surface) and
$B_h(x_\beta,x_\alpha,\tau_c)=\sqrt{B_\alpha^2(x_\beta,x_\alpha,\tau_c)+B_\beta^2(x_\beta,y_\alpha,\tau_c)}$
(horizontal component of the magnetic field). By the same method as in
Section~\ref{subsection:constanttau} we then employ ellipses to
determine the angular averages of these physical parameters as a
function of the normalized radial distance in the sunspot:
$r/R_s$. However, as opposed to the previous section, it is now
possible to determine this radial variations at different optical
depths. The results are presented in Figure~\ref{figure:fig9}, in red
color for the deep photosphere ($\log\tau_c=0$), blue for the
mid-photosphere ($\log\tau_c=-1.5$), and green for the high-photosphere
($\log\tau_c=-3$)\epubtkFootnote{According to Equation~(\ref{equation:tauz}),
  $z$ and $\tau_c$ have opposite signs. This indicates that $\tau_c$
  decreases when $z$ increases and, therefore, $\tau_c$ decreases from
  the photosphere to the corona.}.

Figure~\ref{figure:fig9} shows two distinct regions. The first one
corresponds to the inner part of the sunspot: $r/R_s < 0.5$, where the
total magnetic field strength $B_{\mathrm{tot}}$ (upper-left panel) decreases
from the deep photosphere (red color) upwards. This is caused by an
upwards decrease of the vertical $B_\rho$ (upper-right), and
horizontal $B_h$ (lower-left) components of the magnetic field. Also,
in this region the inclination of the magnetic field vector $\zeta$
(lower-right) remains constant with height. From the middle-half of
the sunspot and outwards, $r/R_s > 0.5$, the situation, however,
reverses. The total magnetic field strength, as well as the vertical
and horizontal components of the magnetic field, increase from the
deep photosphere ($\log\tau_c=0$) to the higher photosphere
($\log\tau_c=-3$). In this region, the inclination of the magnetic
field vector $\zeta$ no longer remains constant with $\tau_c$ but it
decreases towards the higher photospheric layers. The actual values of
the gradients are given in Figure~\ref{figure:dbdz}. These values are
close to the lower limits ($<1\mathrm{\ G\ km}^{-1}$) obtained in
early works \epubtkFootnote{We shall mention here that, in order to provide
  the values of the derivatives in terms of the geometrical height
  instead of the optical depth, we have assumed that hydrostatic
  equilibrium holds (see Section~\ref{subsubsection:tau2z}).}
\citep[][and references
  therein]{kotov1970,makita1976,osherovich1984}. However,
Figures~\ref{figure:fig9} and \ref{figure:dbdz} extend those results
for the three components of the magnetic field vector and not only for its
vertical component $B_\rho$. In addition, these figures show a clear
distinction between the inner and the outer sunspot. Although
Figures~\ref{figure:fig9} and \ref{figure:dbdz} show only the results
for AR~10923, the other analyzed sunspot (AR~10933) presents very
similar features.

Similar studies have been carried out in a number of recent works. For
instance, our results are in very good agreement with those from
\cite{westendorp2001} (see their Figure~9) in the value and sign of
the gradients in the different components of the magnetic field. In
our case, as well as theirs, the total magnetic field strength
decreases towards the deep photosphere for $r/R_s > 0.6$. At the same
time the inclination (with respect to the vertical) $\zeta$ increases
towards deeper photospheric layers. This can be interpreted in terms
of the existence of a \textit{canopy} \citep[see also][]{Leka2003},
and is perfectly consistent with a picture in which sunspots are
vertical flux tubes where the magnetic field lines fan out with
increasing height as they meet a plasma with lower densities. Another
interesting result concerns the fact that, once the physical
parameters are allowed to vary with optical depth $\tau$
(Equation~(\ref{equation:eq2nodes})), the evidence for return-flux ($\zeta
>90\deg$) in the deep photosphere becomes more clear: compare
lower-right panels in Figures~\ref{figure:fig8} and
\ref{figure:fig9}. It is important to note that \cite{westendorp2001}
also employed in their inversions spectropolarimetric data from the
Fe\,{\sc i} line pair at 630~nm.

Other spectral lines, such as the Fe\,{\sc i} line pair at 1564.8~nm
were employed by \cite{shibu2003}, who instead found that the
magnetic field strength increases towards deeper layers in the
photosphere at all radial distances in the sunspot: $dB_{\mathrm{tot}}/d\tau >
0$ (see their Figure~15). In addition, they found that the inclination
of the magnetic field $\zeta$ decreases towards deep layers:
$d\zeta/d\tau < 0$ at all radial distances. These results are, 
therefore, consistent with ours as far as the inner part of the sunspot
is concerned, but they are indeed opposite to ours \citep[and
  to][]{westendorp2001} for the sunspot's outer half. Furthermore,
\cite{SanchezCuberes2005}, as well as \cite{balthasar2008}, analyzed
two Fe\,{\sc i} lines and one Si\,{\sc i} line at 1078.3~nm to study
the magnetic structure of a sunspot. From their spectropolarimetric
analysis (see their Figure~11) they inferred a total magnetic field
strength that was stronger in the deep photospheric layers:
$dB_{\mathrm{tot}}/d\tau > 0$ at all radial distances from the
sunspot's center \citep[in agreement with][]{shibu2003}. As far as the
inclination $\zeta$ of the magnetic field is concerned,
\cite{SanchezCuberes2005} obtained different behaviors depending on
the scheme employed to treat the stray light in the
instrument. However, they lend more credibility to the results
obtained with a constant amount of stray light. In this case, they
concluded that $d\zeta/d\tau \approx 0$ for $r/R_s < 0.5$ and
$d\zeta/d\tau > 0$ for $r/R_s >0.5$, which supports our results and
those from \cite{westendorp2001}, but not \cite{shibu2003}. Results
from all the aforementioned investigations are summarized in
Table~\ref{table:gradients}. It is important to mention that, although
\cite{balthasar2008} did not find $dB_{\mathrm{tot}}/dx_\rho >0$ in the outer
half of the sunspot (considered as evidence for a canopy), they did
indeed find this trend outside the visible boundary of the sunspot.

\begin{table}[htbp]
\caption[Sign of the gradients of the different components of the
  magnetic field vectors]{Sign of the gradients of the different
  components of the magnetic field vectors: total magnetic field
  strength $B_{\mathrm{tot}}$, vertical component of the magnetic field vector
  $B_\rho$, horizontal component of the magnetic field vector $B_h$,
  and inclination of the magnetic field vector with respect to the
  vertical direction on the solar surface $\zeta$. The sign of the
  gradients are split in two distinct regions: inner sunspot $r/R_s <
  0.5$, and outer sunspot $r/R_s > 0.5$. Figure~\ref{figure:dbdz}
  gives the actual values.}
\label{table:gradients}
  \centering

  \begin{tabular}{p{2cm}p{2cm}p{2cm}p{2cm}p{2cm}}
        \multicolumn{5}{c}{This work and \cite{westendorp2001}}\\
        \toprule
	$r/R_s$ & $dB_{\mathrm{tot}}/dx_\rho$ & $dB_\rho/dx_\rho$ & $dB_h/dx_\rho$ & $d\zeta/dx_\rho$\\
	\midrule
        $<0.5$ & $<0$ & $<0$ & $<0$ & $\approx 0$\\ 
        $>0.5$ & $>0$ & $>0$ & $>0$ & $<0$\\ 
        \bottomrule
  \end{tabular}

  \vspace{1cm}
  \begin{tabular}{p{2cm}p{2cm}p{2cm}p{2cm}p{2cm}}
        \multicolumn{5}{c}{\cite{shibu2003}}\\
        \toprule
        $r/R_s$ & $dB_{\mathrm{tot}}/d\tau$ & $d\zeta/d\tau$ & $dB_{\mathrm{tot}}/dx_\rho$ & $d\zeta/dx_\rho$\\
        \midrule
        all & $>0$ & $<0$ & $<0$ & $>0$ \\ 
        \bottomrule
  \end{tabular}

  \vspace{1cm}
  \begin{tabular}{p{2cm}p{2cm}p{2cm}p{2cm}p{2cm}}
        \multicolumn{5}{c}{\cite{SanchezCuberes2005,balthasar2008}}\\
        \toprule
        $r/R_s$ & $dB_{\mathrm{tot}}/d\tau$ & $d\zeta/d\tau$ & $dB_{\mathrm{tot}}/dx_\rho$ & $d\zeta/dx_\rho$\\
        \midrule
        $<0.5$ & $>0$ & $\approx 0$ & $<0$ & $\approx 0$ \\ 
        $>0.5$ & $>0$ & $>0$ & $<0$ & $<0$ \\ 
        \bottomrule
  \end{tabular}
\end{table}

In the light of these opposing results it is critical to ask ourselves
where do these differences come from. One possible source is the
spatial resolution of the
observations. \cite{westendorp2001}, \cite{shibu2003}, and 
\cite{SanchezCuberes2005}
employed spectropolarimetric observations at low spatial resolution
(about 1"). The data employed here (Hinode/SP) possess much better
resolution: 0.32". However, this should not be very influential to the
study of the global properties of the sunspot,  since we are
discussing azimuthally or $\Psi$-averaged quantities. Another possible
explanation lies in the different formation heights of the employed
spectral lines. Since each set of spectral lines samples a slightly
different $\bar{\tau}$ region in the solar photosphere, they might be
sensing slightly different magnetic fields, thereby yielding gradients
(see Table~\ref{table:gradients}). This is a plausible explanation
because the Fe\,{\sc i} line pair at 1564.8~nm sample a deep and
narrow photospheric layer: $\bar{\tau} \in [3,3\times10^{-2}]$, as
compared to $\bar{\tau} \in [1,10^{-3}]$ for the Fe\,{\sc i} lines at
630~nm (see, for example, Figures~3 and 4 in \citealp{shibu2003}, and Figure~3 in \citealp{luis2000}). Indeed, the different
formation heights have been exploited by numerous authors
\citep{luis2002,shibu2003,borrero2004,Borrero2008ApJ} in order to
explain the opposite gradients obtained from different sets of
spectral lines in terms of penumbral flux tubes and the fine structure
of the sunspot (see, also, Sections~\ref{subsubsection:spine_intraspine}
and \ref{subsubsection:filaments}). The role of the sunspot's fine
structure is emphasized by the fact that the scatter bars (produced by
the inversion of individual pixels; see Figures~\ref{figure:fig9} and
\ref{figure:dbdz}) are of the order of, or even larger than, the
differences between the magnetic field at the different atmospheric
layers chosen for plotting. To solve this problem one would like to
analyze, ideally, many different spectral lines formed at different
heights (Section~\ref{subsubsection:formationheights}). This approach
has been already followed by the recent works of
\cite{dani2008} and \cite{beck2010}, where simultaneous and co-spatial
spectrolarimetric observations in Fe\,{\sc i} 630~nm and Fe\,{\sc i}
1564.8~nm where analyzed. Their results further emphasize the role of
the fine structure of the sunspot in the determination of the vertical
gradients of the magnetic field vector.

A final possibility to explain the difference in the gradients
obtained by different authors could be the different treatments
employed to model the scattered light in the instrument. Arguments in
favor of this possibility are given by \cite{SanchezCuberes2005} and
\cite{solanki2003}. Arguments against the results being affected by
the treatment of the scattered light have been presented in
\cite{Borrero2008ApJ}. Moreover, in \cite{dani2006}, \cite{dani2007}, 
and \cite{dani2008}
a careful correction for scattered light was performed, and still the
fine structure of the sunspot had to be invoked to explain the
observed gradients in the magnetic field vector.

\epubtkImage{dbdz_2nodes.png}{%
  \begin{figure}[htbp]
    \centerline{\includegraphics[width=12cm]{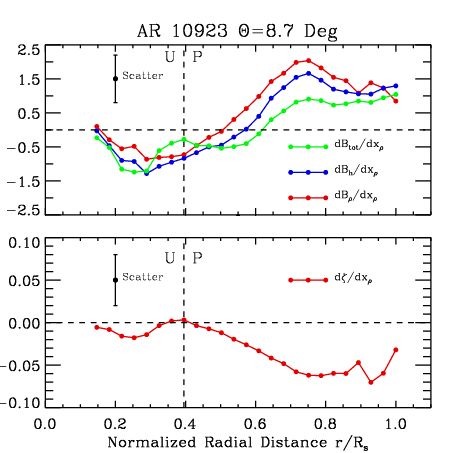}}
    \caption{\emph{Top panel}: vertical derivatives of the different
      components of the magnetic field vector as a function of the
      normalized radial distance in a sunspot: $r/R_s$. Total field
      strength $dB_{\mathrm{tot}}/dx_\rho$ (green), horizontal
      component of the magnetic field $dB_{h}/dx_\rho(r)$ (blue),
      vertical component of the magnetic field $dB_{\rho}/dx_\rho$
      (red). \emph{Bottom panel:} same as above but for the
      inclination of the magnetic field vector with respect to the
      vector perpendicular to the solar surface: $d\zeta/dx_\rho$. The
      vertical dashed line at $r/R_s \simeq 0.4$ represents the
      umbra-penumbra boundary. The vertical solid lines gives an idea
      about the standard deviation (from all pixels across a given
      ellipse in Figure~\ref{figure:fig7}). These results correspond
      to AR~10923, observed on November 14, 2006 at $\Theta$~=~8.7\textdegree.}
    \label{figure:dbdz}
\end{figure}}

\clearpage

\subsection{Is the sunspot magnetic field potential?}
\label{subsection:potential}

The potentiality of the magnetic field vector in sunspots is often
studied by means of the current density vector $\ve{j} =
\frac{1}{\mu_0} \nabla \times \ve{B}$ (in SI units). Theoretical
models for sunspots usually come in two distinct flavors attending to
the vector $\ve{j}$: those where the currents are localized at the
boundaries of the sunspot (current sheets) and the magnetic field
vector is potential elsewhere \citep{simon1970,meyer1977,pizzo1990},
and those where there are volumetric currents distributed everywhere
inside the sunspot \citep{pizzo1986}. From an observational point of
view, in order to evaluate $\ve{j}$ it is necessary to calculate the
vertical derivatives of the three components of the magnetic field
vector: $dB_\alpha/dx_\rho$, $dB_\beta/dx_\rho$, and
$dB_\rho/dx_\rho$. This is not possible through a Milne--Eddington
inversion, because it assumes that the magnetic field vector is
constant with height: $\tau_c$ or $x_\rho$
(Equation~(\ref{equation:tauz})). In this case it is only possible to
determine the vertical component of the current density vector,
$j_\rho$ (aka $j_z$), because it involves only the horizontal
derivatives:
\begin{equation}
\label{equation:jz}
j_\rho = (\nabla \times \ve{B})_z= \frac{1}{\mu_0}\left[\frac{dB_\alpha}{dx_\beta}-\frac{dB_\beta}{dx_\alpha}\right].
\end{equation}

An example of the vertical component of the current density vector,
$j_\rho$, obtained from a ME inversion is presented in
Figure~\ref{figure:jz}. This corresponds to the sunspot observed in
November 14, 2006 with Hinode/SP at $\Theta$~=~8.7\textdegree. The derivatives
in Equation~\ref{equation:jz} have been obtained from
Figures~\ref{figure:fig_amb2} and \ref{figure:fig_amb3}. Because the
magnetic field vector was obtained from a ME inversion, these
derivatives of the magnetic field vector refer to a constant optical
depth $\bar{\tau}$ in the atmosphere. As long as the
$\bar{\tau}(x_\rho)$-surface (Wilson depression) is not very
corrugated (small pixel-to-pixel variations) and that the
vertical-$\tau$ variations of the magnetic field vector are not very
strong (Equation~(\ref{equation:mecondition})), it is justified to assume
that the maps in Figures~\ref{figure:fig_amb2} and
\ref{figure:fig_amb3} also correspond to a constant geometrical height
$x_\rho$. If these assumptions are in fact not valid, artificial
currents in $j_\rho$ might appear as a consequence of measuring the
magnetic field at different heights from one pixel to the next one.

Note that prior to the calculation of currents, the
180\textdegree-ambiguity in the azimuth of the magnetic field vector
must be solved (see Section~\ref{subsubsection:180deg}). The final
results for $j_\rho$ are displayed in Figure~\ref{figure:jz}, where it
can be seen that the vertical component of the current density vector
is highly structured in radial patterns resembling penumbral
filaments. The values of the current density are of the order of
$|j_\rho| < 75\mathrm{\ mA\ m}^{-2}$. This value is consistent with previous
results obtained with different instruments and, therefore, different
spectral lines and spatial resolutions: $|j_\rho| < 50\mathrm{\ mA\ m}^{-2}$
(Figure~10 in \citealp{li2009}; 2" and 
Fe\,{\sc i} 630~nm), $|j_\rho| < 40\mathrm{\ mA\ m}^{-2}$ 
(Figure~8 in \citealp{balthasar2008}; 
$\approx$~0.9" and Si/Fe\,{\sc i} 1078~nm), 
$|j_\rho| < 100\mathrm{\ mA\ m}^{-2}$ 
(Figure~3 in \citealp{shimizu2009}; 
0.32" and Fe\,{\sc i} 630~nm). These various results show a weak tendency for
the current to increase with increasing spatial resolution. However,
this result is to be taken cautiously, since at low spatial
resolutions two competing effects can play a role. On the one hand, a
better spatial resolution can detect larger pixel-to-pixel variations
in the magnetic field and, thus, yields larger values for $j_z$. On the
other hand, a worse spatial resolution can leave certain magnetic
structures unresolved and, in this case, the finite-differences
involved in the Equation~(\ref{equation:jz}) can produce artificial
currents where originally there were none.

A curious effect worth noticing is the large and negative values of
$j_\rho$ in Figure~\ref{figure:jz} around the sunspot center that
describe an oval shape (next to the light bridges). This is an
artificial result produced by an incorrect solution to the
180\textdegree-ambiguity in the azimuth of the magnetic field close to
the umbral center (see Section~\ref{subsubsection:180deg}). An
incorrect choice between $\varphi$ and $\varphi+\pi$ (see for instance
Equation~(\ref{equation:bfieldinlos})) can lead to very large and unrealistic
pixel-to-pixel variations in $dB_\alpha/dx_\beta$ or
$dB_\beta/dx_\alpha$. Thus, regions where large values of $j_\rho$ are
consistently obtained can sometimes be used to identify places where
the solution to the 180\textdegree-ambiguity was not correct. Indeed, many
methods to solve the 180\textdegree-ambiguity minimize $j_\rho$ in
order to choose between the two possible solutions in the azimuth of
the magnetic field vector \citep[][see also
  Section~\ref{subsubsection:180deg}]{Metcalf2006}.

In order to compute the horizontal component of the current density
vector $j_h=\sqrt{j_\alpha^2+j_\beta^2}$, it is necessary to analyze
the spectropolarimetric data employing an inversion code that allows
to retrieve the stratification with optical depth in the solar
atmosphere (see Section~\ref{subsection:verticaltau}). Even in this
case, the derivatives must be evaluated in terms of the geometrical
height $x_\rho$ instead of the optical depth $\tau_c$. Because the
conversion from these two variables assuming hydrostatic equilibrium
is not reliable in sunspots (see Section~\ref{subsubsection:tau2z}),
$j_h$ is not something commonly found in the literature. As a matter
of fact, most inferences of $j_h$ were performed through indirect
means \citep{ji2003,manolis2004}. Very recently, however,
\cite{Puschmann2010b} have been able to determine the full current
density vector $\ve{j}$ from purely observational means (inversion of
Stokes profiles including $\tau_c$-dependence) plus a proper
conversion between $\tau_c$ and $x_\rho$ \citep[][see also
  Section~\ref{subsubsection:tau2z}]{Puschmann2010a}. In the latter
two works, the authors found that the horizontal component of the
current density vector is about 3\,--\,4 times larger than the
vertical one: $j_h \approx 4 j_\rho$. Figure~\ref{figure:jz_klaus}
reproduces Figure~1 from \cite{Puschmann2010b}, which shows the
$\ve{j}$ vector in a region of the penumbra. $j_\rho$ (they refer to
it as $j_z$) also shows radial patterns as in our
Figure~\ref{figure:jz}. More importantly, $j_h$ is strongest in the
vicinity of the regions where $j_\rho$ is large.

\epubtkImage{jz_14nov06.png}{%
  \begin{figure}[htb]
    \centerline{\includegraphics[width=12cm]{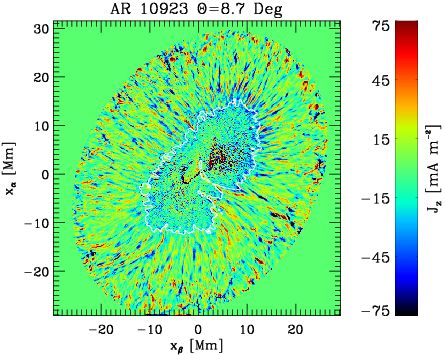}}
    \caption{Same as
      Figures~\ref{figure:fig_amb1}\,--\,\ref{figure:fig_amb5} but for the
      vertical component of the current density vector $j_z$ (or
      $j_\rho$) in sunspot AR~10923.}
    \label{figure:jz}
\end{figure}}

\epubtkImage{jz_klaus_small.png}{%
  \begin{figure}[htb]
    \centerline{\includegraphics[width=10cm]{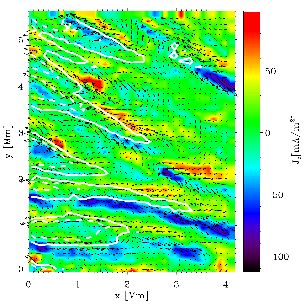}}
    \caption{Same as Figure~\ref{figure:zw_klaus} but for the vertical
      component of the current density vector: $j_z$ (or
      $j_\rho$). The arrows indicate the horizontal component of the
      current density vector: $j_\beta$ and $j_\alpha$. The white
      contours enclose the area where the vertical component of the
      magnetic field, $B_\rho$,  is equal to 650~G (solid) and 450~G
      (dashed) \citep[from][reproduced by permission of the AAS]
      {Puschmann2010b}.}
    \label{figure:jz_klaus}
\end{figure}}

Currents in the chromosphere have also been studied, although to a
smaller extent, by \cite{lagg2003} and \cite{Hector_Ca2005b}. The
latter author finds values for the vertical component of the current
density vector in the chromosphere which are compatible with those in
the photosphere: $|j_z| < 250\mathrm{\ mA\ m}^{-2}$. In addition, the
detected currents are distributed in structures that resemble vertical
current sheets, spanning up to 1.5~Mm in height. The mere presence of
large currents within sunspots clearly implies that the magnetic field
vector is not potential: $\nabla \times \ve{B} \ne 0$.

\clearpage

\subsection[What is the plasma-$\beta$ in sunspots?]{What is the
  plasma-{\boldmath$\beta$} in sunspots?}
\label{subsection:plasmabeta}

In the previous Section~\ref{subsection:potential} we have argued that
the magnetic field vector in sunspots is non-potential. However, in
order to establish its degree of non-potentiality it is important to
develop this statement further. The way this has been traditionally
done is through the study of the plasma-$\beta$ parameter. The
plasma-$\beta$ is defined as the ratio between the gas pressure and
the magnetic pressure:
\begin{equation}
\beta = \frac{P_{g}}{P_{m}} = \frac{8 \pi P_{g}}{B^2} \;\; \text{(in cgs units)}.
\end{equation}

In the solar atmosphere, if $\beta>>1$ the dynamics of the system are
dominated by the plasma motions, which twist and drag the magnetic
field lines while forcing them into highly  non-potential
configurations. If $\beta<<1$ the opposite situation occurs, that is,
the magnetic field is not influenced by the plasma motions. In this
case, the magnetic field will evolve into a state of minimum energy
which happens to coincide with a potential configuration \citep[see
  Chapter~3.4 in][]{priest1984}. Therefore, many works throughout the
literature focus on the plasma-$\beta$ in order to study the
potentiality of the magnetic field. Here, we will employ our results
from the inversion of spectropolarimetric data in
Section~\ref{subsection:verticaltau} to investigate the value of the
plasma-$\beta$ parameter in a sunspot.  Figure~\ref{figure:plasmabeta}
shows the variation of the azimuthally averaged plasma-$\beta$ (along
ellipses in Figure~\ref{figure:fig7}) as a function of the normalized
radial distance in the sunspot $r/R_s$. This figure displays $\beta$
at four different optical depths, from the deep photosphere $\tau_c=1$
(yellow) to the high-photosphere $\tau_c=10^{-3}$ (blue). This figure
shows that $\beta<<1$ above $\tau_c \le 10^{-2}$ and, thus, the
magnetic field can be considered to be nearly potential (or at least
force-free) in these high layers. At $\tau_c=1$ (referred to as
\textit{continuum}) $\beta \ge 1$ and, therefore, the magnetic field is
non-potential. At the intermediate layer of $\tau_c=0.1$ (around 100
kilometers above the continuum) the magnetic field is nearly potential
in the umbra, but it cannot be considered this way in the penumbra:
$r/R_s > 0.4$.

\epubtkImage{plasmabeta.png}{
\begin{figure}[htbp]
\centerline{\includegraphics[width=10cm]{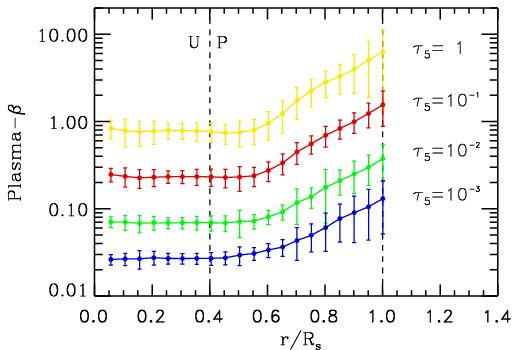}}
\caption{Similar to Figure~\ref{figure:fig8} but for the
  plasma-$\beta$ as a function of the normalized radial distance in
  the sunspot: $r/R_s$. The different curves refer to different
  optical depths in the sunspot: $\tau_c=1$ (yellow), $\tau_c=0.1$
  (red), $\tau_c=10^{-2}$ (green), and $\tau_c=10^{-3}$ (blue). The
  vertical dashed line at $r/R_s \simeq 0.4$ indicates the
  umbra-penumbra boundary.}
\label{figure:plasmabeta}
\end{figure}}

In Figure~\ref{figure:plasmabeta} the gas pressure was obtained under
the assumption of hydrostatic equilibrium
(Section~\ref{subsubsection:tau2z}), which we know not to be very
reliable in sunpots. A more realistic approach was followed by
\citet[][and references therein]{shibu2004}, where an attempt to
consider the effect of the magnetic field in the force balance of the
sunspot was made. Their results for the deep photosphere ($\tau_c=1$)
obtained from the inversion of the Fe\,{\sc i} line pair at 1564.8~nm
are consistent with our Figure~\ref{figure:plasmabeta} (obtained from
the inversion of the Fe\,{\sc i} line pair at 630~nm), with $\beta
\approx 1$ close to the continuum everywhere in the sunspot. Similar
results were also obtained by \citet[][see their
  Figure~4]{Puschmann2010b}, who performed an even more realistic
estimation of the geometrical height scale, considering the three
components of the Lorenz force term ($\ve{j}\times\ve{B}$;
Equation~(\ref{equation:momentum})). In Figure~\ref{figure:plasmabeta_klaus}
we reproduce their results, which further confirm that the $\beta
\approx 1$ in the deep photospheric layers of the penumbra.

\epubtkImage{plasmabeta_klaus.png}{%
  \begin{figure}[htb]
    \centerline{\includegraphics[width=10cm]{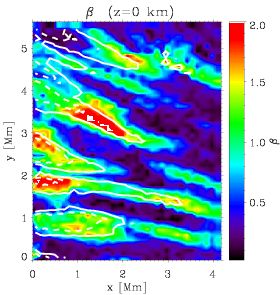}}
    \caption{Same as Figures~\ref{figure:zw_klaus} and
      \ref{figure:jz_klaus} but for the plasma-$\beta$ at $z=0$ in the
      inner penumbra of a sunspot. The white contours are the same as
      in Figure~\ref{figure:jz_klaus}: $B_\rho=650$ (solid white) and
      $B_\rho=450$ (dashed white). This sunspot is AR~10953 observed
      on May 1st, 2007 with Hinode/SP 
      \citep[from][reproduced by permission of the AAS]{Puschmann2010b}.}
    \label{figure:plasmabeta_klaus}
\end{figure}}

These results have important consequences for magnetic field
extrapolations from the photosphere towards the corona, because they
imply that those extrapolations cannot be potential. In addition, as
pointed out by \cite{Puschmann2010b} the magnetic field is not
force-free because in many regions the current density vector $\ve{j}$
and the magnetic field vector $\ve{B}$ are not
parallel. Unfortunately, extrapolations cannot deal thus far with
non-force-free magnetic field configurations. Considering that it has
now become possible to infer the full current density vector $\ve{j}$,
developing tools to perform non-force free magnetic field
extrapolations will be a necessary and important step for future
investigations. These results also have important consequences for
sunspot's helioseismology, because of the deep photospheric location
of the $\beta=1$ region, which is the region  where most of the
conversion from sound waves into magneto-acoustic waves takes place.

In the chromosphere of sunspots, the magnetic field strength is about
half of the photospheric value \citep[see Figure~4
  in][]{Orozco2005}. Therefore, the magnetic pressure in the
chromosphere is only about 25\% of the photospheric value. However,
the density and gas pressure are at least 2\,--\,3 orders of magnitude
smaller. Thus, the chromosphere of sunspots is clearly a low-$\beta$
($\beta << 1$) environment, which in turn means that the magnetic
field configuration is nearly potential.

\subsection{Sunspots' thermal brightness and thermal-magnetic relation}
\label{subsection:wilson}

The Eddington--Barbier approximation can be employed to relate the
observed intensity from any solar structure with a temperature close
to the continuum layer: $\tau=2/3$. This is done by assuming that the
observed intensity is equal to the Planck's function, and solving for
the temperature:
\begin{equation}
I_{\mathrm{obs}} \sim \frac{2hc^2}{\lambda^5} \exp\left\{-\frac{hc}{\lambda K T}\right\}.
\end{equation}

Variations in the observed intensity can be related to a change in the temperature through:
\begin{equation}
\frac{\Delta I_{\mathrm{obs}}}{\Delta T} \sim \frac{d I_{\mathrm{obs}}}{dT} = \frac{hc}{\lambda K T^2}. \\
\label{equation:barbier}
\end{equation}

The observed brightness of a sunspot umbra at visible wavelengths is
about 5\,--\,25\% of the observed brightness of the granulation at the
same wavelength: $I_{\mathrm{umb}} \approx 0.05\mbox{\,--\,}0.25
I_{\mathrm{qs}}$. In the penumbra this number is about 65\,--\,85\% of
the granulation brightness: $I_{\mathrm{pen}} \approx
0.65\mbox{\,--\,}0.85 I_{\mathrm{qs}}$. Assuming that the temperature
at $\tau=2/3$ for the quiet Sun is about 6050~K, the numbers we obtain
from Equation~(\ref{equation:barbier}) are: $T_{\mathrm{umb}}(\tau=2/3) \approx
4800\mathrm{\ K}$ and $T_{\mathrm{pen}}(\tau=2/3) \approx
5650\mathrm{\ K}$.

At infrared wavelengths the difference in the brightness between quiet
Sun and umbra or penumbra is greatly reduced: $I_{\mathrm{umb}}
\approx 0.4-0.6 I_{\mathrm{c,qs}}$ and $I_{\mathrm{pen}} \approx
0.7-0.9 I_{\mathrm{c,qs}}$ \citep[see Figure~1 in][]{shibu2003}. This
happens as a consequence of the behavior of the Planck's function
$B(\lambda,T)$, whose ratio for two different temperatures decreases
towards larger wavelengths. All numbers mentioned thus far are
strongly dependent on the spatial resolution and optical quality of
the instruments. For example, large amounts of scattered light tend to
reduce the intensity contrast and, therefore, temperature differences
between different solar structures.

In Figure~\ref{figure:thermal}, we present scatter plots showing the
relationship between the sunspot's thermal brightness and the
components of the magnetic field vector. These plots have been adapted
from Figure~4 in \cite{shibu2004}. They show $T(\tau=1)$: vs.\ $B$
(total magnetic field strength; upper-left), vs.\ $\zeta$ (zenith
angle -- Equation~(\ref{equation:zeta}) -- upper-right), vs.\ $B_z$ (or our
$B_\rho$ vertical component of the magnetic field; lower-left), and
vs.\ $B_r$ (or our $B_h$ horizontal component of the magnetic field; 
lower-right). As expected, the thermal brightness anti-correlates with the
total field strength $B$ since the latter is larger (see
Figures~\ref{figure:fig_amb1} and \ref{figure:fig8}) in the darkest
part of the sunspot: the umbra. However, the inclination of the
magnetic field $\zeta$ correlates well with the thermal
brightness. Again, this was to be expected (see
Figures~\ref{figure:fig_amb4} and \ref{figure:fig8}) since the
inclination of the magnetic field increases towards the penumbra,
which is brighter (see Figures~\ref{figure:sunspotmap1} and
\ref{figure:sunspotmap2}). As we will discuss intensively throughout
Section~\ref{section:smallscale}, these trivial results have important
consequences for the energy transport in sunspots.

\epubtkImage{thermal_mag_mathew.png}{%
  \begin{figure}[htbp]
    \centerline{\includegraphics[width=\textwidth]{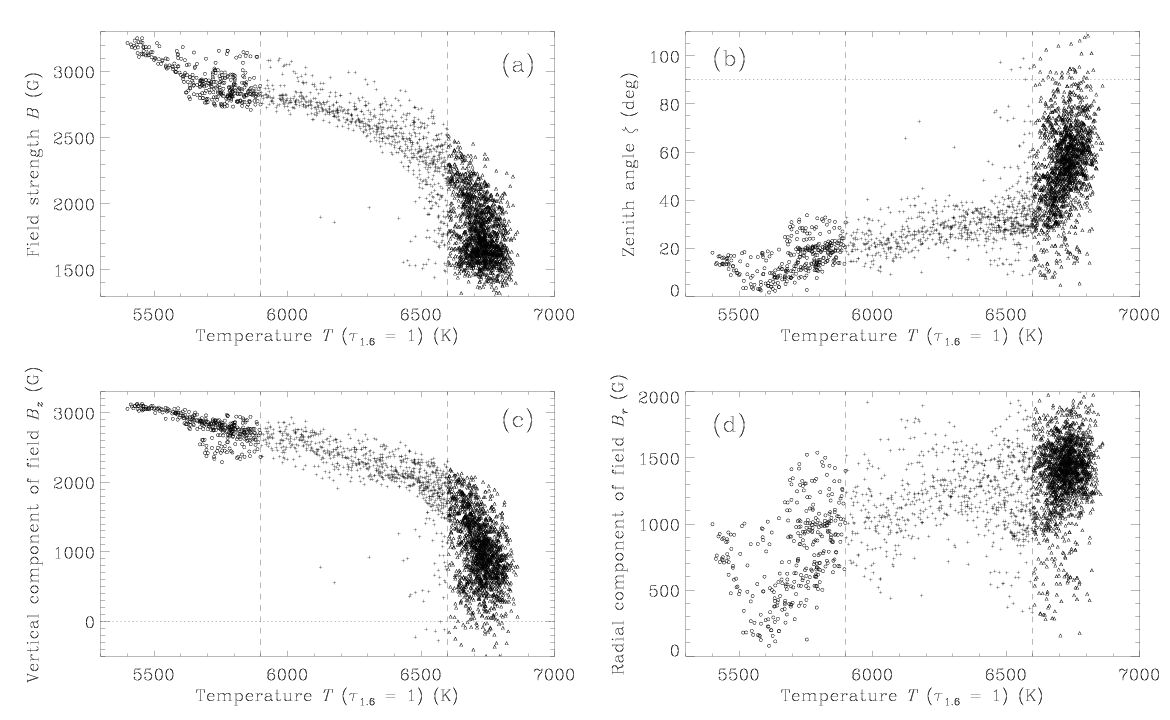}}
    \caption{\emph{Upper-left:} scatter plot of the total field
      strength vs.\ temperature at $\tau_c=1$. \emph{Upper-right:}
      inclination of the magnetic field with respect to the vertical
      direction on the solar surface ($\zeta$; see
      Equation~(\ref{equation:zeta})) versus temperature at
      $\tau_c=1$. \emph{Bottom-left:} vertical component of the
      magnetic field $B_z$ (called $B_\rho$ in our
      Section~\ref{subsection:constanttau}) vs.\ temperature at
      $\tau_c=1$. \emph{Bottom-right:} horizontal component of the
      magnetic field $B_r$ (called $B_h$ in
      Section~\ref{subsection:constanttau}) versus the temperature at
      $\tau_c=1$. In all these panels circles represent umbral points,
      whereas crosses and triangles correspond to points in the
      umbra-penumbra boundary and penumbral points, respectively 
      \citep[from][reproduced by permission of the ESO]{shibu2004}.}
    \label{figure:thermal}
\end{figure}}

\
\clearpage
\subsection{Twist and helicity in sunspots' magnetic field}
\label{subsection:twist}

Let us define the angle of twist of a sunspot's magnetic field,
$\Delta$, as the angle between the magnetic field vector $\ve{B}$ at a
given point of the sunspot and the radial vector that connects that
particular point with the sunspot's center, $\ve{r}$
(Equation~(\ref{equation:opou})):
\begin{equation}
\label{equation:twist}
\Delta = \cos^{-1}\left[\frac{\ve{B}\ve{r}}{|\ve{ B}||\ve{r}|} \right].
\end{equation}

Note that in Section~\ref{subsubsection:180deg} this angle $\Delta$
is precisely the quantity that was being minimized when solving the
180\textdegree-ambiguity (Equation~(\ref{equation:parallel})). However,
minimizing it does not guarantee that $\Delta$ will be zero. This is,
therefore, the origin of the twist: a deviation from a purely radial
(i.e., parallel to $\ve{r}$) magnetic field in the
sunspot. Figure~\ref{figure:fig_amb6} shows maps of the twist angle
$\Delta$ for two different sunspots at two different heliocentric
angles. These two examples illustrate that the magnetic field vector
is radial throughout most of the sunspot, but there are regions  where
significant deviations are observed. These deviations could be
already seen in the arrows in Figures~\ref{figure:fig_amb2} and
\ref{figure:fig_amb3} representing the magnetic field vector in the
plane of the solar surface. In addition, in these two examples the
sign of the twist (wherever it exists) remains constant for the entire
sunspot.

Twisted magnetic fields in sunspots have been observed for a very long
time, going back to the early works of 
\cite{hale1925,hale1927} and \cite{richardson1941}, who observed them in
$\mathrm{H}_\alpha$ filaments. They established what is known as \textit{Hale's
  rule}, which states that sunspots in the Northern hemisphere have a
predominantly counter-clockwise rotation, whereas it is clockwise in
the Southern hemisphere. However, sunspots violating Hale's rule are
common if we attend only at $\mathrm{H}_\alpha$ filaments
\citep{nakagawa1971}. A better estimation of the twist in the magnetic
field lines can be obtained from spectropolarimetric observations. To
our knowledge, the first attempts in this direction were performed by
\cite{stepanov1965}.

Twist in sunspots can also be studied by means of the
$\alpha$-parameter in non-potential force-free magnetic
configurations: $\nabla \times \ve{B} = \alpha \ve{B}$. Another
commonly used parameter is the helicity: $H = \int_{V} \ve{A} \cdot
\ve{B}\,dV$, where $\ve{B}$ is the magnetic field vector and $\ve{A}$
represents the magnetic vector potential. As demonstrated by
\cite{tiwari2009a} the value of $\alpha$ corresponds to twice the
degree of twist per unit axial length. In addition, $\alpha$ and $H$
posses the same sign. Thus, any of these two parameters can be also
employed to study the sign of the twist in the magnetic field
vector. Using these parameters \cite{pevtsov1994} and \cite{abramenko1996} 
found
a good correlation (up to 90\%) between the sign of the twist and the
hemisphere where the sunspot appear (Hale's rule).

Recent works, however, find large deviations from Hale's rule
\citep{pevtsov2005,tiwari2009b}. It has been hypothesized that these
deviations from Hale's rule might indicate a dependence of the twist
with the solar cycle \citep{choudhuri2004}. Other possible
explanations for the twist of the magnetic field in sunspots, in terms
of the solar rotation and Coriolis force, have been offered by
\cite{peter1996} and \cite{yuhong2000}. In particular, the former work
is also able to explain the deviations from Hale's rule observed in
$\mathrm{H}_\alpha$ filaments. However, a definite explanation is yet
to be identified. This might be more complicated than it seems at
first glance because different twisting mechanisms might operate in
different regimes and atmospheric layers. This is supported by recent
spectropolarimetric observations that infer a twist in the magnetic
field that can change sign from the photosphere to the chromosphere
\citep{Hector_Ca2005a}.

\epubtkImage{twist_14nov06-twist_09jan07.png}{%
  \begin{figure}[htbp]
  \centerline{\includegraphics[width=10cm]{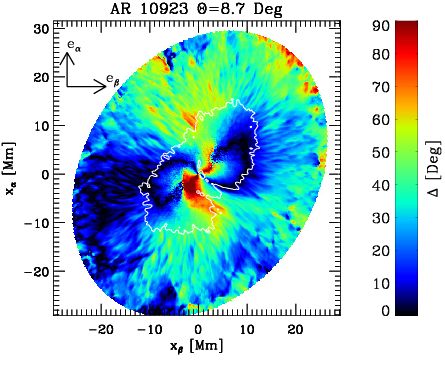}}
  \centerline{\includegraphics[width=10cm]{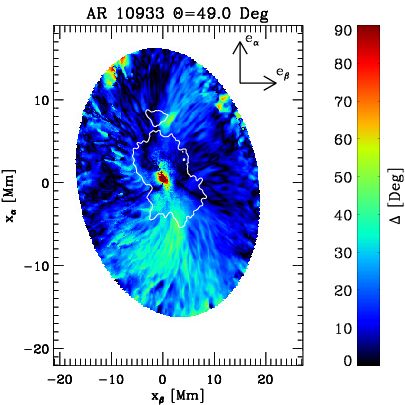}}
    \caption{Same as Figure~\ref{figure:fig_amb4} but for twist angle
      of the magnetic field in the plane of the solar surface:
      $\Delta$ (Equation~(\ref{equation:twist})).}
    \label{figure:fig_amb6}
\end{figure}}
\newpage
\section{The Era of 0.1\,--\,0.5'' Resolution: Small-Scale Magnetic Structures in Sunspots}
\label{section:smallscale}

In Section~\ref{section:global}, we focused on sunspot's
global magnetic structure. To that end we studied the radial variation
of the azimuthally averaged magnetic properties: three components of
the magnetic field vector, plasma-$\beta$, potentiality, currents,
etcetera. In this section, we will investigate the small-scale
structure of the magnetic field. This will help us understand and
identify some of the basic building blocks of the sunspot's magnetic
field, as well as the fundamental physical processes that occur in
sunspots. Another difference with Section~\ref{section:global}, where
only the magnetic field structure was discussed, is that in this
section we will also address the velocity field since they are both
intimately linked at small scales \citep[e.g., Evershed flow;][]{evershed1909}.

In this section spectropolarimetric observations at the highest
spatial resolution will be employed and, instead of discussing
averaged quantities, we will focus mostly in their pixel-to-pixel
variations. In the first part of this section we will address the fine
structure of the umbral magnetic field, whereas the second will be
devoted to the penumbral magnetic field. This division is somewhat
artificial because the current paradigm points towards a clear
relationship between the small-scale structure in these two different
regions \citep{Rimmele2008}. However, there is one important
difference between these two regions (umbra and penumbra), and it has
to do with the mean inclination of the ambient magnetic field $\zeta$:
in the umbra the magnetic field is mostly vertical, whereas in the
penumbra is highly inclined (see Figure~\ref{figure:fig8}). This
difference leads to a somewhat different interaction between the
convective motions and the magnetic field.

\epubtkImage{sst_ud.png}{%
  \begin{figure}[htbp]
    \centerline{\includegraphics[width=10cm]{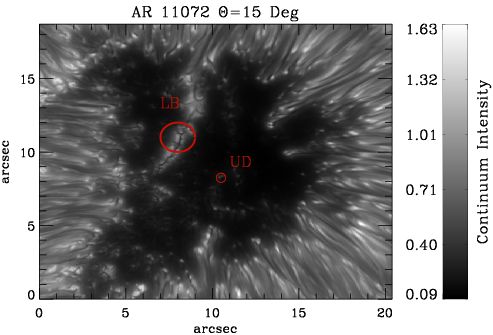}}
    \caption{Map of a sunspot (AR~11072) umbra and inner penumbra
      obtained with the Swedish 1-m Solar Telescope (SST). This sunspot
      was observed on May 23, 2010 at $\Theta$~=~15\textdegree. The
      image was taken with a 10~{\AA} filter located between the
      Ca~\textit{H} and Ca~\textit{K} spectral lines. It was
      subsequently restored using Multi-Object Multi-Frame Blind
      Deconvolution (MOMFBD) technique. The red circles surround a
      local intensity enhancement in the umbral core: \emph{umbral
        dot} (UD), and a portion of a \emph{light bridge} 
      (LB) (adapted from Henriques \textit{et al.}, 2011; in
      preparation).}
    \label{figure:sst_ud}
\end{figure}}

As we saw in Section~\ref{subsection:wilson}, typical temperatures at
$\tau=2/3$ in the umbra and penumbra are approximately 4500~K and
6000~K, respectively. Plasma heated up to this temperature loses
energy in the form of radiation. If the brightness of the umbra and
penumbra is to remain constant, the energy loses due to radiation must
be compensated by some other transport mechanism that will bring
energy from the convection zone into the photosphere. The mechanism
usually invoked is convection. However, the strong magnetic field
present in the sunspots (see Figure~\ref{figure:fig8}) inhibits
convective flows \citep{cowling1953}. This inhibited convection is,
therefore, the reason why umbra and penumbra posses a reduced
brightness compared to the granulation. How do  these convective
movements take place? In the umbra the answer to this question is to be
found in the so-called \textit{umbral dots}, whereas in the penumbra
convection occurs within the \textit{penumbral filaments}.

\subsection{Sunspot umbra and umbral dots}
\label{subsection:umbra}

\subsubsection{Central and peripheral umbral dots}
\label{subsubsection:description_dots}

Umbral dots appear as small-scale regions of enhanced brightness
within the umbral core (see Figure~\ref{figure:sst_ud}). Sizes and
lifetimes of umbral dots have been extensively discussed in the
literature. The current consensus points towards a large selection
bias. Although it is clear that umbral dots are detected at spatial
scales smaller than 1" and temporal scales larger than 2 minutes, it
is not well established whether they posses a \textit{typical} size or
lifetime,  since more and more are detected as the spatial resolution
of the observations increases \citep{sobotka2005,tino2008b}.

Traditionally, umbral dots have been sub-categorized in central (CUDs)
and peripheral umbral dots (PUDs)
\citep{loughhead1979,doerth1986}. This distinction is based upon the
location of the umbral dots: CUDs appear mostly close to the darkest
region of the umbra, whereas PUDs appear commonly at the umbral and
penumbral boundary. Although sometimes disputed 
\citep[see, e.g.,][]{sobotka1997b}, there are many works that claim that these
two families of umbral dots posses very different proper motions
\citep{molowny1994, sobotka1995, tino2008b, watanabe2009}, with the
peripheral ones exhibiting the largest velocities and apparently being
related to inner bright penumbral grains. The physical similarities
between peripheral umbral dots and penumbral grains have been studied
by \cite{Sobotka2009}.

\subsubsection{Thermal and magnetic structure of umbral dots}
\label{subsubsection:magnetic_dots}

The large continuum intensities, as compared to the umbral dark
surroundings, immediately implies (see, for example,
Section~\ref{subsection:wilson}) that the temperature in umbral dots
at $\tau_c=2/3$ is larger than the temperature at the same layer in
the umbral background. Old and current estimates all coincide in a
temperature difference that ranges from 500~K
\citep{doerth1986,tino2008a} up to 1500~K
\citep{ali1997,hector2004}. This temperature difference almost
vanishes about 200\,--\,250~km above $\tau_c=1$: see Figure~8 in
\cite{hector2004}, and Figure~4 in \cite{tino2008a}, which is
reproduced here (in Figure~\ref{figure:tino_umbraldots}).

The strength of the magnetic field inside umbral dots has been a
somewhat controversial subject, with some works finding no large
differences between umbral dots and the umbral background
\citep{lites1989,ali1997}, and other works finding a clear reduction
of the field strength both in central and peripheral umbral dots
\citep{wiehr1993,hector2004,tino2008a}. However, as pointed out by the
the latter works this could again be $\tau$-dependent, with the
differences in the magnetic field being small a few hundred kilometers
above $\tau_c=1$, but fairly large close to this level. Here the
difference can be such that the magnetic field inside the umbral dot
is only a few hundred Gauss (see
Figure~\ref{figure:tino_umbraldots}). The inclination of the magnetic
field $\zeta$ has been found to be only slightly larger than in the
mean umbral background (see Figure~8 in \citealp{hector2004}, and Figure~4 in \citealp{lokesh2009}), which is itself very much
vertical (see Figures~\ref{figure:fig_amb4} and
\ref{figure:fig8}). This will be a recurrent topic in future sections
(Sections~\ref{subsubsection:filaments},
\ref{subsubsection:unified_model}, and \ref{subsubsection:ncp}) when
discussing the differences/similarities between umbral dots, penumbral
filaments, and light bridges.

The smaller field strengths inside umbral dots leads to an enhanced
gas pressure as compared to the surrounding umbra. This is consistent
with the larger temperatures found inside UDs. These numbers can be
employed to derive a Wilson depression of about 100\,--\,200~km, that
is, the $\tau_c=1$ level is formed about 100\,--\,200~km higher in
umbral dots than in the surrounding umbra \citep{hector2004}. This
value is similar to the height difference for the continuum level
between penumbral spines and intraspines (see
Figure~\ref{figure:zw_klaus}). This has important consequences because
the measured differences in the thermal and magnetic structure
correspond to $\tau_c=1$. If the continuum level is actually formed
higher (in the geometrical height scale) inside UDs than in the umbra,
this means that the differences, if measured at the same geometrical
height, would be much larger than the numbers previously cited. This
effect applies indeed, not only to umbral dots, but also in any other
structure in the solar photosphere that is elevated with respect to
its surroundings. Note also that the Wilson depression between the
umbra and umbral dots can be inferred from purely geometrical
considerations of sunspot observations close to the limb
\citep{lites2004,watson2009}.

\epubtkImage{fig_tino1-fig_tino2.png}{%
  \begin{figure}[htbp]
    \centerline{\includegraphics[width=6cm]{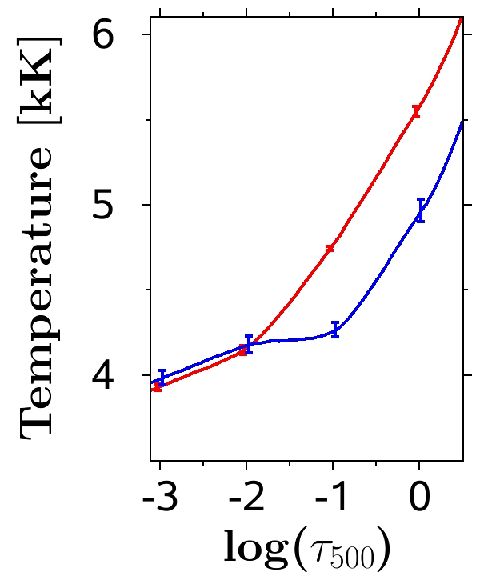}\qquad
      \includegraphics[width=6cm]{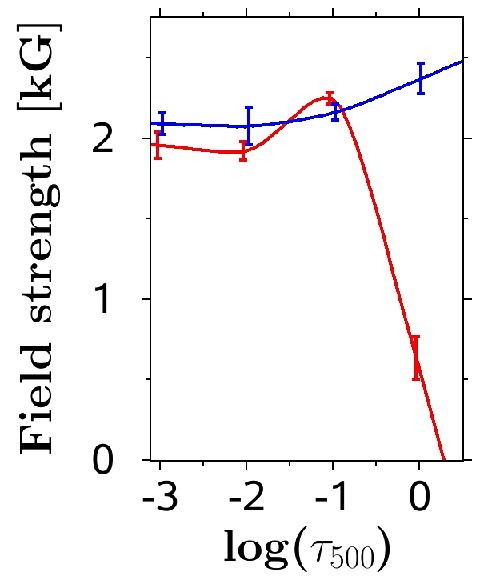}}
    \caption{Vertical stratification (in optical depth $\tau_c$-scale)
      of the temperature (left panel) and the total magnetic field
      strength (right panel). The blue curves shows the stratification
      for the diffuse umbral background, where the red curves
      correspond to the vertical stratification along an umbral
      dot. Close to the continuum, $\tau_c=1$, the umbral dot is much
      hotter and possesses a weaker magnetic field than the
      umbra. These differences disappear about 100\,--\,200~km higher
      in the photosphere: $\log\tau_c \sim -2$
     \citep[from][reproduced by permission of the AAS]{tino2008a}.}
    \label{figure:tino_umbraldots}
\end{figure}}

Note that, the presence of regions inside the sunspot umbra where the
magnetic field is strongly reduced and the temperature and gas
pressure enhanced around $\tau_c=1$, goes along the same lines as
Section~\ref{subsection:plasmabeta}, where we concluded that close to
the continuum level the plasma-$\beta$ is larger than unity. As
explained in Sect~\ref{subsection:potential} this leads to
non-potential configurations for the sunspot magnetic field because
the convective motions are strong enough to drag and twist the
magnetic field lines.

\subsubsection{Signatures of convection in umbral dots}
\label{subsubsection:convection_dots}

As mentioned in Section~\ref{subsection:umbra}, there must exist some
form of convection operating in the umbra of sunspots. The main
candidate for this are the umbral dots. This was motivated by the fact
that umbral dots show enhanced brightness with respect to the umbral
background and, therefore, must be heated more efficiently. In addition,
numerical simulations of umbral magneto-convection \citep{manfred2006}
predict the existence of upflows at the center of umbral dots and
downflows at its edges. As it occurs in the case of penumbral
filaments (see Section~\ref{subsubsection:vertical_motions}), the
search for convective-like velocity patterns in umbral dots has been
hindered by the limited spatial resolution of the observations. For
instance, while upflows ranging from 0.4\,--\,1.0~\kms at the center
of umbral dots have been known for quite some time
\citep{rimmele2004,hector2004,Watanabe2009ud}, downflows have been
much more difficult to detect. However, in the past few years there
have been a few positive detections of downflows at the edges of
umbral dots \citep{lokesh2007,ada2010}. The latter work presents
evidence that supports the numerical simulations of umbral convection
in great detail, with umbral dots that show upflows along their
central dark lane and strong downflows at the footpoints of the dark
lanes \citep[see Figure~3 in][]{ada2010}. This agreement is evident if
we compare the observations from \cite{ada2010} in
Figure~\ref{figure:ada_umbraldots} with the simulations from
\cite{manfred2006} in Figure~\ref{figure:msch_umbraldots}.

The lower magnetic field inside umbral dots mentioned in
Section~\ref{subsubsection:magnetic_dots} is a direct consequence of
the convective motions described here. In the sunspot umbra,
convective motions push the magnetic field lines towards the boundary
of the convective cell, thereby creating a region where the vertical
component of the magnetic field vector is strongly reduced. Since the
ambient magnetic field is vertical, this automatically yields a very
small field inside the umbral dot. At the top of the convective cell
the magnetic field forms a cusp or canopy, preventing the material
from continuing to flow upwards. The pile-up of material at this point
creates a region of locally enhanced density, which is responsible for
the appearance of the central dark lane inside umbral dots
\citep{manfred2006}.

\epubtkImage{ada_ud1-ada_ud2.png}{%
  \begin{figure}[htbp]
    \centerline{\includegraphics[height=8cm]{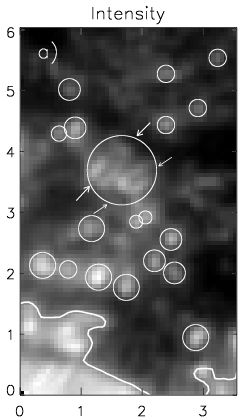}\qquad
      \includegraphics[height=8cm]{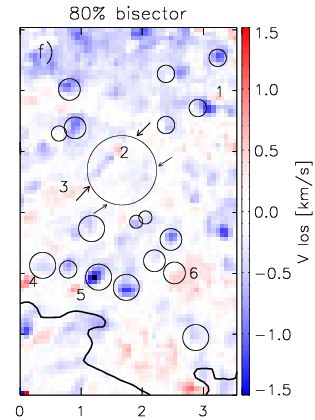}}
    \caption{Results from spectropolarimetric observations. \emph{Left
        panel}: continuum intensity map inside the umbra of a
      sunspot. The circles denote the location of several umbral dots
      (see as intensity enhancements; see also
      \ref{figure:sst_ud}). The largest circle encircles two large
      umbral dots that show prominent central dark lanes. \emph{Right
        panel:} map of the line-of-sight velocity in deep layers. This map shows
      an upflow (blueshift) along the central dark lane and downflows
      (redshift) at the footpoints of the dark lane
      \citep[from][reproduced by permission of the AAS]{ada2010}.}
    \label{figure:ada_umbraldots}
\end{figure}}

\epubtkImage{f1a-f1b_msch.png}{%
  \begin{figure}[htbp]
    \centerline{\includegraphics[width=6cm]{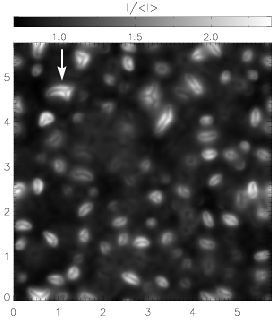}\qquad
      \includegraphics[width=6cm]{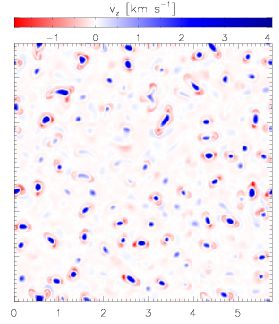}}
    \caption{Results from 3D MHD simulations. \emph{Left panel}:
      continuum intensity in the umbra of a sunspot. \emph{Right
        panel:} map of the line-of-sight velocity. This panel shows upflows
      (blueshift) along the central dark lane of umbral
      dots. Downflows (redshift) are also visible all around the
      central dark lane, although they are stronger at the footpoints
      of the dark lane
      \citep[from][reproduced by permission of the AAS]{manfred2006}.}
    \label{figure:msch_umbraldots}
\end{figure}}

\subsubsection{Light bridges}
\label{subsubsection:light_bridges}

Besides umbral dots, the most striking manifestation of convection in
the umbra appears in the form of \textit{light bridges}. These are
elongated bright features that often split the umbra in two sections
connecting two different sides of the penumbra (see
Figure~\ref{figure:sst_ud}). Light bridges and umbral dots share many
similarities. For instance, both feature a central dark lane and
bright edges. Indeed, light bridges can be considered as an extreme
form of elongated umbral dots. Their larger sizes have actually
allowed for the detection of both blue and redshifted velocities with
only a moderate spatial resolution of 1"
\citep{sobotka1995,leka1997,rimmele1997}.

Recent observations at much better spatial resolution have also been
able to establish a clear connection between upflows and the central
dark lane in light bridges, as well as downflows and the bright edges
of the light bridge \citep{hirzberger2002,berger2003,luc2010}. In
addition, as it also occurs with umbral dots (see
Section~\ref{subsubsection:magnetic_dots}), the magnetic field is
weaker and slightly more inclined in light bridges as compared to the
surrounding umbra \citep{beckers1969,ruedi1995,Jurcak2006}. The nature
of the central dark lane in light bridges is the same as in umbral
dots (Section~\ref{subsubsection:convection_dots}).

\subsubsection{Subsurface structure of sunspots: cluster vs.\ monolithic models}
\label{subsubsection:spageti_vs_monolithic}

The presence of several convective features in the umbra of sunspots
immediately posses the question of whether the convective upflows and
downflows in umbral dots and light bridges extend deep into the solar
interior or, on the contrary, are only a surface effect. Two distinct
theoretical models are usually cited to showcase these two
possibilities: the \textit{cluster} model \citep{parker1979} and the
\textit{monolithic} model \citep{gokhale1972,meyer1974,meyer1977}. In
both cases, convective upflows at the plume's center reach the
photosphere, where they lose their energy via radiative cooling and
sink back into the Sun at the edges of the umbral dots or light
bridges. In the monolithic model the vertical extension of the plumes
is small, leading to a situation in which the plume is completely
surrounded by the sunspot's magnetic field. However, in the cluster
model convective plumes reach very deep into the solar interior,
connecting with field-free convection zone below the sunspot. In the
latter model, what appears as a single flux tube in the photosphere
splits into many smaller flux tubes deeper down, leaving intrusions of
field-free plasma in between the smaller tubes. Inside this intrusions
is where the convection takes place.

It is not possible to distinguish between these two models employing
spectropolarimetric observations because, below $\tau_c=1$, the plasma
is so opaque that no photon can travel from that depth without being
absorbed. Currently, the only observational tool at our disposal that
can allow us to infer the subsurface structure of sunspots is local
helioseismology \citep{Gizon2005,moradi2010}. Although this technique
is still under development, it will hopefully shed some light on this
subject in the near future.

An alternative way of studying the subsurface structure of sunspots is
by means of numerical simulations of solar magneto-convection. Some
recent studies \citep{manfred2006} show that convection can occur in
the umbra in the form of plumes that do not reach more than 1~Mm
beneath the solar surface. These convective plumes are completely
surrounded by the sunspot's magnetic field and manifest themselves in
the photosphere in the form of umbral dots. Furthermore, they also
transport sufficient amounts of energy as to account for the observed
umbral brightness (see Sections~\ref{subsection:wilson} and
\ref{subsection:umbra}; see also Figure~\ref{figure:msch_umbraldots}):
10\,--\,30\% of the quiet Sun. At first glance these simulations seem
to lend support to the monolithic sunspot model. However, the depth of
the simulation box in \cite{manfred2006} is only 1.6~Mm. New
simulations with deeper domains have been presented by
\cite{Rempel2011} and \cite{mark2010}, with boxes of 6.1 and 8.2~Mm
depth, respectively. In these new simulations, umbral dots present a
very similar topology as with shallower boxes. However, light bridges
appear to be rooted very deep, with convective plumes that reach more
than 2~Mm into the Sun \citep[see, for example, Figure~12
  in][]{mark2010}. Further work is, therefore, needed since the current
simulations are not sufficient to completely rule out the cluster
model.

\subsection{Sunspot penumbra and penumbral filaments}
\label{subsection:penumbra}

\subsubsection{Spines and intraspines}
\label{subsubsection:spine_intraspine}

The filamentary structure of sunspot penumbra was recognised early in
the 19th century in visual observations \citep[see review
  by][]{ThomasWeiss2008}. The progress of observational techniques to
attain higher spatial resolution revealed that the sunspot penumbra
consists of radially elongated filaments with a width of 0.2\,--\,0.3"
as seen in continuum images
\citep[e.g.,][]{Danielson1961obs,Muller1976}. Resolving the structure
of the magnetic field with such high resolution is much more difficult
because polarimetric measurements require multiple images taken in
different polarization states, and a longer exposure time in a narrow
wavelength band to isolate the Zeeman signal in a spectral line. For
this reason, until recently many of the investigations of the magnetic
field in the penumbra have reported contradictory results.

A hint of fluctuation in the magnetic field, in association with 
the penumbral filamentary structure, was first reported by 
\cite{BeckersSchroter1969}, who reported that the magnetic 
field was stronger and more horizontal in dark regions of the penumbra.
\cite{WiehrStellmacher1989}, however, found no general relationship
between brightness and the strength of the magnetic field.

Advancement of large solar telescopes at locations with a good seeing
conditions made it possible to better resolve the penumbral
filamentary structure in spectroscopic and polarimetric observations,
and a number of papers on the small-scale magnetic field structures in
sunspot penumbra were published in early 1990s. \cite{Lites1990},
using the Swedish 1-m Solar Telescope (SST) in La Palma, found a
rapid change in the inclination of the magnetic field between some
dark and light filaments near the edge of the penumbra, while the
field strength showed only a gradual variation across the
filaments. \cite{DegenhardtWiehr1991}, using the Gregory Coud\'e Telescope
in Tenerife,  found fluctuations in the inclination of the magnetic
field vector in the penumbra by 7\,--\,14\textdegree,  with steeper
(more vertical) regions having a stronger magnetic field.

\cite{Schmidt1992}, using the German Vacuum Tower Telescope (VTT) in
Tenerife, found more horizontal field lines in dark filaments, while
the strength of the magnetic field did not differ between bright and
dark penumbral filaments. \cite{title1993}, using a series of
Dopplergrams and line-of-sight magnetograms taken by a tunable
narrowband filter equipped on SST, found variations in the
inclination of the magnetic field of about
\textpm~18\textdegree\ across penumbral filaments. \cite{lites1993},
using the Advanced Stokes Polarimeter (ASP) on the Dunn Solar
Telescope at Sacramento Peak, identified radial narrow lanes in the
penumbra where the magnetic field is more vertical and stronger,
thereby naming such regions as \textit{spines}. Their results
indicated that spines feature an azimuthal expansion of the magnetic
field towards the sunspot's border. In addition, they found no clear
evidence for a spatial correlation between spines and brightness. The
correlation between the magnetic field strength and field inclination
(i.e., stronger field in spines) was confirmed by
\cite{stanchfield1997} using ASP data.

With a highly resolved spectrum in the Fe\,{\sc i} 684.3~nm spectral
line, which is formed in the deep photosphere, \cite{Wiehr2000} found
that darker penumbral lanes correlate with a stronger and more
horizontal magnetic field, though the slit of the spectrograph sampled
only a portion of the penumbrae. Better defined polarization maps of
spines were taken with the Swedish 1-m Solar Telescope employing
adaptive optics \citep{Langhans2005}, demonstrating that spines are
regions with stronger and more vertical magnetic field, and that they
are associated with bright penumbral filaments.

\epubtkImage{spot_dc.png}{
\begin{figure}[htb]
\centerline{\includegraphics[width=0.9\textwidth]{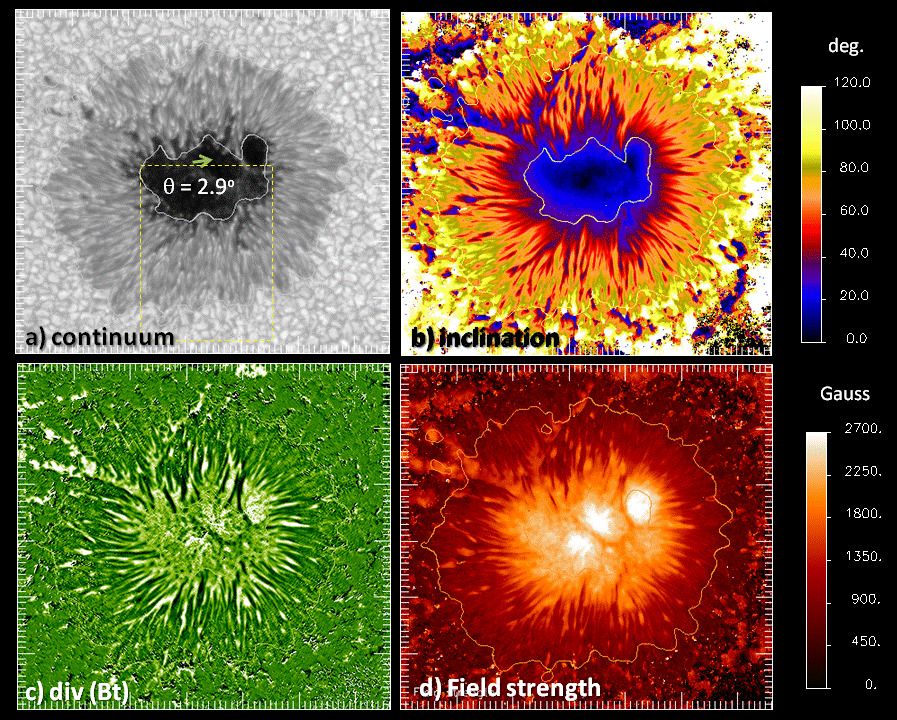}}
\caption{Sunspot AR~10933 observed at $\Theta$~=~2.9\textdegree\ on
  January~5, 2007 with the spectropolarimeter SOT/SP on-board
  Hinode. Displayed are: \emph{a)} continuum intensity $I_c$, \emph{b)}
  magnetic field inclination $\gamma$, \emph{c)} divergence of the
  horizontal component of the magnetic field vector $\nabla \cdot
  \ve{B}_h$, and \emph{d)} total field strength $B$. All parameters
  were obtained from a Milne--Eddington inversion of the recorded
  Stokes spectra. The green arrow in panel \emph{a} indicates the
  direction of the center of the solar disk. The yellow box surrounds
  the sunspot region displayed in Figure~\ref{figure:pen_updown}.}
\label{figure:spot_dc}
\end{figure}}

High quality vector magnetograms with a high spatial resolution are
now routinely obtained by the spectropolarimeter (SP) on-board
Hinode. Figure~\ref{figure:spot_dc} (panels \emph{a} and \emph{b})
show the continuum intensity and the inclination of the magnetic field
for a sunspot observed on January~5, 2007 (AR~10933), located very
close to the center of the solar disk ($\Theta \approx 2.9\deg$). The
field inclination was derived by a Milne--Eddington inversion
(Section~\ref{subsection:tools}) as the angle between the magnetic
field vector and the line-of-sight, $\gamma$ (see
Equation~\ref{equation:bfieldinlos}), but because of the proximity of
the sunspot to the disk center, the inclination can be regarded as the
inclination of the magnetic field, $\zeta$
(Equation~\ref{equation:zeta}), with respect to the local normal to
the solar surface: $\ep$ (see Figure~\ref{figure:earthsun}). It is
obvious in the inclination map that the penumbrae consists of radial
channels that have alternative larger and smaller field inclination. A
close comparison with the continuum image shows that more horizontal
field channels in panel \emph{b} (also called \textit{intraspines})
tend to be bright filaments in inner penumbra but to be dark filaments
in outer penumbra. Panels \emph{c} and \emph{d} in
Figure~\ref{figure:spot_dc} show the divergence of transverse
component of the magnetic field vector $(\nabla \cdot \ve{B}_h)$ and
the total field strength $B$, respectively, obtained from the
Milne--Eddington inversion. It is confirmed that \textit{spines} have
stronger field than \textit{intraspines}, as well as a positive field
divergence. Also noticeable is the presence of a number of patches
that have opposite  polarity to the sunspot around the outer border of
the penumbra (see also
Figures~\ref{figure:fig_amb1} and \ref{figure:fig_amb4}).

Thus, the penumbral magnetic field consists of two major components:
\textit{spines} where the magnetic field is stronger and more
vertical with respect to the direction perpendicular to the solar
surface, and \textit{intraspines} where the magnetic field is weaker
and more horizontal. Whereas the magnetic field of the spines possibly
connect with regions far from the sunspot to form coronal loops over
the active region, the magnetic field in the intraspines turns back
into the photosphere at the outer border of the sunspot or extend over
the photosphere to form a canopy \citep{Solanki1992canopy,
  Rueedi1998}. The filamentary structure of the penumbra persists even
after averaging a time series of continuum images over
2\,--\,4.5~hours \citep{Balthasar1996, Sobotka1999}. This suggests
that the two magnetic field components are more or less exclusive to
each other \citep{ThomasWeiss2004,Weiss2006} except for a possible
interaction through reconnection at the interface between them in the
photosphere \citep{Katsukawa2007}. Such structure of the penumbral
magnetic field, i.e., magnetic fields with two distinct inclinations
interlaced with each other in the azimuthal direction, is referred to
as \textit{uncombed penumbra} \citep{SolankiMontavon1993} or
\textit{interlocking comb} structure \citep{ThomasWeiss1992}. The fact
that the magnetic field is weakened in the \textit{intraspines}, as
compared with the \textit{spines}, can also be employed to deduce
through total pressure balance considerations (as we already did in
the case of umbral dots and light bridges; see
Section~\ref{subsubsection:magnetic_dots}) that the
\textit{intraspines} are elevated with respect to the \textit{spines}.

To account for the filamentary structure of penumbra with the uncombed
magnetic fields, some distinguished models, that are under a hot
discussion nowadays, were proposed. One of these models, the
\textit{embedded flux tube model} is an empirical model proposed by
\cite{SolankiMontavon1993}, in which nearly horizontal magnetic flux
tubes forming the intraspines are embedded in more vertical background
magnetic fields (spines) in the penumbra
(Figure~\ref{figure:pen_models}, left panel). The \textit{downward
  pumping mechanism} \citep{Thomas2002} was proposed to explain the
origin of field lines that return back into the solar surface at the
outer penumbra (Figures~\ref{figure:fig_amb1}, \ref{figure:fig_amb4}, and
\ref{figure:spot_dc}). In this scenario, submergence of the outer part
of flux tubes occurs as a result of the downward pumping by the
granular convection outside the sunspots, and such magnetic fields
form the low-laying horizontal flux tubes. Another idea to account for
the penumbral filaments is the \textit{field-free gap model} initially
proposed by \cite{Choudhuri1986} and later refined by
\cite{SpruitScharmer2006} and \cite{ScharmerSpruit2006}. Here, the
penumbral bright filaments are regarded as manifestations of the
protrusion of non-magnetized, convecting hot gas into the background
oblique magnetic fields of the penumbra. Due to the continuity
condition of the normal component of the magnetic field across the
boundary between the background field and the protruding
non-magnetized gas, the vertical component of the magnetic field
vector, $B_\rho$, in the background magnetic field must vanish right
on top of the non-magnetic gas. This immediately yields a region,
above the penumbral filaments, where the magnetic field is almost
horizontal (i.e., intraspines).

All the aforementioned models attempt to explain, with different
degrees of success, the configuration of the magnetic field in the
penumbra. However, the appearance of a penumbra is always associated
with a distinctive gas flow, i.e., the Evershed flow and, therefore,
this must also be taken into account by these models. In the next
section we will address this issue.

\epubtkImage{pen_models.png}{
\begin{figure}[htbp]
\centerline{\includegraphics[width=\textwidth]{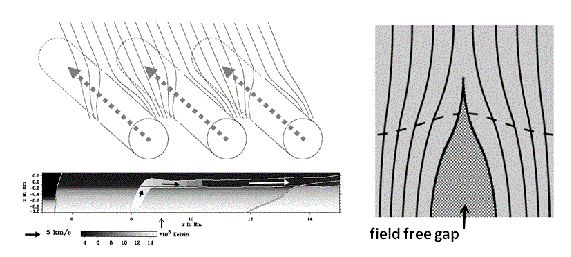}}
\caption{Models for explaining the uncombed penumbral structure. 
    Upper-left: embedded flux tube model \citep[from][reproduced by
    permission of the ESO]{SolankiMontavon1993}; lower-left: rising flux tube
    model \citep[from][reproduced by permission of the
    ESO]{Schlichenmaier1998AA}; right: field-free gap model 
   \citep[from][reproduced by permission of the ESO]{SpruitScharmer2006}.}
\label{figure:pen_models}
\end{figure}}

\subsubsection{Relation between the sunspot magnetic structure and the Evershed flow}
\label{subsubsection:evershed}

The Evershed flow was discovered in 1909 by John Evershed at the
Kodaikanal Observatory in India as red and blue wavelength shifts in
the spectra of absorption lines in the limb-side and disk-center-side
of the penumbra, respectively. This feature can be explained  by a
nearly horizontal outflow in the photosphere of the penumbra
\citep{evershed1909}. Under an insufficient spatial resolution, it
appears as a stationary flow with typical speeds of 1\,--\,2~\kms,
where the magnitude of the flow velocity increases with optical depth
$\tau_c$ \citep[towards the deep
  photosphere;][]{BrayLoughhead1979sunspot}.

An outstanding puzzle about the Evershed flow lies in the relation
between the velocity vector and the magnetic field vector in the
penumbra. Since the averaged magnetic field in the penumbra has a
significant vertical component, with an angle with respect to the
normal vector on the solar surface between $\zeta \approx
40\mbox{\,--\,}80\deg$ (see Figure~\ref{figure:fig8} in
Section~\ref{subsection:constanttau}), and the Evershed flow is
apparently horizontal, this would mean that the flow would move across
the magnetic field. Under these circumstances, the sunspot's magnetic
field (which is frozen-in to the photospheric gas) would be removed
away within a few hours.

It is highly plausible that there is a close relationship between the
Evershed flow and the filamentary structure of the penumbra. Indeed, it
was recognized in the 1960s that the flow is not spatially uniform but
concentrated in narrow channels in penumbra; e.g., \cite{Beckers1968}
reported that the flow originates primarily in dark regions between
bright penumbral filaments. Two models were proposed to account for
the nature of penumbral filaments and the Evershed flow before 1990.
One is the elevated dark filament model in which the penumbral dark
regions are regarded as elevated fibrils with nearly horizontal
magnetic field overlaying the normal photosphere and carry the
Evershed flow in them
\citep{Moore1981,Cram1981,Thomas1988,Ichimoto1988}. The other is the
rolling convection model in which penumbral filaments are regarded as
convective elements radially elongated by a nearly horizontal magnetic
field in penumbra and where the Evershed flow is confined in dark
lanes that are analogous to the intergranular dark lanes
\citep{Danielson1961roll,Galloway1975}. Both models assume a nearly
horizontal magnetic field in the penumbra and, therefore, contradict
the observational fact that a significant fraction of sunspot's
vertical magnetic flux comes out through the penumbra
\citep{SolankiSchmidt1993}.

\epubtkImage{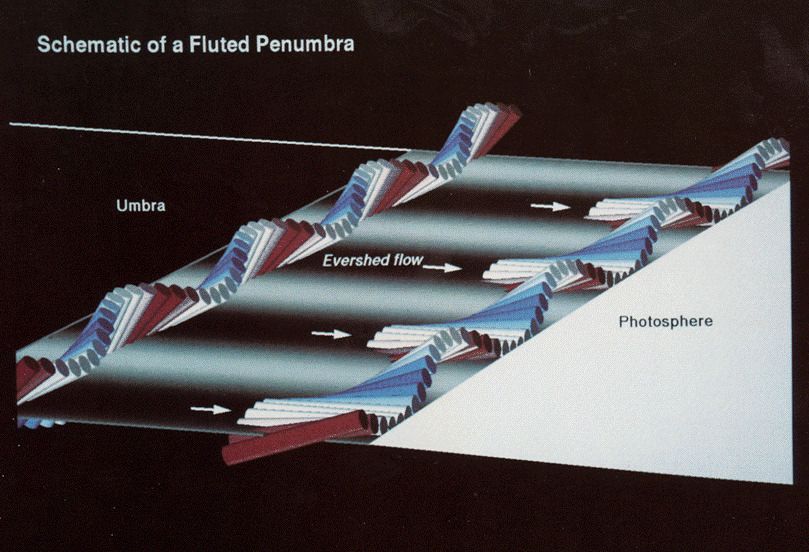}{
\begin{figure}[htbp]
\centerline{\includegraphics[width=0.9\textwidth]{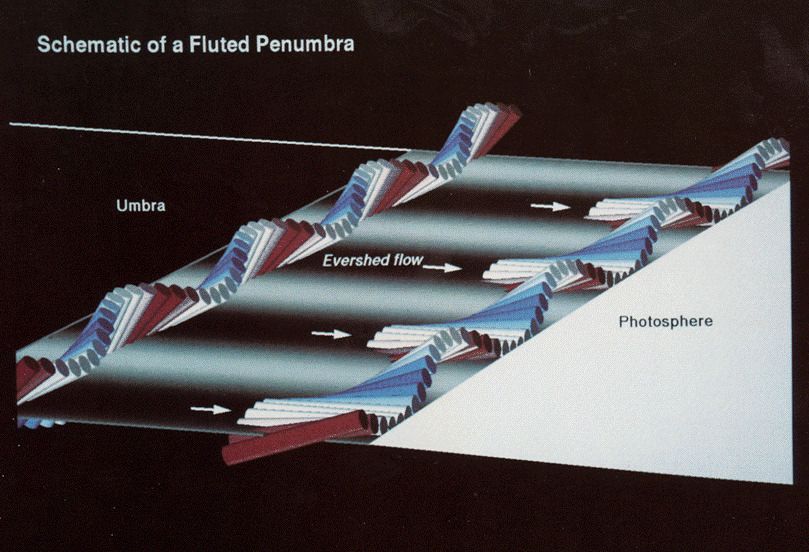}}
\caption{Geometry of magnetic field and Evershed flow in
  penumbra. Magnetic field lines are shown by inclined and colored
  cylinders, while the Evershed flow is indicated by white arrows in
  dark penumbral channels. Note that the Evershed flow concentrates
  along the more horizontal magnetic field lines (white
  cylinders)
  \citep[from][reproduced by permission of the AAS]{title1993}.}
\label{figure:title}
\end{figure}}

The long-lasting enigma on the Evershed flow was finally solved by the
discovery of the interlocking comb structure of the penumbral magnetic
field (Section~\ref{subsubsection:spine_intraspine}). Under this
scenario, the Evershed flow is confined in nearly horizontal magnetic
field channels in penumbra (i.e., \textit{intraspines}), while out of
the flow channels (i.e., in the \textit{spines}) the magnetic field is
more vertical. Both components, when averaged together, make the
spatially averaged magnetic field far from completely horizontal
\citep[Figures~\ref{figure:fig8} and \ref{figure:title}; see
  also][]{title1993}. The relationship between the Evershed flow and
the horizontal magnetic field in the penumbra has been highlighted in
many works in the past: \citet[][Figure~7]{stanchfield1997} or
\citet[][Figure~12]{shibu2003}. The latter two works were obtained
with spectropolarimetric data at 1" resolution. A more updated result,
employing Hinode/SP data with 0.3" resolution, has been presented by
\citet[][see
  Figure~\ref{figure:spine_intraspine_borrero}]{Borrero2008ApJ}. This
figure demonstrates that the Evershed flow (seen as large positive or
redshifted line-of-sight velocities; middle panel) is concentrated
along the intraspines: regions where the magnetic field is horizontal
($\gamma \approx 90\deg$; bottom panel) and weaker (upper panel).

\epubtkImage{spine_intraspine_borrero.png}{
\begin{figure}[ht]
\centerline{\includegraphics[scale=0.75, angle = 0]{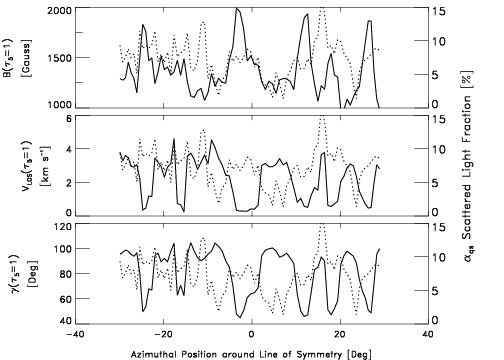}}
\caption{Variation of the physical parameters at $\tau_c=1$
  (continuum) along an azimuthal cut around the limb-side penumbra
  (i.e., along one of the blue ellipses in
  Figure~\ref{figure:fig7}). From top to bottom: magnetic field
  strength $B$, line-of-sight velocity $V_{\mathrm{los}}$, and
  inclination of the magnetic field $\gamma$. Dotted curves in each
  panel show the scattered light fraction obtained from the inversion
  algorithm. Note that the velocity (Evershed flow) is strongest in
  the regions where the magnetic field is weak and horizontal
  (\emph{intraspines}), while it avoids the regions with more less
  inclined and stronger magnetic field (\emph{spines})
  \citep[from][reproduced by permission of the AAS]{Borrero2008ApJ}.}
\label{figure:spine_intraspine_borrero}
\end{figure}}

In the \textit{embedded flux-tube model} \citep{SolankiMontavon1993},
the Evershed flow is supposed to be confined in the horizontal
magnetic flux tubes embedded in more vertical background magnetic
field of the penumbra. In such picture, the siphon flow mechanism was
proposed as the driver of the Evershed flow
\citep{MeyerSchmidt1968,Thomas1988,Degenhardt1991,MontesinosThomas1993}:
a difference in the magnetic field strength between two footpoints of
a flux tube causes a difference of gas pressure, and drives the flow
in a direction towards the footpoint with a higher field strength
(i.e., the footpoint outside the sunspot to account for the Evershed
outflow). \cite{Schlichenmaier1998ApJ} and \cite{Schlichenmaier1998AA}
investigated the dynamical evolution of a thin magnetic flux
tube\epubtkFootnote{\label{footnote:thin}\textit{Thin} in this context
  means that the thin flux-tube approximation \citep{spruit1981} can
  be applied. This reduces the problem to a 1D problem: it does so by
  assuming that the flux tube's radius is much smaller than the
  pressure scale height.} embedded in a penumbral stratification
\citep{JahnSchmidt1994} and proposed the \textit{hot rising flux tube
  model}, in which a radial and thin flux tube containing hot plasma
raises towards the solar surface due to buoyancy. As the flux tube
reaches the $\tau_c=1$-level it cools down due to radiation, thereby
producing a gradient on the gas pressure along the flux tube and, thus,
driving the Evershed flow along the tube's axis (i.e., radial
direction in the penumbra).

By performing an inversion of Stokes profiles of three infrared spectral lines at 1565~nm and
using a two component penumbral model in which two different magnetic atmospheres are 
interlaced horizontally, \cite{BellotRubio2004} found a perfect alignment of the magnetic 
field vector and the velocity vector in the component that contains the Evershed flow.
This picture was supported by \cite{borrero2004} who also performed Stokes inversions of the same 
infrared lines. With a further elaborated analysis, \cite{Borrero2005} found that the penumbral flux tubes 
are hotter and not completely horizontal in the inner part of the penumbra, while they become 
gradually more horizontal and cooler with increasing radial distance. This is accompanied by 
an increase in the flow velocity and a decrease of the gas pressure 
difference between flux tube and the background component, with the flow speed
eventually exceeding the critical value to form a shock front at large radial distances (\V~\textgreater~6\,--\,7~\kms).
They argued that these results strongly support the siphon flow as the physical mechanism responsible for the 
Evershed flow.

Until recently the relationship between the Evershed flow and the
brightness of the penumbral filaments has been somewhat
controversial. Many authors \citep{Beckers1968, title1993, Shine1994,
  Rimmele1995, Balthasar1996, stanchfield1997, Voort2002} have
presented evidence that the Evershed flow is concentrated in dark
filaments, while some studies claimed that there is no correlation
\citep{WiehrStellmacher1989, Lites1990,
  Hirzberger2001}. \cite{Rimmele1995} showed that the correlation
becomes better when one compares the intensity and velocity
originating from the same height, and also gave a hint that the
correlation is different between inner and outer
penumbra. \cite{Schlichenmaier2005}, \cite{BellotRubio2006}, and
\cite{Ichimoto2007PASJ} presented evidence that the Evershed flow
takes place preferentially in bright filaments in the inner penumbra,
but in dark filaments in the outer  penumbra.

\epubtkImage{ev_corr.png}{
\begin{figure}[htb]
\centerline{\includegraphics[scale=0.45, angle=0]{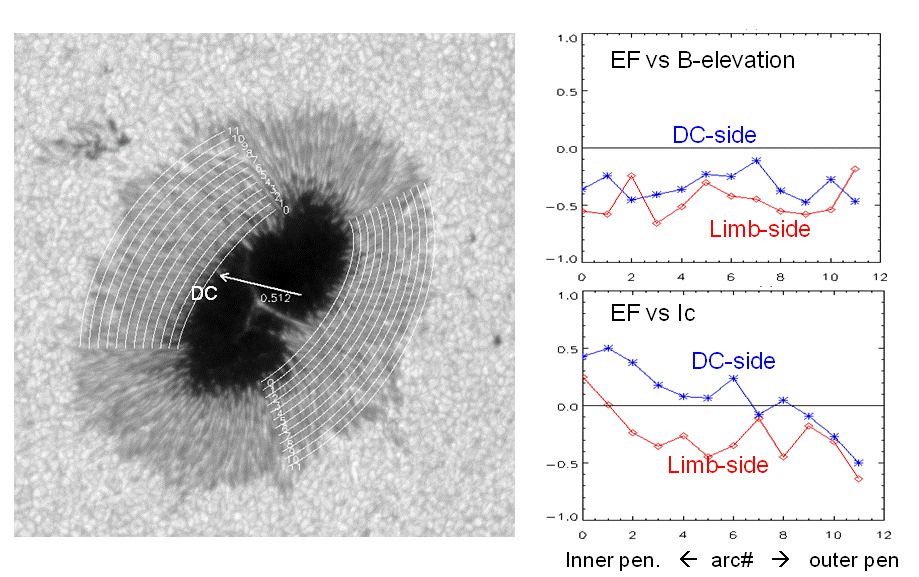}}
\caption{Spatial correlation between penumbral filaments and the
  Evershed flow. Correlation coefficients between the Evershed flow
  and brightness (lower-right), and between the Evershed flow and the
  elevation angle of magnetic field from the solar surface
  (upper-right) are shown as a function of radial distance from the
  sunspot center. Arc-segments along which the correlation is
  calculated are shown in the left panel. The sunspot was located at
  the heliocentric angle of $\Theta$~=~31\textdegree. The direction to
  the  center of the solar disk is shown by an arrow in the left
  panel. The data employed here was recorded with the
  spectropolarimeter on-board Hinode (SOT/SP). Red lines show the
  results for the limb-side penumbra, whereas blue corresponds to the
  center-side penumbra.}
\label{figure:ev_corr}
\end{figure}}

Figure~\ref{figure:ev_corr} presents the spatial correlation between
penumbral filaments and the Evershed flow. The correlation coefficient
between the Doppler shift ($V_{\mathrm{los}}$) and the elevation angle
of magnetic field vector from the solar surface\epubtkFootnote{The
  \textit{elevation} angle is defined as ($1-\zeta$) (see
  Equation~\ref{equation:zeta}). Thus, small elevation angles
  correspond to horizontal magnetic fields (contained in the solar
  surface) and large elevation angles indicate magnetic fields that
  are rather vertical (perpendicular to the solar surface).} as a
function of the radial position in the penumbra, is displayed in the
upper-right panel, whereas the correlation between the Doppler shift
($V_{\mathrm{los}}$) and continuum intensity is shown in the
lower-right panel. In these plots, the results for limb-side penumbra
are shown in red color and for disk-center-side are shown in blue
color. The abscissa in this figure spans from the umbra-penumbra
boundary (left) to the outer border of the penumbra (right). The
curves along which the correlation coefficients are obtained are shown
for both limb-side and disk-center-side penumbra in the left
panels. The data employed for this figure was obtained by  SOT/SP when
the sunspot was located at the heliocentric angle of $\Theta$~=~31\textdegree,
thus, the Doppler shift is mainly produced by the horizontal Evershed
flow. In this plot, line-of-sight velocities are taken in absolute
value such that there is no difference between the redshifts in the
limb-side and the blueshifts in the center side that are
characteristic of the Evershed flow. In Figure~\ref{figure:ev_corr},
we notice that the Evershed flow correlates with  more horizontal
magnetic fields throughout the entire penumbra, while it correlates
with bright filaments in the inner penumbra but with dark filaments in
the outer penumbra. These results are consistent with the idea that
penumbral filaments, which harbor a nearly horizontal magnetic field,
are brighter in inner penumbra but darker in outer penumbra. The
correlation between Doppler shift and intensity shows an asymmetric
distribution between the disk center-side and limb-side penumbra
(lower right). This suggests that overposed to the Evershed flow,
which is mainly horizontal, there exists a vertical component in the
velocity vector in the penumbra. This vertical component will be
discussed in detail in Sections~\ref{subsubsection:heating} and
\ref{subsubsection:vertical_motions}.

So far we have discussed only investigations that were carried out
with ME-inversion codes and, therefore, referred only to the physical
parameters of the sunspot penumbra at a constant $\tau$-level (see
Sections~\ref{subsubsection:formationheights} and 
\ref{subsection:constanttau}). In order to investigate the
depth dependence of the line-of-sight velocity and magnetic field
vector in the penumbra, $\tau$-dependent inversion codes (see
Sections~\ref{subsubsection:formationheights} and
\ref{subsection:verticaltau}) must be applied to observations of the
polarization signals in spectral lines. This has been addressed by a
number of authors, such as \cite{Jurcak2007}, who applied the SIR
inversion code \citep{basilio1992} to the spectropolarimetric data
obtained by SOT/SP on Hinode and found that a weaker and more
horizontal magnetic field is associated with an increased
line-of-sight velocity in the deep layers of the bright filaments in
the inner penumbra. In the outer penumbra, however, stronger flows and
more horizontal magnetic fields tend to be located in dark filaments
\citep{Jurcak2008}. With a further application of the SIR inversion on
SOT/SP data, \cite{Borrero2008AA} found that the magnetic field in the
spines wraps around the horizontal filaments (i.e., intraspines). Some
results from the latter two works are presented in
Figures~\ref{figure:pen_depth_jurcak} and \ref{figure:pen_depth_borrero}.

\bigskip

\epubtkImage{Jurcak.png}{
\begin{figure}[htbp]
\centerline{\includegraphics[width=0.9\textwidth]{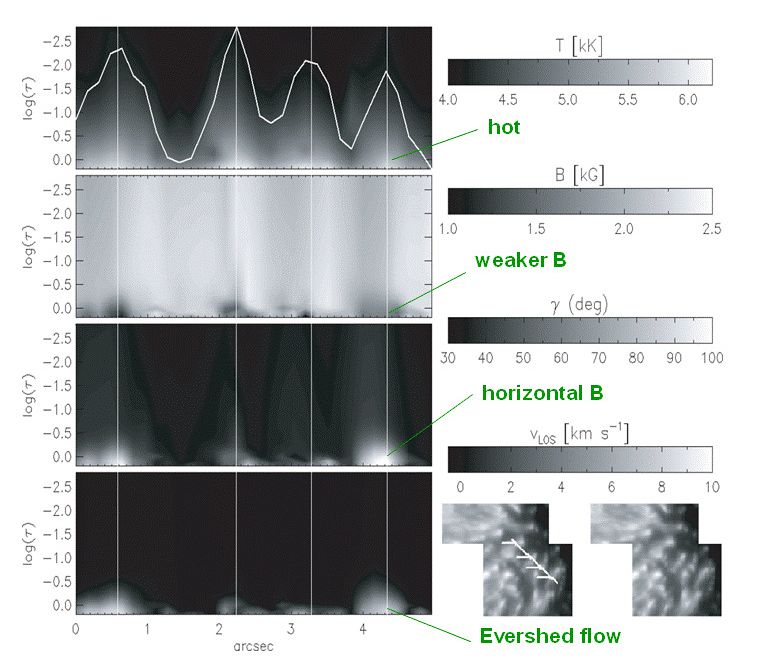}}
\caption{Depth structure of penumbra derived from Stokes inversions of
  spectro-polarimetric data. Showns are vertical cuts across the
  penumbral filaments. On the left, from top to bottom, are temperature
  $T$, field strength $B$, field inclination $\gamma$, and
  line-of-sight velocity $V_{\mathrm{los}}$
  \citep[from][reproduced by permission of the PASP]{Jurcak2007}.}
\label{figure:pen_depth_jurcak}
\end{figure}}

\epubtkImage{borrero_fil1-fil2-fil3-fil4.png}{%
  \begin{figure}[htbp]
    \centerline{\includegraphics[width=7cm]{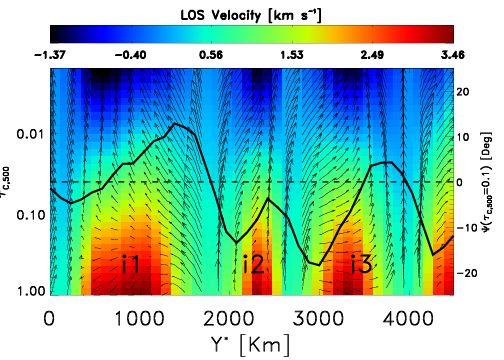}\qquad
      \includegraphics[width=7cm]{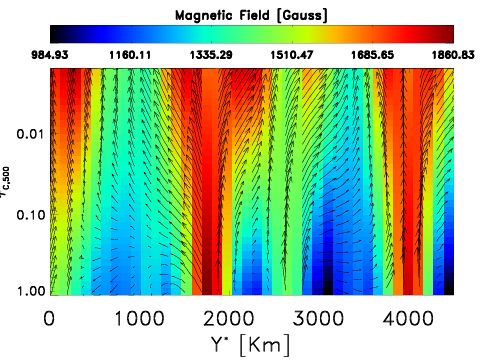}}
     \centerline{\includegraphics[width=7cm]{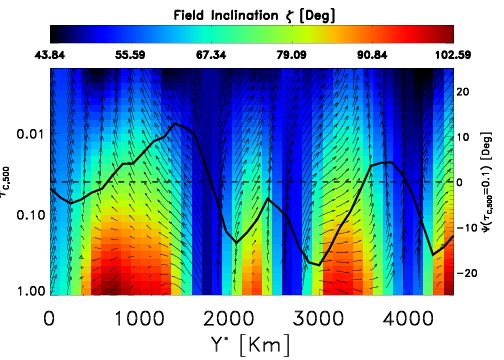}\qquad
      \includegraphics[width=7cm]{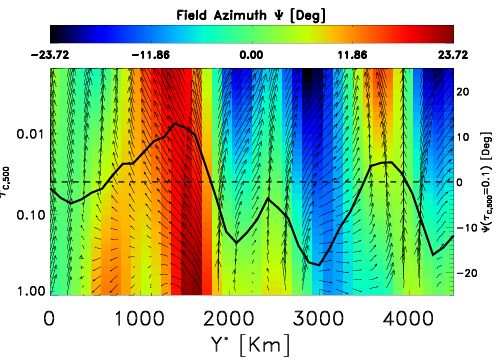}}
    \caption{Vertical stratification (optical depth $\tau_c$) of the
      physical parameters in the penumbra. The horizontal axis is the
      azimuthal direction around the penumbra and, therefore, it is
      perpendicular to the radial penumbral
      filaments. \emph{Upper-left panel}: line-of-sight velocity
      $V_{\mathrm{los}}$. \emph{Upper-right}: total magnetic field
      strength $B$. \emph{Lower-left}: inclination of the magnetic
      field vector with respect to the normal vector to the solar
      surface $\zeta$ (see Equation~\ref{equation:zeta}).
      \emph{Lower-right}: azimuth of the magnetic field vector $\Psi$
      (Equation~\ref{equation:psi}). This plot demonstrates that the
      strong and vertical magnetic field of the \emph{spines} extends
      above the \emph{intraspines} (indicated by the index $i$), where
      the Evershed flow is located where the magnetic field is
      rather horizontal and weak. It also shows that the azimuth of
      the magnetic field changes sign above the intraspines,
      indicating that the magnetic field of the spines wraps around
      the intraspines. The arrows in this figure show the direction of
      the magnetic field in the plane perpendicular to the axis of the
      penumbral filaments
     \citep[from][reproduced by permission of the ESO]{Borrero2008AA}.}
    \label{figure:pen_depth_borrero}
\end{figure}}

\clearpage

\subsubsection{The problem of penumbral heating}
\label{subsubsection:heating}

One important issue that needs to be addressed to understand the
origin of the penumbra is how the energy transport takes
place. Whatever mechanism exists, it must supply enough energy to
maintain the penumbral surface brightness to a level of 70\,--\,80\%
of the quiet Sun granulation (see Sections~\ref{subsection:wilson} and
\ref{section:smallscale}). Since the most efficient form of
energy transport in the solar photosphere is convection, the key
question is therefore to identify how convective motions in the
presence of a rather strong, $B \approx 1500\mathrm{\ G}$, and
horizontal, $\zeta \approx 40\mbox{\,--\,}80\deg$, magnetic field (see
Figure~\ref{figure:fig8}) occur.

\epubtkImage{convection_new2.png}{
\begin{figure}[htbp]
\centerline{\includegraphics[width=0.8\textwidth]{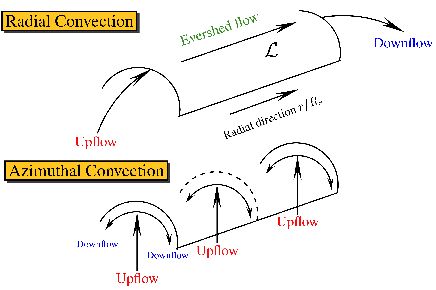}}
\caption{Possible patterns of convection present in the sunspot
  penumbra. The upper panel corresponds to a pattern of radial
  convection, where upflows are presented at the inner footpoints of
  the penumbral filaments and downflows at the outer footpoints. This
  pattern is predicted by the \emph{embedded flux-tube model} and the
  \emph{hot rising flux-tube model}. The lower panel shows a pattern
  of azimuthal or overturning convection, where the upflows/downflows
  alternate in the direction perpendicular to the filaments'
  axis. This is the flow pattern predicted by the \emph{field-free gap
    model}.}
\label{figure:convection_penumbra}
\end{figure}}

In search for these convective motions we have to examine the
predictions that the different models (see
Section~\ref{subsubsection:evershed}) make about vertical flows in the
penumbra. The \textit{hot rising flux-tube model} (see
Section~\ref{subsubsection:evershed}) predicts the presence of upflows
at the inner footpoints of the flux tubes and downflows at their
outer footpoints \citep{Schlichenmaier2002AN}. Provided that the flux
tubes are evenly distributed, this would yield a preference for upflows
in the inner penumbra, whereas downflows would dominate at large
radial distances. Because of these features, we will refer to this
form of convection as \textit{radial convection}, with convective
flows occurring along the penumbral filament (see upper panel in
Figure~\ref{figure:convection_penumbra}). \cite{SchlichenmaierSolanki2003}
examined the possible heat transport in the context of this model and
found that the heat supplied by this model is sufficient only if the
upflowing hot plasma at the inner flux tube's footpoint travels only a
small radial distance $\mathcal{L}$ before tuning into a downflow (see
upper panel in Figure~\ref{figure:convection_penumbra}), with a new
flux tube appearing immediately after. This implies that there should
be a significant magnetic flux and mass flux returning to and emerging
from the photosphere in penumbra. This is a natural consequence of the
rapid cooling that the hot rising plasma suffers once it reaches the
$\tau_c=1$-level \citep{schlichenmaier1999}.

Contrary to the aforementioned models, the \textit{field-free gap
  model} provides a very efficient heat transport since here
convective motions are present over the entire length along the bright
penumbral filaments, with upflows at the center of the filaments and
downflows at the filaments' edges. This strongly resembles to the
convective motions discussed in
Sections~\ref{subsubsection:convection_dots} and
\ref{subsubsection:light_bridges} in the context of umbral dots and
light bridges and, therefore, provides a connection between the
different small-scale features in sunspots. The \textit{field-free
  gap} model does not predict any particular radial preference for
upflows and downflows in the penumbra. It however predicts that
alternating upflows/downflows should be detected in the direction
perpendicular to the filaments (i.e., azimuthally around the
penumbra). Because of this feature we will refer to this type of
convection as \textit{azimuthal convection} or \textit{overturning
  convection} (see lower panel in
Figure~\ref{figure:convection_penumbra}). Note that the
\textit{field-free gap} model does not readily offer an explanation
for the Evershed flow. This is an important point that will be
addressed in Section ~\ref{subsubsection:unified_model}.

\subsubsection{Vertical motions in penumbra and signature of convection}
\label{subsubsection:vertical_motions}

In order to distinguish between the different proposed models that
attempt to explain the heating mechanism in the penumbra
(Section~\ref{subsubsection:heating}) we need to address the origin
of the Evershed flow, as well as identifying the sources and sinks
associated with this flow and its mass balance. Since its discovery,
the Evershed flow has been recognized as a horizontal motion of the
photospheric gas. However, as already mentioned in
Section~\ref{subsubsection:evershed}, the Evershed flow is not purely
horizontal as it possesses a vertical component. A clear evidence for
the vertical motions appeared only after the 1990s when high spatial
resolution became available in spectroscopic observations.

Since then, a number of observations have been reported regarding the
vertical component of the flow in the penumbra. On the one hand,
upflows in the penumbra have been reported by \cite{Johannesson1993}, 
\cite{SchlichenmaierSchmidt1999,SchlichenmaierSchmidt2000}, and 
\cite{BellotRubio2005},
 and with much higher spatial resolution by
\cite{RimmeleMarino2006}. On the other hand, downflows have been
observed in and around the outer edge of penumbra by, among others,
\cite{Rimmele1995ApJ}, \cite{WestendorpPlaza1997}, \cite{ToroIniesta2001}, 
\cite{Schlichenmaier2004}, \cite{BellotRubio2004}, and
\cite{SanchezCuberes2005}. Both down- and upflows have been
simultaneously observed in the penumbra by
\cite{SchmidtSchlichenmaier2000}, \cite{SchlichenmaierSchmidt2000}, 
\cite{WestendorpPlaza2001}, \cite{Tritschler2004}, \cite{SanchezAlmeida2007}, 
\cite{Ichimoto2007PASJ}, and \cite{Franz2009}.
Figure~\ref{figure:WestenRimmele} highlights some
observations that clearly show the downflow patches around a sunspot
with an opposite polarity \citep[][left panel]{WestendorpPlaza1997} and
upflow patches at the leading edge of penumbral bright filaments
\citep[][right panel]{RimmeleMarino2006}. These last features correspond
to the solar called \emph{bright penumbral grains} and they are
related to the peripheral umbral dots \citep{Sobotka1999}.

\epubtkImage{westenrimmele.png}{
\begin{figure}[htb]
  \centerline{\includegraphics[width=0.9\textwidth]{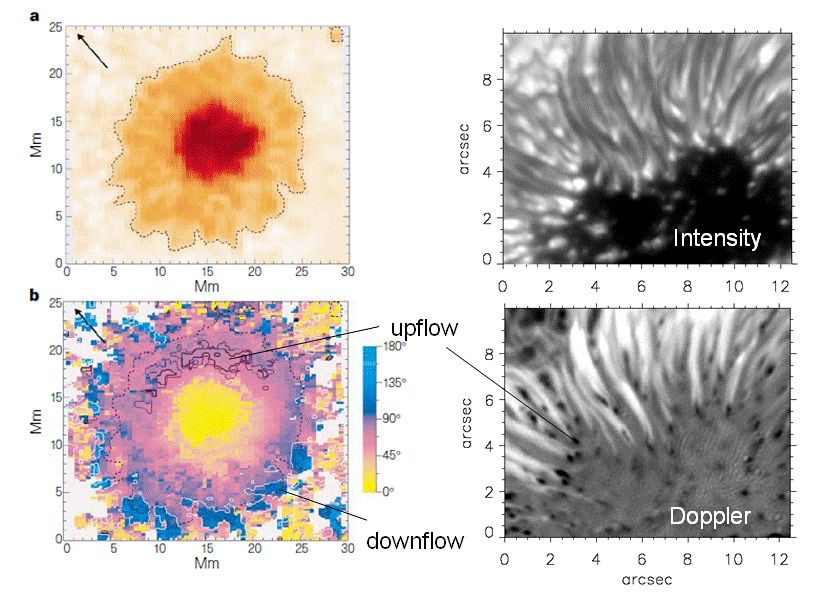}}
  \caption{Selected observations of vertical motions in
    sunspots. \emph{Left panels}: Discovery of downflows around the
    outer border of a sunspot. The sunspot is located near the center
    of the solar disk. Top is the continuum image and bottom is the
    magnetic field inclination overlaid with velocity contours. Blue
    regions have a magnetic field polarity opposite to the sunspot,
    while white contours associated with these regions show downflows
    with +~3~\kms \citep[from][reproduced by permission of Macmillan 
    Publishers Ltd: Nature]{WestendorpPlaza1997}. \emph{Right
    panels}: Close-up of the inner part of a limb-side penumbra. Top
    and bottom are filtergram (intensity) and Dopplergram
    ($V_{\mathrm{los}}$) in the Fe\,{\sc i} 5576~{\AA} line. Each Evershed flow
    channel (white filaments in the Dopplergram) is associated with a
    bright grain and upflow (dark point in the Dopplergram)
    \citep[from][reproduced by permission of the AAS]{RimmeleMarino2006}.}
\label{figure:WestenRimmele}
\end{figure}}

According to the picture drawn by the present observations, most
material carried by the Evershed flow is, thus, supposed to flow back
into the photosphere at the downflow patches
\citep{WestendorpPlaza2001}, while some fraction ($\sim$~10\%) of the
material may continue to flow across the penumbral outer edge along
the elevated magnetic field to form a canopy \citep{solanki1994,
  Solanki1999}. The mass flux balance between up- and downflows in a
sunspot observed near the disk center was also inferred under the
MISMA hypothesis, though individual flow regions were not spatially
resolved \citep{SanchezAlmeida2005}. This model postulates that the
magnetic field varies rapidly (in all three directions) at scales much
smaller than the mean free path of the photon
\citep{SanchezAlmeida1996a, SanchezAlmeida1996b}.

The configuration of the magnetic field is affected by the
aforementioned vertical flows. In fact, some of the magnetic field
lines plunge back into the deep photosphere at the outer edge of the
sunspot and its surroundings (see Figures~\ref{figure:fig_amb1} and
\ref{figure:fig_amb4}). The relationship between vertical motions and
the magnetic field vector in the penumbra is clearly demonstrated by
spectropolarimetric data of a sunspot near disk
center. Figure~\ref{figure:spot_dcv} shows maps of Stokes~\V (circular
polarization) at $\Delta\lambda=\pm (100,300)\mathrm{\ m{\AA}}$ away
from the line center of the Fe\,{\sc i} 6302.5~{\AA} spectral
line. The sunspot in this figure is the same one as in
Figure~\ref{figure:spot_dc} ($\Theta$~=~2.9\textdegree). The sign of Stokes~\V
is reversed for $\Delta\lambda=(100,300)\mathrm{\ m{\AA}}$. If there are
no mass motions in the sunspot, Stokes~\V maps in the blue and red wings
are expected to be identical since the Zeeman effect produces
anti-symmetric Stokes~\V profiles around the line center. This is the
case of the main lobes of the Stokes~\V profiles at
\textpm~100~m{\AA}. However, the maps in \textpm~300~m{\AA}\ are
remarkably different from each other: a number of small and elongated
structures with the same polarity of the sunspot are visible in the
--300~m{\AA}\ \V map (middle-left panel in
Figure~\ref{figure:spot_dcv}) over the penumbra, but with a slight
preference to appear in inner penumbra, whereas a number of patches
with the opposite polarity of the sunspot are seen in the
+300~m{\AA}\ \V map (middle-right panel in
Figure~\ref{figure:spot_dcv}), preferentially in the mid and outer
parts of the penumbra. As is confirmed by the Dopplergram in the
line-wing of Stokes~\I (bottom-left panel in
Figure~\ref{figure:spot_dcv}), the former features are associated with
upward motions while the later correspond to strong downflows. The
typical line-of-sight velocities of the blueshifted regions and the
redshifted regions are approximately 1~\kms and $\sim$~4\,--\,7~\kms,
respectively. The presence of very fast downflows in the mid and outer
regions of the penumbra has been reported previously by
\cite{ToroIniesta2001} and \cite{BellotRubio2004}, who suggest that
many of those downflow patches (where the magnetic field also turns
back into the solar photosphere) harbor supersonic velocities. 

\epubtkImage{spot_dcv.png}{
\begin{figure}[htbp]
\centerline{\includegraphics[scale=0.45]{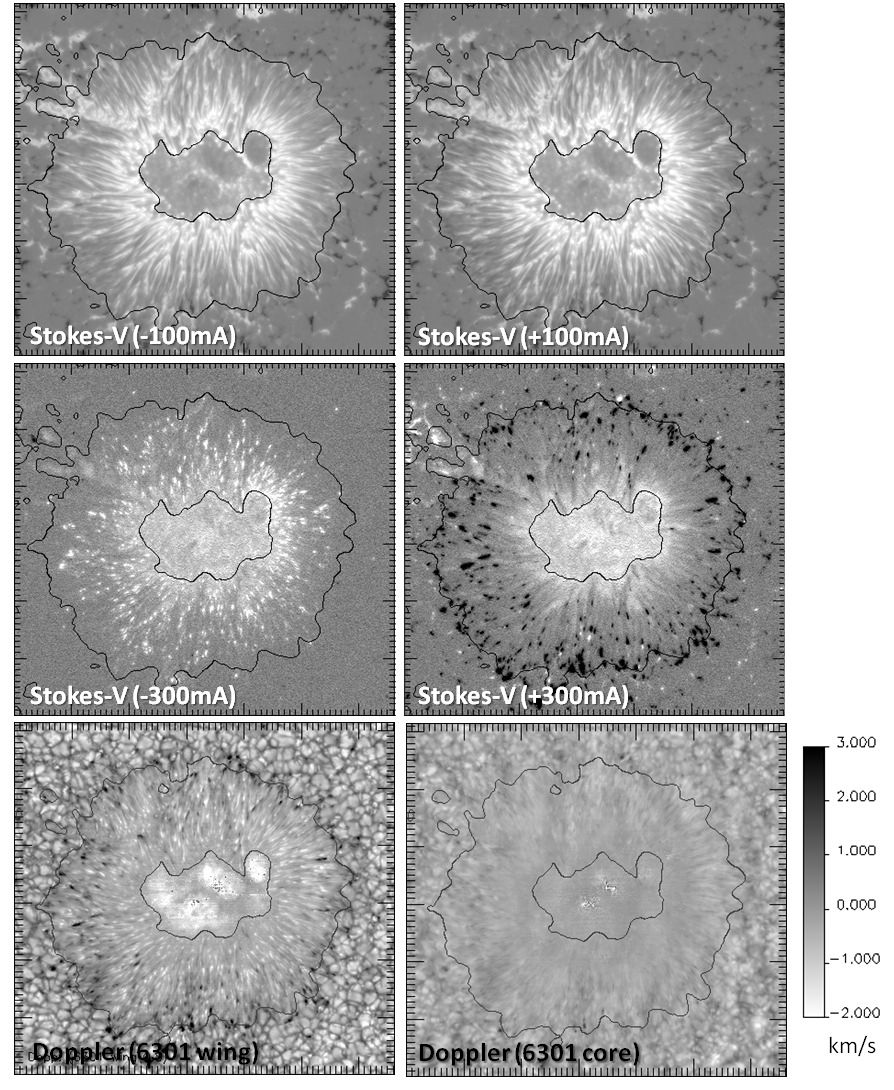}}
\caption{\emph{Upper and middle panels}: Stokes~\V maps of a sunspot near the
  solar disk center ($\Theta$~=~2.9\textdegree; same sunspot as in
  Figures~\ref{figure:spot_dc} and \ref{figure:pen_updown}) at two
  different wavelengths (shown in each panel) from the center of the
  Fe\,{\sc i} 6302.5~{\AA} spectral line. The sign of Stokes~\V is
  reversed for +(100,300)~m{\AA}. \emph{Bottom panels}: line-of-sight
  velocity  (Doppler velocity) measured in the wings (left) and on the
  core (right) of the spectral line.}
\label{figure:spot_dcv}
\end{figure}}

\epubtkImage{pen_updown.png}{
\begin{figure}[htbp]
\centerline{\includegraphics[scale=0.5, angle=0, trim = 0 0 0 0, clip]{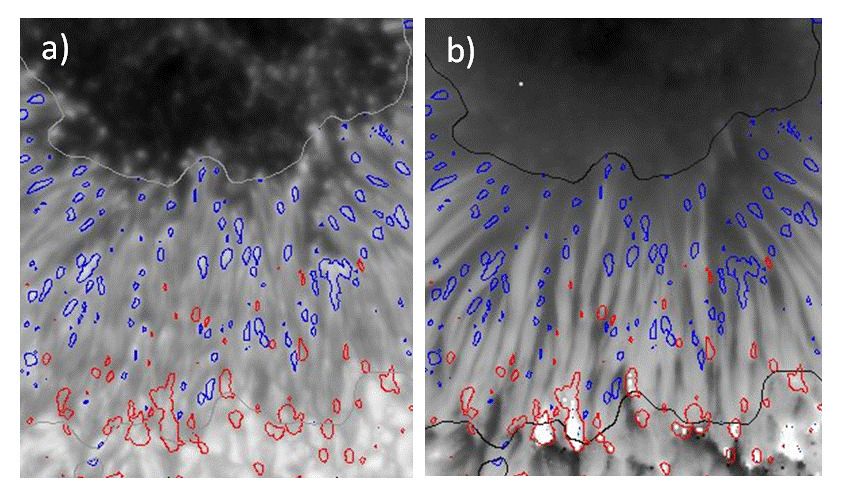}}
\caption{Continuum intensity $I_c$ (\emph{panel a}) and field
  inclination $\gamma$ (\emph{panel b}) in the penumbral region shown
  as a yellow box in Figure~\ref{figure:spot_dc}a. Overlaid are
  contours for upflow regions with 0.8~\kms (blue) and downflow
  regions with $V/I_c = 0.01$ in the far red wing of Fe\,{\sc i}
  6302.5~{\AA}\ line (red). The sunspot shown here was located almost
  at disk center: $\Theta$~=~2.9\textdegree.}
\label{figure:pen_updown}
\end{figure}}

Figure~\ref{figure:pen_updown} shows enlargement of a penumbral region
indicated by a box in Figure~\ref{figure:spot_dc}a, where contours for
upflow and downflow regions are overlaid on continuum intensity (panel
\emph{a}) and field inclination $\gamma$ (panel \emph{b}) maps. It is
obvious in panel \emph{b} that the upflow and downflow patches are
aligned with nearly horizontal field channels (filaments with light
appearance in the inclination map) that carry the Evershed flow, and
that small-scale upflows are preferentially located near the inner
penumbra, while downflows dominate at the outer ends of the horizontal
field channels (see also Figure~1 in \citealp{Ichimoto2010} and Figure~5 in \citealp{Franz2009}). Thus, the upflow and downflow patches seen here
can be regarded as the sources and sinks of the elementary Evershed
flow embedded in deep penumbral photosphere. When all the
aforementioned results and observations for the velocity and the
magnetic field vector are put together, the picture of the penumbra
that emerges is that of Figure~\ref{figure:pen_cartoon}.

\epubtkImage{pen_cartoon.png}{
\begin{figure}[htbp]
\centerline{\includegraphics[scale=0.4, angle=0]{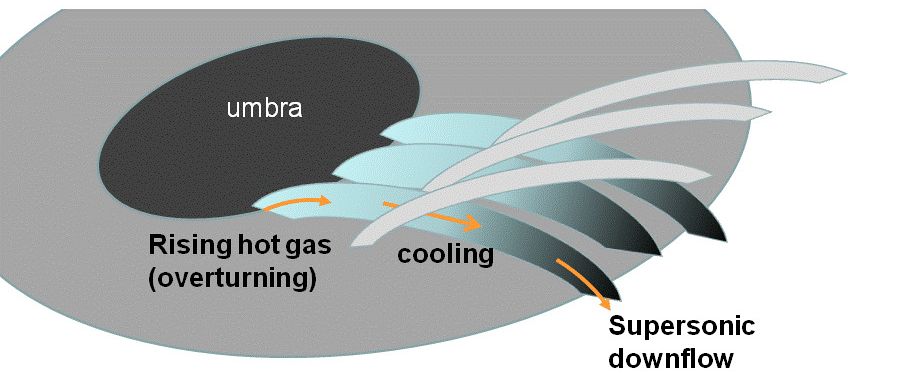}}
\caption{Cartoon of penumbral magnetic field and the Evershed flow structure.}
\label{figure:pen_cartoon}
\end{figure}}

These results seem to support the idea that convective motions occur
in a radial pattern along the penumbral filaments (\textit{radial
  convection}; see upper panel in
Figure~\ref{figure:convection_penumbra}). In principle, this lends a
strong support to the  \textit{hot rising flux-tube model} (see
Section~\ref{subsubsection:spine_intraspine}). However, a closer look
at Figure~\ref{figure:pen_updown} reveals that the radial distance
$\mathcal{L}$ between upflows in the inner penumbra and downflows in
the outer penumbra is typically several megameters. This would imply
that the energy carried by the upflows is not sufficient to heat the
penumbra according to the argument of
\cite{SchlichenmaierSolanki2003}. Interestingly, \cite{RuizCobo2008}
revisited this problem with \textit{the embedded flux-tube model}, and
argued that the significant portion of brightness of the penumbra can
be explained with the hot Evershed flow taking place at the inner
footpoints of rather thick flux tubes (\textgreater~200~km; see
Footnote~\ref{footnote:thin}) even for large values of $\mathcal{L}$.

Notice, however, that the fact that one type of convection is detected,
does not immediately rule out the existence of the other type,
\textit{azimuthal/overturning convection} which is proposed by the
\textit{field-free gap model} (lower panel in
Figure~\ref{figure:convection_penumbra}). Indeed, the search for an
azimuthal convective pattern, i.e., upflows at the center of penumbral
filaments and downflows at their edges, has intensified in the past few
years. Some works, employing continuum images, have provided
compelling evidence that the \textit{azimuthal/overturning convection}
does indeed also exist \citep[][see also 
Section~\ref{subsubsection:filaments}]{Marquez2006,Ichimoto2007Sci,Bharti2010},
at least in the inner penumbra. Unfortunately, results based on
spectroscopic measurements have been contradictory so far. Whereas
some works \citep{Rimmele2008,Zakharov2008} report on positive
detections of such downflows and upflows (of up to 1~\kms), others
claim that at the present resolution that convective pattern does not
exist \citep{Franz2009,BellotRubio2010}. It is, therefore, of the
uttermost importance to provide a conclusive detection (or ruling out)
of an azimuthal/overturning convective flow in the penumbra that might
help us settle once and for all the problem of the penumbral
heating. A number of reasons have been put forward in order to explain
the lack of evidence supporting an \textit{azimuthal/overturning
  convection}. One of the reasons is the lack of sufficient spatial
resolution to resolve the velocity fields inside the penumbral
filaments. However, this does not explain why
\textit{azimuthal/overturning convective} motions are already detected
in umbral dots at the present resolution (see
Section~\ref{subsubsection:convection_dots}) but not in penumbral
filaments. Another reason that has been advocated, within the context
of the \textit{field-free gap model}, has been that the $\tau_c=1$
level is formed above the convective flow rendering it invisible to
spectropolarimetric observations. This argument, however, fails to
explain why is then the \textit{azimuthal/overturning convective}
pattern seen in umbral dots since there the $\tau_c=1$ level should
be formed even higher above the convective flow than in penumbral
filaments \citep{Borrero2009Ch}. The most adequate explanation,
therefore, for the lack of evidence supporting an
\textit{azimuthal/overturning convective} flow pattern  (if it exists)
lies in the large magnitude of the Evershed flow, which overshadows
the contribution from the convective up/downflows on the measured
line-of-sight velocity as soon as the observed sunspot is slightly
away from disk-center ($\Theta$~=~0\textdegree). 

Regardless of which form of convection takes place, it is very clear
that this is indeed the mechanism that is responsible for the energy
transport in the penumbra. This is emphasized by the very close
relationship existing between upflows and bright grains in the
penumbra as seen in Figure~\ref{figure:pen_updown}. In this figure we
display the continuum intensity $I_c$ (panel \emph{a}) and the
inclination of the magnetic field vector $\gamma$ 
(panel \emph{b}) for the southern part of the sunspot shown in
Figure~\ref{figure:spot_dc} overlaid with contours showing the
vertical motions. The blue contours show blueshifts equal or larger  
than 0.8~\kms~in the wing of Stokes~\I of Fe\,{\sc i} 6301.5~{\AA},
while the red contours show $V/I_{c} > 0.01$ at Fe\,{\sc i}
6302.5~{\AA}~+~0.365~{\AA}\ representing strong downflow regions with
opposite magnetic polarity to the spot. The fact that upflows (blue
contours) correlate so well with bright penumbral regions, strongly
suggest that the vertical component of the  Evershed flow supplies the
heat to maintain the penumbral brightness even though a quantitative
evaluation of the heat flux is not available
\citep{Ichimoto2007PASJ}. \cite{Puschmann2010a} supports this scenario
in a more quantitative manner based on their 3D empirical penumbral
model derived from the Stokes inversion of the Hinode/SP data, i.e.,
the penumbral brightness can be explained by the energy transfer of
the ascending mass carried by the Evershed flow if the obtained
physical quantities are extrapolated to slightly deeper layer below
the observable depth ($\tau_c = 1$).

\epubtkImage{pen_stokes2.png}{
\begin{figure}[htbp]
\centerline{\includegraphics[scale=0.4, angle=0]{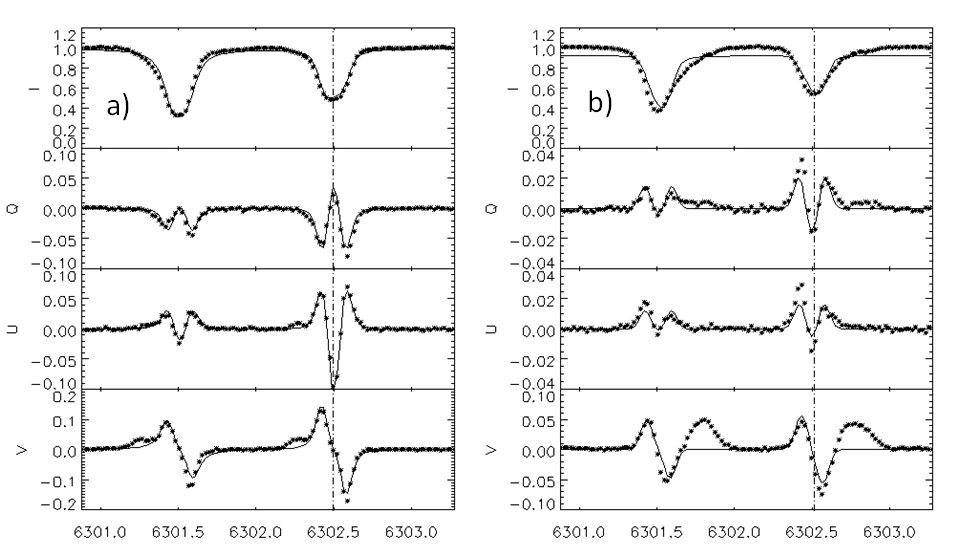}}
\caption{Stokes profiles (observed with Hinode/SP) of Fe\,{\sc i}
  6301.5~{\AA}\ and 6302.5~{\AA}\ spectral lines in an upflow 
  (panel \emph{a}) and downflow (panel \emph{b}) regions in the 
  penumbra. Solid
  curves show results of a Milne--Eddington fitting algorithm (see
  Section~\ref{subsection:tools}.). These profiles 
  correspond to the sunspot observed very close to disk center
  ($\Theta$~=~1.1\textdegree) on February 28, 2007 (AR~10944).}
\label{figure:pen_stokes2}
\end{figure}}

Figure~\ref{figure:pen_stokes2} shows typical Stokes profiles in
upflow (panel \emph{a}) and downflow (panel \emph{b}) regions in the
penumbra, respectively. These profiles correspond to the sunspot
AR~10944 on February 28, 2007 very close to the center of the solar
disk: $\Theta$~=~1.1\textdegree. It is noticeable that the upflow region shows
a blue hump in Stokes~\V with the same polarity of the main lobe in
the blue wing, while the downflow region shows a strong third lobe with
opposite polarity in the far red wing of Stokes~\V. These asymmetric
Stokes~\V profiles imply the presence of a strong velocity (and
magnetic field) gradient along the line-of-sight:
$V_{\mathrm{los}}(\tau_c)$ and $\ve{B}(\tau_c)$. The solid curves show
the best-fit profiles produced by a Milne--Eddington (ME) inversion
algorithm. A ME-inversion assumes that the physical parameters are
constant with optical depth (see Section~\ref{subsection:tools}) and,
therefore, it always produces anti-symmetric Stokes~\V profiles,
thereby failing to properly fit the highly asymmetric observed
circular polarization signals. \cite{SanchezAlmeida2009} reproduced
the red-lobe profiles using the MISMA model (see
Section~\ref{subsubsection:vertical_motions}), and suggested that
reverse polarity patches result from aligned magnetic field lines and
mass flows that bend over and return to the solar interior at very
small scales all throughout the penumbra. While this scenario does not
help to distinguish between a radial or azimuthal/overturning form of
convection (Figure~\ref{figure:convection_penumbra}) it certainly
emphasizes the presence of small-scale convection, which in turn is
needed to sustain the penumbral brightness. Other works have also
pointed out the relationship between the polarity of the vertical
component of the magnetic field and the upflow/downflow regions in the
penumbra. For instance, \cite{DaldaRubio2008} found small-scale,
radially elongated, bipolar magnetic structures in the mid-penumbra
aligned with intraspines. They move radially outward and were
interpreted by these authors as manifestations of the sea-serpent
field lines that harbor the Evershed flow
\citep{Schlichenmaier2002AN} and, eventually, leave the spot to form
moving magnetic features. \cite{MartinezPillet2009} found a
continuation of such magnetized Evershed flow outside sunspots at
supersonic speeds.

\subsubsection{Inner structure of penumbral filaments}
\label{subsubsection:filaments}

Improvements in the spatial resolution in ground-based optical
observations revealed further details about the rich variety of
fine-scale structures in the penumbra. \cite{Scharmer2002} discovered
a notable feature in penumbral filaments at 0.1" resolution with the
Swedish 1-m Solar Telescope at La Palma, i.e., bright penumbral
filaments in the inner penumbra often show internal substructure in
the form of two bright edges separated by a central dark core
(Figure~\ref{figure:scharmer2002}). The temporal evolution of these
structures shows that the dark core and lateral bright edges move
together as a single entity. Some of the dark features are not in
parallel to penumbral filaments, but they form oblique dark streaks
crossing penumbral filaments. These streaks make the filaments look as
if they are twisting with several turns along their length.

\epubtkImage{scharmer2002_html.png}{
\begin{figure}[htbp]
\centerline{\includegraphics[angle=-90, width=0.9\textwidth]{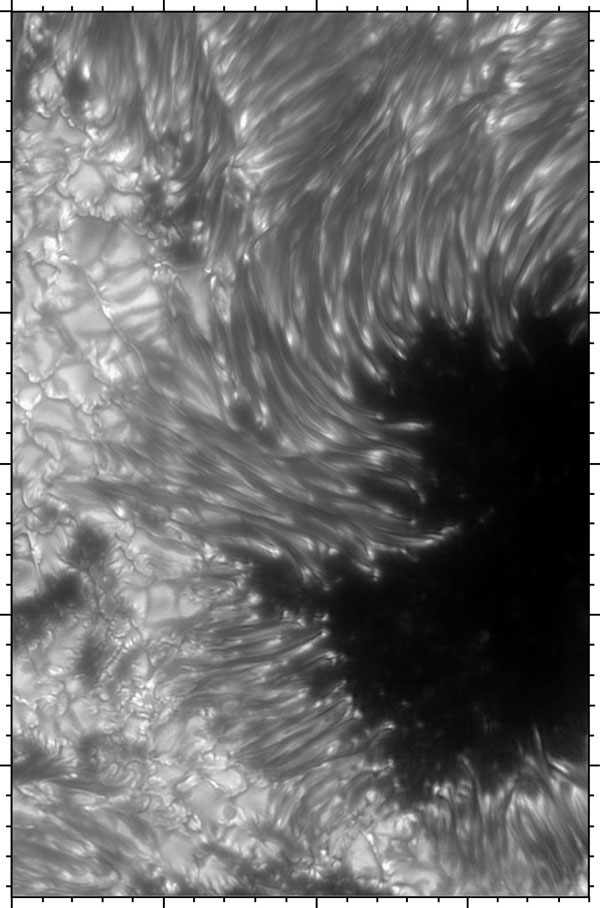}}
\caption{Bright penumbral filaments showing a dark central core. 
Image was taken in G-band at 430.5~nm with the Swedish 1-m Solar Telescope.
Tickmarks have a scaling of 1000~km on the Sun
\citep[from][reproduced by permission of Macmillan Publishers Ltd: Nature]{Scharmer2002}.}
\label{figure:scharmer2002}
\end{figure}}

The visibility of the dark cores in the filaments is not uniform over the penumbra when the sunspot 
is located outside the disk center: dark cores are more clearly identified in disk center-side penumbra
while they are hardly seen in limb side penumbra \citep{Sutterlin2004, Langhans2007}. In addition, dark 
cores are better defined in G-band images than in continuum, which suggests that dark cores are structures 
that are elevated above the continuum formation height $\tau_c=1$ \citep{Rimmele2008}.

The typical lifetime of dark-cored penumbral filaments was estimated as $<$~45 minutes
\citep{Sutterlin2004} while some dark cores last longer than 90~minutes \citep{Langhans2007}.
The first spectroscopic observation of dark cores was reported by \citet{BellotRubio2005}, who 
found a significant enhancement of the Doppler shift which they interpreted as an upflow in the
dark cores. It is also found by spectroscopic \citep{BellotRubio2007} and filtergram \citep{Langhans2007}
observations, that dark-cored filaments are more prominent in polarized light than in continuum intensity, 
and that dark cores are associated with a weaker and a more horizontal magnetic field than their lateral
brightenings and harbor an enhanced radial Evershed outflow. These features are to be considered on top
of the already weak and horizontal magnetic field that characterizes the penumbral intraspines 
(see Section~\ref{subsubsection:spine_intraspine}).

Based on a stratified atmosphere consisting of nearly horizontal magnetic flux tubes embedded in a stronger 
and more vertical field \citet{Borrero2007AA}, as well as \citet{RuizCobo2008}, performed radiative transfer
calculations to show that these models reproduce the appearance of the dark-cored penumbral filaments.
In these models, the origin of the dark cores is attributed to the presence of
the higher density region inside the tubes, which shifts the surface of optical depth unity 
towards higher (cooler) layers.

\epubtkImage{pen_twist.png}{
\begin{figure}[htbp]
\centerline{\includegraphics[scale=0.45, angle=0]{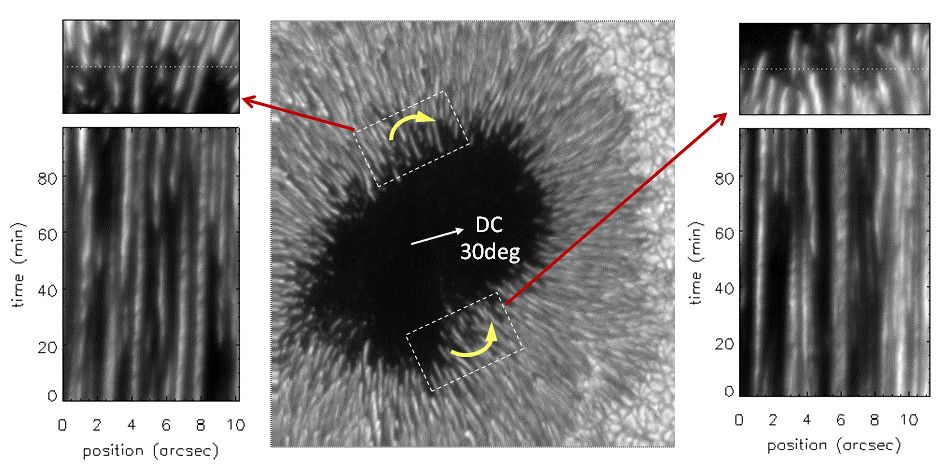}}
\caption{A sunspot located at $\Theta$~=~30\textdegree, east-ward from
  the center of the solar disk. Space-time plots along the slits
  across inner penumbral filaments are shown on both sides. The
  position of the slits are indicated at the top of each space-time
  plot with partial images whose locations are shown by dashed lines
  on the sunspot image. Twist (or turning motion) of penumbral bright
  filament is seen as helical structures of bright filaments in the
  space-time plot.}
\label{figure:pen_twist}
\end{figure}}

From time series of continuum images taken by Hinode/SOT, \cite{Ichimoto2007Sci}
found a number of penumbral bright filaments revealing twisting motions around their axes.
As it also happens with the dark core at the center of the penumbral filaments, the twisting 
motions are well observed only in particular portions in the penumbra but, in this case, the locations 
in which penumbral filaments are oriented parallel to the limb. The direction of the twist (lateral 
motions of dark streaks that run across filaments diagonally) 
is always from the limb-side to disk-center side (see Figure~\ref{figure:pen_twist}). Therefore, the twisting feature is not 
likely a real twist or turn of filaments but, rather, are a manifestation of their dynamical nature, so that their 
appearance depends on the viewing angle. Overturning/azimuthal convection (see Figure~\ref{figure:convection_penumbra}) at
the source region of the Evershed flow \citep{Ichimoto2007Sci,Zakharov2008} has been proposed as the origin of such
features. Such picture with overturning/azimuthal convection causing the observed twisting 
motions is supported by a positive correlation between the speed of twisting motion and the brightness of penumbral filament 
in space and time \citep{Bharti2010}. \cite{Spruit2010} interpret the oblique striations that propagate outward to produce 
the twisting appearance of the filaments as a corrugation of the boundary between the convective flow inside the bright filament
 and the magnetic field wrapping around it. On the other hand, there are some arguments that some of filaments have intrinsic 
twist originated from the screw pinch instability \citep{Ryutova2008, Su2010}.

\subsubsection{The Net Circular Polarization in sunspots}
\label{subsubsection:ncp}

The net circular polarization (NCP) is defined as
$\mathcal{N}=\int{V(\lambda)\,d\lambda}$ with the Stokes~\V signal
integrated over a spectral line. $\mathcal{N}=0$ in a perfectly
anti-symmetric Stokes~\V profile, as the area of the blue lobe
compensates the area of the red lobe (see solid lines in
Figure~\ref{figure:pen_stokes2}). However, the Stokes~\V profiles
deviate from purely anti-symmetric (see dots in the same figure) and,
therefore, $\mathcal{N} \ne 0$ if there exists a gradient of plasma
motion along the line-of-sight: $V_{\mathrm{los}}(\tau_c)$. In
general, $V_{\mathrm{los}}(\tau_c)$ can produce only small amounts of
the NCP. For the large values observed in sunspots, a coupling of a
velocity gradient and a gradient in the magnetic field vector
$\ve{B}(\tau_c)$ within the line-forming region $\bar{\tau}$ (see
Section~\ref{subsubsection:formationheights}) is required
\citep{AuerHeasley1978,Landolfi1996}. Consequently, the NCP provides a
valuable tool to diagnose the magnetic field and velocity structures
along the optical depth $\tau_c$ in the sunspot's
atmosphere. Observation of the NCP in sunspots were first reported by
\citet{Illing1974a} and \citet{Illing1974b}, and were followed by
\citet{HensonKemp1984} and \citet{MakitaOhki1986}. From these early
observations a number of basic features and properties of the NCP at
low spatial resolution were inferred \cite[see, e.g.,][]{MartinezPillet2000}:

\begin{enumerate}
\item The largest NCP occurs in the limb-side penumbra around the apparent magnetic
neutral line with the same sign as the umbra's blue lobe of the Stokes~\V profile.
\item The disk center-side penumbra also shows NCP but in the opposite sign to that of the 
limb-side penumbra and with less magnitude.
\item The penumbra of sunspots at disk center show a NCP with the same sign to that of the limb-side
penumbra.
\end{enumerate}

Besides a gradient in $V_{\mathrm{los}}(\tau_c)$, which is a necessary condition to produce
a non-vanishing NCP, the works from \cite{SanchezAlmeida1992} and \cite{Landolfi1996} show that
a gradient in any of the three components of the magnetic field vector will also enhance the amount
of NCP. These gradients are often referred to as the $\Delta B$, $\Delta\gamma$ and $\Delta\varphi$ mechanisms,
with $\Delta$ indicating a variation of the physical quantity with optical depth $\tau_c$. The NCP in sunspots was first 
interpreted in terms of the $\Delta B$-effect by \citet{Illing1975}, who employed a magnetic field strength and line-of-sight
velocity that increased with optical depth in the penumbra. \cite{Makita1986} interpreted the NCP in sunspots by
means of the $\Delta\varphi$-effect, i.e., the sunspot's magnetic field is twisted and unwound along its axis, and 
has azimuthal rotation along the line-of-sight in the penumbra.

\epubtkImage{muller_ncp.png}{
\begin{figure}[htbp]
\centerline{\includegraphics[width=0.9\textwidth]{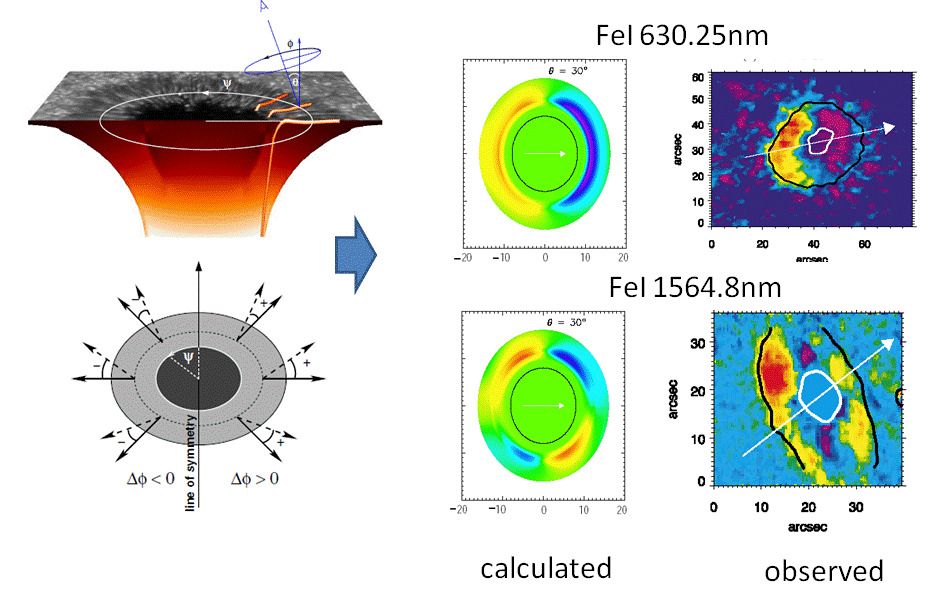}}
\caption{Spatial distribution of the net circular polarization in sunspots reproduced by the
\emph{embedded flux tube model} and observed in Fe\,{\sc i} 6302.5~{\AA} and Fe\,{\sc i} 15648~{\AA} spectral lines
\citep[from][reproduced by permission of the ESO]{Muller2002, Muller2006}.}
\label{figure:muller_ncp}
\end{figure}}

Nowadays, the most successful scenario to reproduce the NCP of sunspots is based on 
the $\Delta\gamma$-effect.  \cite{SanchezAlmeida1992} explained the NCP employing a penumbral model 
in which the Evershed flow increases with depth and where magnetic field lines become
progressively more horizontal with depth in the penumbra. They argued, however, that the large
gradient in the inclination of the magnetic field needed to explain the observed NCP
was not consistent with a sunspot's magnetic field in magnetohydrostatic equilibrium.
\cite{SolankiMontavon1993} addressed this problem, and proposed that the needed
gradients to reproduce the NCP could be achieved without affecting the sunspot's
equilibrium if they assumed the presence of a horizontal flux tube carrying the Evershed
flow embedded in a more vertical background that wraps around it: \textit{embedded flux-tube model}. In this model,
the gradients in $V_{\mathrm{los}}$ and $\gamma$ are naturally produced as the line-of-sight
crosses the boundary between the background and the horizontal flux tube.
Those works were followed by more elaborated models by \cite{MartinezPillet2000} and \cite{Borrero2006}. 
In order to explain the NCP observed in penumbra at disk-center, 
\cite{SolankiMontavon1993} and \cite{MartinezPillet2000} assumed an upflow in the background
magnetic field to make the $\Delta\gamma$-effect to operate with the deeply
embedded flux tubes that carries the horizontal Evershed flow.
\cite{Schlichenmaier2002ncp}, \cite{Muller2002}, and \cite{Muller2006} further developed this
idea and provided three-dimensional penumbral models in which horizontal flux tubes are embedded 
in a more vertical penumbral background magnetic field, to successfully account for the observed 
azimuthal (i.e., variation around the sunspot) distribution of the NCP at low spatial resolution ($\simeq$~1") over the 
penumbra located outside the disk center (see Figure~\ref{figure:muller_ncp}).
In these models, the $\Delta\varphi$-effect also plays an important role to reproduce the 
asymmetric distribution of the NCP around the line connecting the disk center and the sunspot's
center, in particular for the signals observed in the Fe\,{\sc i} 15648~{\AA}\ spectral line 
\citep{Schlichenmaier2002}. Additional improvements were implemented by
\cite{Borrero2007}, who incorporated a more realistic configuration of the flux tubes 
and surrounding magnetic fields. The azimuthal distribution and center-to-limb variation of NCP
in Fe\,{\sc i} 6302.5~{\AA}\ and Fe\,{\sc i} 15648~{\AA}\ lines were again reproduced successfully.
\cite{BorreroSolanki2010} further studied the effect of \textit{azimuthal/overturning} convective motions 
in penumbral filaments on the NCP, and found that these convective motions are less significant 
than Evershed flow for the generation of net circular polarization.

High resolution observations ($<$~0.5") of the NCP in sunspots were first reported by \cite{Tritschler2007}. 
They demonstrated the filamentary distribution of NCP in the penumbra, although the spatial correlation 
with the Evershed flow channels was not conclusive. Using Hinode/SP data \cite{Ichimoto2008} found that, 
as expected, the NCP with the same sign as the umbral blue-lobe is associated 
with the Evershed flow channels in limb-side penumbra. Remarkably these authors also found that the Evershed flow channels 
in the disk-center-side penumbra show again the same sign of the NCP, 
whereas the opposite sign was observed in disk-center-side penumbra in the inter-Evershed flow channels 
(\textit{spines}; see Section~\ref{subsubsection:spine_intraspine}). This is indicated in Figure~\ref{figure:sot_ncp}.

When the sunspot is close to disk center, the NCP in both upflow and
downflow regions is associated with the same sign of the umbral
blue-lobe (see panels \emph{a} and \emph{b} in
Figure~\ref{figure:sot_ncp}). These results appear to be inconsistent
with the current explanation of the NCP by means of the
$\Delta\gamma$-effect associated with the presence of the Evershed
flow in the deep layers of the penumbra. It rather suggests a positive
correlation between the magnetic field strength and the flow velocity
as the cause of the NCP, and also serves as a strong evidence for the
presence of gas flows in inter-Evershed flow channels
(\textit{spines}). The presence of plasma motions in \textit{spines}
inferred from the net circular polarization informs us that the
current simple two component penumbral models consisting of Evershed
flow channels and the spines (with no mass motion) are not compatible
with the observations, and strongly suggests that there are dynamic
features in penumbra that remain unresolved with the current observations.

\epubtkImage{sot_ncp.png}{
\begin{figure}[htb]
\centerline{\includegraphics[width=0.9\textwidth]{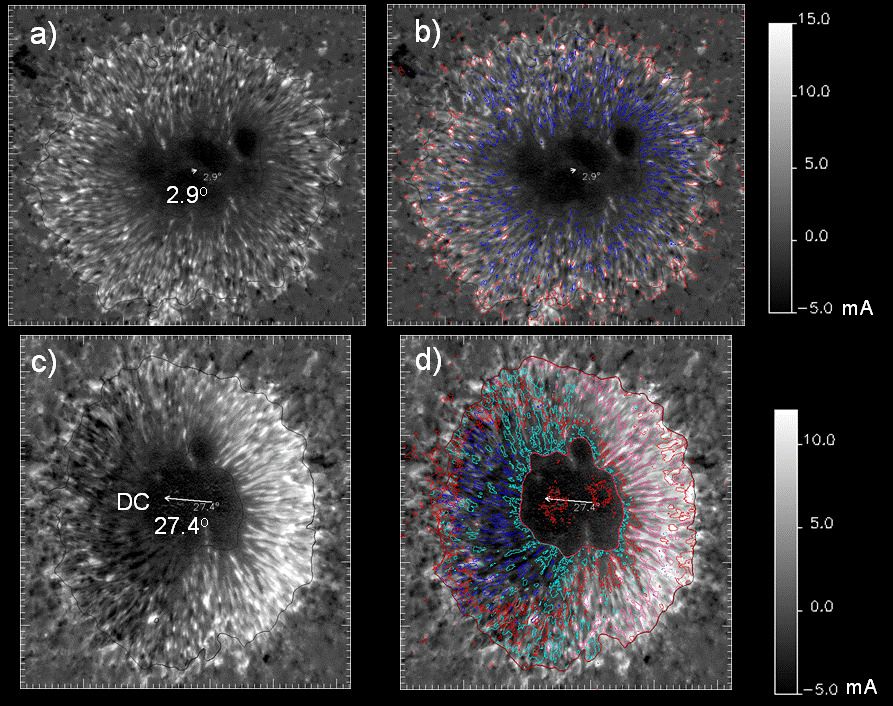}}
\caption{\emph{Upper panels}: spatial distribution of NCP observed by
  SOT/SP in the Fe\,{\sc i} 6302.5~{\AA} spectral line for a sunspot
  close to disk center ($\Theta$~=~2.9\textdegree; same as sunspot in
  Figure~\ref{figure:spot_dc}). \emph{Lower panels}: same as above but
  for a sunspot at $\Theta$~=~27.4\textdegree. In panels b) and d),
  the velocity contours are plotted over the orginal NCP distributions
  shown in panels a) and c), respectively. Color contours indicate
  velocities of --1.8~\kms (blue), --0.6~\kms (green), 0.6~\kms
  (pink), and 1.8~\kms (red), with negative and positive values
  indicating up- and downflows, respectively.}
\label{figure:sot_ncp}
\end{figure}}

It is also noticeable that the NCP associated with the upflow regions in disk-center-side penumbra 
is created by a hump in blue wing of Stokes~\V profiles (see panel \emph{a} in Figure~\ref{figure:pen_stokes2}).
Such Stokes~\V profiles obviously infer that the upflows in the inner penumbra posses a strong magnetic field
compared with that in the surrounding penumbral atmosphere, and is apparently inconsistent with numerical simulations 
that infer upflowing gas with a weaker magnetic field (see Section~\ref{subsubsection:unified_model}). Similar 
conclusions have been reached by \cite{Tritschler2007} and \cite{Borrero2008ApJ} for the outer penumbra.

\subsubsection{Unified picture and numerical simulations of the penumbra}
\label{subsubsection:unified_model}

As described in previous sections, two opposing ideas have been proposed to account for the 
penumbral uncombed structures (see Section~\ref{subsubsection:spine_intraspine}): 
\textit{the embedded flux-tube model} and \textit{the field-free gap model}.
The \textit{embedded flux-tube model}, or the \textit{rising hot flux tube} with the dynamic evolution of the flux tube,
explains a number of observational aspects about the fine scale features of the penumbra such as the origin of Evershed flow, 
inward migration of penumbral grains, and asymmetric Stokes profiles observed in penumbra (see Section~\ref{subsubsection:ncp}),  
but it faces difficulties when attempting to explain the heat transport to the penumbral surface (see Section~\ref{subsubsection:heating}).
Some observational results support the finite vertical extension of the flux tube; i.e., 
Doppler shifts in multiple spectral lines formed at different height
infer elevated Evershed flow channels \citep{Ichimoto1987,Rimmele1995,stanchfield1997}, 
and some SIR-like inversions\epubtkFootnote{SIR-like inversions refer to inversions of spectropolarimetric data
where the physical parameters are allowed to vary with optical depth $\ve{X}(\tau_c)$ (Equation~\ref{equation:x}).
This is discussed in some detail in Section~\ref{subsection:tools}.} of Stokes profiles in spectral lines claim to have detected 
the lower boundary of the flux tube \citep{Borrero2006}. However, these results do not necessary provide 
a concrete evidence of the presence of thin and elevated flux tubes in the inner or middle penumbra 
and, actually, most observations suggest a monotonic increase in the magnitude of the Evershed flow towards the 
deeper photospheric layers, while finding no evidence for a lower discontinuity of the magnetic field in the observable layers 
(see also Figures~\ref{figure:pen_depth_jurcak} and \ref{figure:pen_depth_borrero}). 
The flux-tube models by \citet{Borrero2007} and \citet{RuizCobo2008}
also suggest that penumbral flux tubes are not necessarily thin since the $\tau_c=1$ level 
is located inside the tubes and, therefore, the lower boundary will not be visible.
Thus, there is no definite observational evidence for the presence of a lower boundary in the flux tubes, at least in inner 
and middle penumbra, and the concept of narrow, elevated flux tubes embedded in the penumbra is not a scheme with strong observational bases.

In the \textit{field-free gap penumbral model}, the gap is formed by a convecting hot and field-free gas protruding upward
into the background oblique magnetic fields of the penumbra, and is supposed to be the region that harbors the Evershed flow.
Contrary to the previous flux-tube model, the \textit{field-free gap} penumbral model 
\citep{SpruitScharmer2006,ScharmerSpruit2006} has an advantage in explaining the
heat transport to penumbral surface and, possibly, in explaining the twisting appearance
of penumbral bright filaments. It does so thanks to the \textit{azimuthal/overturning convective pattern} described in 
Section~\ref{subsubsection:heating} (see also lower panel in Figure~\ref{figure:convection_penumbra}).
However, it does not address the origin of the Evershed flow nor physical nature 
of the inner and outer ends of penumbral filaments. Furthermore, it is obvious from the highly Doppler-shifted 
polarization signals in spectral lines in penumbra that the flowing gas is not field-free.
From the SIR inversion of a spectro-polarimetric data, \cite{Borrero2008ApJ} argued that the magnetic field strength 
in the Evershed flow channels (intraspines) increases with the depth below $\tau_c=0.1$ level, 
and there exist strong magnetic fields near the continuum formation level that is not
compatible with the \textit{field-free gap model}.

Thus, both the embedded flux-tube model and the field-free gap model have 
their own advantages but also have considerable shortcomings. It would be natural, therefore, to modify 
these two penumbral models as follows; in the flux-tube model, we may consider vertically elongated flux tubes 
(or slabs) rather than the round cross section, and add an \textit{azimuthal/overturning} convective flow pattern inside the
penumbral filaments in addition to the horizontal Evershed flow. This would allow this model  to transport 
sufficient energy through convection as to explain the penumbral brightness. In the \textit{field-free gap}
model we propose to add a rather strong, $\simeq$~1000~G (yet still weaker than in the \textit{spines}),
and nearly horizontal magnetic field inside the field-free gap. This has the consequence that the
presence of rather strong horizontal magnetic field inside the gap allows this model
to explain the observed net circular polarization in sunspots \cite[see Section~\ref{subsubsection:ncp}; see also][]{BorreroSolanki2010}
as well as featuring a magnetized Evershed flow.

After the proposed modifications, we find that there are no fundamental differences between the two pictures
as far as geometry of the inner penumbra is concerned. Note that in both the \textit{rising flux-tube model} and 
the \textit{field-free gap model}, the rising motions of hot gas in the Evershed flow channels are 
driven by the buoyancy force in the superadiabatic stratification of the penumbral atmosphere.
Such unified picture has been already discussed by \citet{Scharmer2008ApJ}, \citet{Borrero2009Ch}, and 
\citet{Ichimoto2010}. In this concept, the Evershed flow could be understood as a consequence of 
the thermal convection with the gas flow deflected horizontally outward under a strong and inclined magnetic field.

Although the proposed modifications are observationally driven, recent 3D MHD simulations of sunspots
\citep{Heinemann2007,Rempel2009,Rempel2009b,Kitiashvili2009,Rempel2011} present a magnetic field
configuration that closely follows the inner structure for penumbral filaments that we have proposed above.
The results from these simulations are able to reproduce the radial filamentary structure of the penumbra 
as seen in continuum images, the uncombed structure of the magnetic field, Evershed outflows along the filaments 
with a nearly horizontal magnetic field, and overturning convective motions in upwelling plumes. In addition, a detailed inspection of the 
numerical simulations provides great insights on the physical processes taking place in the penumbra.
According to \cite{Rempel2011}, the Evershed flow is driven by vertical 
pressure forces in upflows that are deflected into the horizontal direction through the
Lorentz-force generated by the horizontally stretched magnetic fields in flow channels, 
and the radial flow velocity reaches up to 8~\kms at the depth of $\tau_c=1$ with
a rapid decline toward the higher atmospheric layers.

\enlargethispage{\baselineskip}
Figure~\ref{figure:rempel2011a} shows a vertical cross section of the filaments in the inner penumbra from the MHD simulations
by \cite{Rempel2011}. Remarkable features are the sharp enhancement of the radial component of
the magnetic field around $\tau_c=1$ level, where the upflow of convective gas protrudes and creates a narrow boundary 
layer with a concentration of a strong horizontal Lorenz force that acts as the engine that drives the horizontal Evershed flow.
The connectivity, or the presence of the outer footpoints, of the magnetic field in the flowing channel are rather 
a consequence of the fast outflow than its cause as is assumed in the siphon-flow picture.

\epubtkImage{rempel2011a_html.jpg}{
\begin{figure}[htb]
\centerline{\includegraphics[width=0.9\textwidth]{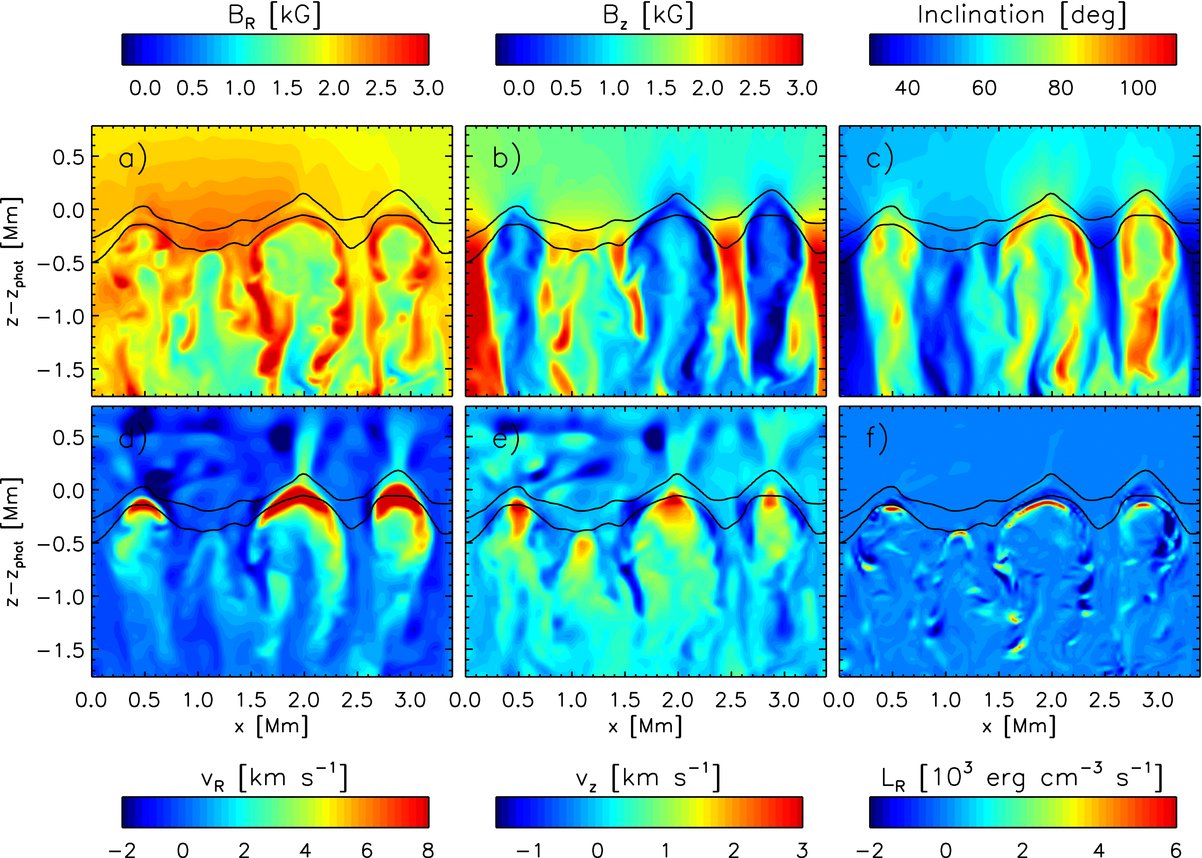}}
\caption{Vertical variation, according to the MHD simulations by
  \cite{Rempel2011}, of the physical parameters across a cut
  perpendicular to the penumbral filaments in the inner
  penumbra. Displayed are: a) radial and b) vertical components of the
  magnetic field vector, c) inclination of the magnetic field vector
  with respect to the vertical direction \textit{z}. The bottom panels show:
  d) radial and e) vertical components of the velocity vector, and
  f) the energy conversion by the component, along the direction of
  the filaments, of the Lorentz force. The two solid lines indicate
  the $\tau_c=1$ and $\tau_c=0.01$ levels
  \citep[from][reproduced by permission of the AAS]{Rempel2011}.}
\label{figure:rempel2011a}
\end{figure}}

Thus, the recent MHD simulations have begun to reproduce many details of fine scale dynamics 
and structure of the magnetic field observed in the penumbra. The most essential physical processes 
that form the penumbra take place near or beneath the $\tau_c=1$. Therefore, the detection of the vertical 
gradients of the magnetic field and velocity vectors in the deep layers of the penumbra is an important target 
for future observations. Another important target for observations is to find the downflows or returning (inward) flows
that could be associated with the \textit{azimuthal/overturning convection} but still not have been detected yet at the spatial resolutions
of Hinode \citep{Franz2009} and of the Swedish 1-m Solar Telescope \citep{BellotRubio2010}. This had been already
discussed in Section~\ref{subsubsection:vertical_motions}.

\enlargethispage{\baselineskip}
The dynamic interaction between magnetic fields and granular convection around outer edge of penumbra is discussed 
as the formation mechanism of the interlocking comb structure of penumbra in the simulations by \citet{Thomas2002}, \citet{Weiss2004}, and 
\citet{Brummell2008}. In this picture, the magnetic fields in the intraspines, which plunge below the solar surface 
near the edge of the spot (see Section~\ref{subsubsection:spine_intraspine}), are created as a consequence of the submergence 
of the magnetic field lines due to the \textit{downward pumping}-mechanism by small-scale granular convection outside the sunspot.
Stochastic flows could be driven along such magnetic fields by the convective collapse caused by the the localized 
submergence of the magnetic fields. This scenario may capture an essential point of the dynamical convective process,
but it is questionable if the entire penumbral structure is controlled by such processes taking place outside the sunspots.

\newpage



\section{Acknowledgements}
\label{section:acknowledgements}

The authors are very grateful to K.D.~Leka and R.H.~Cameron,
as well as to two anonymous referees, for carefully reading the
manuscript and their many suggestions that helped improve this
paper. This work has made extensive use of Hinode/SOT
data. Hinode is a Japanese mission developed and launched by
ISAS/JAXA, collaborating with NAOJ as a domestic partner, NASA and
STFC (UK) as international partners. Scientific operation of the
Hinode mission is conducted by the Hinode science team organized at
ISAS/JAXA. This team mainly consists of scientists from institutes in
the partner countries. Support for the post-launch operation is
provided by JAXA and NAOJ (Japan), STFC (U.K.), NASA, ESA, and NSC
(Norway). This research has made use of NASA's \textit{Astrophysics
  Data System}. Finally, we would like to thank our colleagues and
\textit{Astrophysical Journal}, \textit{Astronomy and Astrophysics},
\textit{Annual Reviews in Astronomy and Astrophysics}, and
\textit{Nature} for granting us the rights to reproduce many
previously published figures.

\newpage
\section{Appendix: Coordinate Transformation}
\label{section:appendix}

In this Appendix we describe the coordinate transformation that
allows us to solve the 180\textdegree-ambiguity in the azimuth
$\varphi$ of the magnetic field vector (see
Section~\ref{subsubsection:180deg}). To that end let us define three
different reference frames: $\{\ex,\ey,\ez\}$, $\{\ea,\eb,\ep\}$, and
$\{\els,\exs,\eys\}$. The first frame is a Cartesian frame centered at
the Sun's center. The second frame is a curvilinear frame located at
the point of observation P, where $\ep$ is the unit vector that is
perpendicular to the plane tangential to the solar surface at this
point. This plane contains the unit vectors $\ea$ and $\eb$. This is
the so-called \textit{local reference frame}. The third reference
frame is the \textit{observer's reference frame}. It is centered at
the point of observation P, with the unit vector $\els$ referring to
the line-of-sight. As mentioned in Section~\ref{subsection:tools} the
inversion of the radiative transfer equation~(\ref{equation:rte})
yields the magnetic field vector in the observer's reference frame:
\begin{eqnarray}
\label{equation:bfieldinlos}
\ve{B} &=& B \cos\gamma \els + B\sin\gamma\cos\varphi\exs + B\sin\gamma\sin\varphi\eys = \notag \\ 
&& \left(\begin{array}{ccc} B\cos\gamma & 0 & 0 \\
0 & B\sin\gamma\cos\varphi  & 0 \\ 
0 & 0 & B\sin\gamma\sin\varphi \end{array}\right)
\left(\begin{array}{c} \els \\ 
\exs \\ \eys  
\end{array}\right) = \hat{\mathcal{B}} \left(\begin{array}{c} \els \\ 
\exs \\ 
\eys \end{array}\right).
\end{eqnarray}
The key point to solve the 180\textdegree-ambiguity is to find the
coordinates of the magnetic field vector into the local reference
frame, where we will apply the condition that the magnetic field
vector in a sunspot must spread radially outwards. In order to do so,
we will obtain the coordinates of $\{\ea,\eb,\ep\}$ and
$\{\els,\exs,\eys\}$ into the reference frame at the Sun's center:
$\{\ex,\ey,\ez\}$.

\begin{enumerate}

\item Let us focus first on $\{\ex,\ey,\ez\}$: the unit vectors defining a coordinate system at the Sun's center.
The unit vector $\ez$ connects the Earth with the Sun's center, while $\ey$ corresponds to the proyection of the Sun's
rotation axis on the plane perpendicular to $\ez$ (see Figure~\ref{figure:earthsun}). In this coordinate system, the vector 
connecting the observer's and the Sun's center is $\ve{OE}$ (E means Earth):
\begin{equation}
\ve{OE} = A \ez,
\end{equation}
where $A=1.496 \times 10^{11}\mathrm{\ m}$ is the Astronomical
Unit. Let us now suppose that we observe a point P on the solar
surface. The coordinates of this point are usually given in the
reference frame of the observer with the values $X_{c}$ and $Y_{c}$
(usually in arcsec). These two values refer to the angular distances
of the point P measured from the center of the solar disk as seen from
the Earth (see Figure~\ref{figure:earthsun}). From this figure we can
extract two triangles: $\widetriangle{\mathrm{OEP}_1}$ and
$\widetriangle{\mathrm{OEP}_2}$ (see Figure~\ref{figure:earthsun})
that can be employed to obtain the values of $\alpha$ and $\beta$
given $X_{c}$ and $Y_{c}$. From the latter triangle we obtain:
\begin{equation}
\label{equation:sinalfa}
\sin\alpha = \frac{A}{R_{\odot}} \tan Y_{c},
\end{equation}

where $R_{\odot}=6.955 \times 10^8\mathrm{\ m}$ is the Sun's
radius. Once $\alpha$ has been obtained, we can employ the cosine
theorem in the triangle $\widetriangle{\mathrm{OEP}_1}$ to determine $\beta$
as:
\begin{eqnarray}
\label{equation:cosbeta1}
\cos\beta &=& \frac{-b \pm \sqrt{b^2-4ac}}{2a}, \\
\label{equation:cosbeta2}
a &=& \frac{R_{\odot}^2\cos^2\alpha}{\sin^2 X_c},\\
\label{equation:cosbeta3}
b &=& -2 R_{\odot}A\cos\alpha,\\
\label{equation:cosbeta4}
c &=& A^2-\frac{R_{\odot}^2\cos^2\alpha}{\sin^2 X_c}.
\end{eqnarray}
Note that the obtained values for $\alpha$ and $\beta$ will have to be
modified depending upon the signs of $X_c$ and $Y_c$ (which determine
the quadrant on the solar disk). Equation~(\ref{equation:cosbeta1}) shows
that there are two possible values for $\beta$, however, one of them
always corresponds to an angle $| \beta | > \pi/2$ and, therefore, can
be neglected. It is also important to bear in mind that in
Figure~\ref{figure:earthsun} the points labeled as $P_1$ and $P_2$ are
the projections of the observed point on the solar surface, $P$, onto
the planes $y=0$ and $z=0$ respectively. In fact, once that $\beta$
and $\alpha$ are known, the coordinates of $P$ in the reference frame
of $\{\ex,\ey,\ez\}$ (the vector $\ve{OP}$) can be written as:
\begin{equation}
\label{equation:pointp}
\ve{OP} = R_{\odot}\cos\alpha\sin\beta\ex+R_{\odot}\sin\alpha\ey+R_{\odot}\cos\alpha\cos\beta\ez.
\end{equation}
Another vector that will be useful later is the unit vector from the
Earth to the observation point P. This can be written as follows:
\begin{equation}
\label{equation:los}
\ve{PE}= -R_{\odot}\cos\alpha\sin\beta\ex-R_{\odot}\sin\alpha\ey+[A-R_{\odot}\cos\alpha\cos\beta]\ez.
\end{equation}

\epubtkImage{coordinates.png}{%
  \begin{figure}[htb]
    \centerline{\includegraphics[width=10cm]{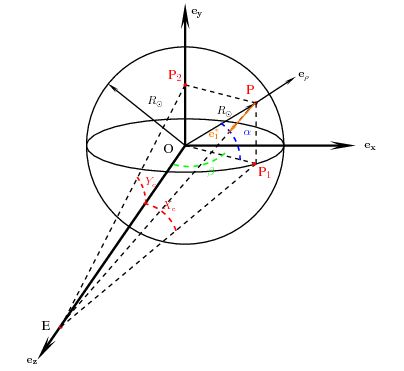}}
    \caption{Sketch showing the geometry of the problem and the different
      reference frames employed in
      Section~\ref{subsubsection:180deg}. The reference frame
      $\{\ex,\ey,\ez\}$ is centered at the Sun's center `O'. The
      observed point at the Sun's surface is denoted by 'P'. The
      observer is located at the point `E', denoting the Earth. The
      vector $\ve{OP}$ is parallel to $\ep$, while $\ve{EP}$ is parallel
      to $\els$.}
    \label{figure:earthsun}
\end{figure}}

\item The next step is to realize that $\ve{OP}$
  (Equation~(\ref{equation:pointp})) is parallel to the radial vector in
  the solar surface and, therefore, it is perpendicular to the
  tangential plane on the solar surface at the point of
  observation. This means that the vector $\ep$ (see
  Figure~\ref{figure:earthsun}) belonging to the \textit{local frame}
  can be obtained as:
\begin{eqnarray}
\label{equation:normalvec}
\ep &=& \frac{\ve{OP}}{|OP|} = \cos\alpha\sin\beta\ex+\sin\alpha\ey+\cos\alpha\cos\beta\ez \notag \\
    &=& e_{\rho x} \ex + e_{\rho y} \ey + e_{\rho z} \ez.
\end{eqnarray}
With this data we can now define the \textit{local reference frame} as
$\{\ea,\eb,\ep\}$. This coordinate system is defined on the plane that
is tangential to the solar surface at the point P, with $\ep$ being
perpendicular to this plane, and $\ea$ and $\eb$ being contained in
this plane. $\ea$ and $\eb$ are the tangential vectors to the
$\beta=$cons and $\alpha=$cons curves, respectively. The relation
between the \textit{local reference frame} and the one located at the
Sun's center can be easily derived from Figure~\ref{figure:earthsun}:
\begin{eqnarray}
\label{equation:trans2}
\left(\begin{array}{c} \ea \\ \eb \\ \ep  \end{array}\right) = \left( \begin{array}{ccc} -\sin\alpha\sin\beta & \cos\alpha & 
-\sin\alpha\cos\beta \\ \cos\beta & 0 & -\sin\beta \\ \cos\alpha\sin\beta & \sin\alpha & \cos\alpha\cos\beta \end{array}\right)
 \left(\begin{array}{c} \ex \\ \ey \\ \ez \end{array}\right) = \hat{\mathcal{M}} \left(\begin{array}{c} \ex \\ \ey \\ \ez 
\end{array}\right).
\end{eqnarray}

\item The third reference frame we have mentioned is the
  \textit{observer's reference frame}: $\{\els,\exs,\eys\}$. Note that
  the first of the unit vectors can be directly obtained (see
  Figure~\ref{figure:earthsun} and Equation~(\ref{equation:los})) as:
\begin{equation}
\label{equation:losunit}
\els = \frac{\ve{PE}}{|PE|} = e_{lx}^{*}\ex + e_{ly}^{*}\ey  + e_{lz}^{*}\ez.
\end{equation}

The heliocentric angle $\Theta$ is defined as the angle between the
normal vector to the solar surface at the point of observation P (aka
$\ep$) and the line-of-sight (aka $\els$). Therefore, the scalar
product between these two already known vectors
(Equations~(\ref{equation:los}) and (\ref{equation:normalvec})) yields the
heliocentric angle $\cos\Theta = \ep \cdot \els$:
\begin{equation}
\label{equation:helioangle}
\cos\Theta = \frac{A\cos\alpha\cos\beta-R_{\odot}}{|PE|}.
\end{equation}

The other two vectors of this coordinate system, i.e., $\exs$ and $\eys$
must be perpendicular to the line-of-sight. The first one, $\exs$, can
be obtained from Equations~(\ref{equation:los}), (\ref{equation:normalvec}),
and (\ref{equation:losunit}) as:
\begin{eqnarray}
\label{equation:exs}
\exs = \frac{\ep \times \els}{|\ep \times \els|} &=& \frac{1}{|\ep \times \els|} \bigg{\{} [e_{\rho y} e_{lz}^{*}-
 e_{\rho z} e_{ly}^{*}]\ex + [e_{\rho z} e_{lx}^{*}-e_{\rho x} e_{lz}^{*}]\ey+
[e_{\rho x} e_{ly}^{*}-e_{\rho y} e_{lx}^{*}]\ex \bigg{\}}\notag \\
 &=& e_{xx}^{*}\ex + e_{xy}^{*}\ey + e_{xz}^{*}\ez.
\end{eqnarray}
Finally, the vector $\eys$ can be obtained employing the following conditions:
\begin{eqnarray}
\left\{\begin{array}{c}
 \exs \cdot \eys = 0 \\
 \els \cdot \eys = 0 \\ 
 |\eys|=1 
\end{array}\right. \notag \\
 \eys = e_{yx}^{*}\ex + e_{yy}^{*}\ey + e_{yz}^{*} \ez.
\end{eqnarray}
This represents a system of three equations with three unknowns:
$e_{yx}^{*},e_{yy}^{*},e_{yz}^{*}$. These represent the proyections of
the $\eys$ unit vector on the base located at the Sun's center:
$\{\ex,\ey,\ez\}$. Solving the previous system of equations yields the
following solutions:
\begin{eqnarray}
\label{equation:comp1}
e_{yz}^{*} = \left[ \left(\frac{e_{xz}^{*}e_{ly}^{*}-e_{xy}^{*}e_{lz}^{*}}{e_{xy}^{*}e_{lx}^{*}-e_{xx}^{*}e_{ly}^{*}}\right)^2
+\left(\frac{e_{xz}^{*}e_{lx}^{*}-e_{xx}^{*}e_{lz}^{*}}{e_{xy}^{*}e_{lx}^{*}-e_{xx}^{*}e_{ly}^{*}}\right)^2+1\right]^{-\frac{1}{2}}, 
\end{eqnarray}

\begin{eqnarray}
\label{equation:comp2}
e_{yx}^{*} = e_{yz}^{*} \left(\frac{e_{xz}^{*}e_{ly}^{*}-e_{xy}^{*}e_{lz}^{*}}  {e_{xy}^{*}e_{lx}^{*}-e_{xx}^{*}e_{ly}^{*}}\right),
\end{eqnarray}

\begin{eqnarray}
\label{equation:comp3}
e_{yy}^{*} = -e_{yz}^{*} \left(\frac{e_{xz}^{*}e_{lx}^{*}-e_{xx}^{*}e_{lz}^{*}}  {e_{xy}^{*}e_{lx}^{*}-e_{xx}^{*}e_{ly}^{*}}\right).
\end{eqnarray}
All the three components of $\eys$ -- Equations~(\ref{equation:comp1}),
(\ref{equation:comp2}), and (\ref{equation:comp3}) -- can be obtained through
Equations~(\ref{equation:losunit}) and (\ref{equation:exs}). We are,
therefore, able to construct the matrix $\hat{\mathcal{N}}$, which
transforms the \textit{observer's reference frame} into the reference
frame at the Sun's center:
\begin{eqnarray}
\label{equation:trans1}
\left(\begin{array}{c} \els \\ \exs \\ \eys  \end{array}\right) = \left( \begin{array}{ccc} e_{lx}^{*} & e_{ly}^{*} & 
e_{lz}^{*} \\ e_{xx}^{*} & e_{xy}^{*} & e_{xz}^{*} \\ e_{yx}^{*} & e_{yy}^{*} & e_{yz}^{*} \end{array}\right)
 \left(\begin{array}{c} \ex \\ \ey \\ \ez \end{array}\right) = \hat{\mathcal{N}} \left(\begin{array}{c} \ex \\ \ey \\ \ez 
\end{array}\right).
\end{eqnarray}
An important point to consider is that the \textit{local reference
  frame}, $\{\ea,\eb,\ep\}$, and the \textit{observer's reference
  frame}, $\{\els,\exs,\eys\}$, vary with the observed point P on the
solar surface. This means that each of these coordinates systems must
be recalculated for each point of an observed 2-dimensional map.

\item We can now express the magnetic field vector in
  Equation~(\ref{equation:bfieldinlos}), in the \textit{local reference
    frame}:
\begin{eqnarray}
\label{equation:bfieldinloc}
\ve{B} = \underbrace{\hat{\mathcal{B}} \left(\begin{array}{c} \els \\ \exs \\ \eys  \end{array}\right)}_{\mathrm{Eq.~(\ref{equation:bfieldinlos})}} 
= \hat{\mathcal{B}} \underbrace{\hat{\mathcal{N}}\left(\begin{array}{c} \ex \\ \ey \\ \ez  \end{array}\right)}_{\mathrm{Eq.~(\ref{equation:trans1})}} 
= \hat{\mathcal{B}} \hat{\mathcal{N}} \underbrace{\hat{\mathcal{M}}^{-1} \left(\begin{array}{c} \ea \\ \eb \\ \ep  \end{array}\right)}_{\mathrm{Eq.~(\ref{equation:trans2})}}
= B_{\alpha} \ea + B_{\beta} \eb + B_{\rho} \ep.
\end{eqnarray}
Note that because of the 180\textdegree-ambiguity in the determination of
$\varphi$, the two possible solutions in the magnetic field vector
also exist (in an intricate manner) in the local reference frame
$\{\ea,\eb,\ep\}$: $\ve{B}(\varphi)$ and $\ve{B}(\varphi+\phi)$
(Equation~(\ref{equation:bfieldinloc})). In order to distinguish which
one of the two is the correct one, we will consider that the magnetic
field in a sunspot is mostly radial from the center of the
sunspot. This means that we will take the solution that minimizes
Equation~(\ref{equation:parallel}). To evaluate that equation we already
know the coordinates of the magnetic field vector $\ve{B}$ in the
\textit{local reference frame}. However, we must still find the
coordinates, in this same reference frame, of the vector $\ve{r}$
which, as already mentioned in Section~\ref{subsubsection:180deg}, is
the vector that connects the center of the umbra (denoted by U) with
the point P of observation: 

\begin{eqnarray}
\label{equation:opou}
\ve{r} = \ve{OP}-\ve{OU} &=& R_{\odot}[(\cos\alpha_p\sin\beta_p-\cos\alpha_u\sin\beta_u)\ex+(\sin\alpha_p-\sin\alpha_u)\ey+ \notag \\ 
                          && (\cos\alpha_p\cos\beta_p-\cos\alpha_u\cos\beta_u)\ez] = r_x \ex + r_y \ey + r_z \ez \;,
\end{eqnarray}
where we have distinguished between ($\alpha_p$,$\beta_p$) and
($\alpha_u$,$\beta_u$) to differentiate the coordinates of the point of
observation P and the umbral center U, respectively. For convenience,
we re-write Equation~(\ref{equation:opou}) as follows:
\begin{eqnarray}
\label{equation:opou2}
\ve{r} = \underbrace{\left( \begin{array}{ccc} r_z & 0 & 0 \\ 0 & r_y & 0 \\ 0 &  0 & r_z \end{array}\right)}_{\hat{\mathcal{R}}} \left(\begin{array}{c} \ex \\ \ey \\ \ez 
\end{array}\right) = \hat{\mathcal{R}} \left(\begin{array}{c} \ex \\ \ey \\ \ez \end{array}\right) \;.
\end{eqnarray}
Note that Equations~(\ref{equation:opou}) and (\ref{equation:opou2}) refer
to the reference frame at the Sun's center. However, in order to
calculate its scalar product with the magnetic field vector
(Equation~(\ref{equation:parallel})) we need to express it in the
\textit{local reference frame} $\{\ea,\eb,\ep\}$ at the point of
observation P:
\begin{eqnarray}
\label{equation:opou3}
\ve{r} = \hat{\mathcal{R}} \left(\begin{array}{c} \ex \\ \ey \\ \ez \end{array}\right) = \hat{\mathcal{R}} \underbrace{\hat{\mathcal{M}}^{-1} \left(\begin{array}{c} 
\ea \\ \eb \\ \ep \end{array}\right)}_{\mathrm{Eq.~(\ref{equation:trans2})}} \;,
\end{eqnarray}
where the inverse matrix $\hat{\mathcal{M}}^{-1}$
(Equation~(\ref{equation:trans2})) must be obtained employing $\alpha_p$
and $\beta_p$ (coordinates on the solar surface for the observed point
P). Once $\ve{B}$ and $\ve{r}$ are known in the \textit{local
  reference frame} (Equations~(\ref{equation:bfieldinloc}) and
(\ref{equation:opou3}), respectively), it is now possible to
evaluate Equation~(\ref{equation:parallel}). This allows us to determine
which solution for the azimuth of the magnetic field, $\varphi$ or
$\varphi+\pi$, yields a magnetic field vector which is closer to be
radially aligned.

\end{enumerate}
\newpage



\bibliography{refs}

\end{document}